\documentclass[1p]{elsarticle}
\makeatletter
\def\Hline{%
\noalign{\ifnum0=`}\fi\hrule \@height 1.5pt \futurelet
\reserved@a\@xhline}
\makeatother
\usepackage[bookmarks=false]{hyperref}
\usepackage{graphicx}
\usepackage{float}
\usepackage{subfigure}% subcaptions for subfigures
\usepackage{subfigmat}% matrices of similar subfigures,
\usepackage{amsmath}
\usepackage{enumitem}
\usepackage{lipsum}

\setlist[itemize]{leftmargin=*}
\usepackage{amssymb}
\usepackage{xcolor}
\usepackage{rotating} % Rotating the figures but keeps the page in portrait mode
\usepackage{pdflscape} % Rotates single page into Landscape mode
\usepackage{silence}
\usepackage{ulem,cancel}

\usepackage{setspace,lipsum}
\begin{document}

\begin{frontmatter}


% use the tnoteref command within \title for footnotes;
% use the tnotetext command for theassociated footnote;
% use the fnref command within \author or \address for footnotes;
% use the fntext command for theassociated footnote;
% use the corref command within \author for corresponding author footnotes;
% use the cortext command for theassociated footnote;
% \author{Name\corref{cor1}\fnref{label2}}
% \address{Address\fnref{label3}}

\title{{\color{black}First order hyperbolic approach for Anisotropic Diffusion equation}}

% Group authors per affiliation:
\author[AA_address]{Amareshwara Sainadh Chamarthi\fnref{fn1} \cortext[cor1]{Corresponding author. \\ 
E-mail address: s.chamarthi@al.t.u-tokyo.ac.jp (Amareshwara Sainadh  Ch.).}}
\author[BB_address]{Hiroaki Nishikawa\fnref{fn1}}
\author[AA_address]{Kimiya Komurasaki}
\address[AA_address]{Department of Aeronautics and Astronautics, The University of Tokyo, 7-3-1 Hongo, Bunkyo, Tokyo 113-8656, Japan}
\address[BB_address]{National Institute of Aerospace, 100 Exploration Way, Hampton, VA 23666, USA}
\fntext[fn1]{These authors contributed equally to the work.}
\begin{abstract}
In this paper, we present a high order finite difference solver for anisotropic diffusion problems based on the first-order hyperbolic system method. In particular, we demonstrate that the construction of a uniformly accurate fifth-order scheme that is independent of the degree of anisotropy is made straightforward by the hyperbolic method with an optimal length scale. We demonstrate that the gradients are computed simultaneously to the same order of accuracy as that of the solution variable by using weight compact finite difference schemes. Furthermore, the approach is extended to improve further the simulation of the magnetized electrons test case previously discussed in Refs.[J. Comput. Phys., 284 (2015) 59-69 and 374 (2018) 1120-1151]. Numerical results indicate that these schemes are capable of delivering high accuracy and the proposed approach is expected to allow the hyperbolic method to be successfully applied to a wide variety of  linear and nonlinear problems with anisotropic diffusion.
\end{abstract}

\end{frontmatter}

%\def\columnseprulecolor{\color{black}} %Column color
%    \title{First order hyperbolic approach for Anisotropic diffusion on non-aligned coordinates}
%\begin{document}
%   \author{Amareshwara Sainadh Ch.}
%    \date{\today}  
%    \maketitle
    
\section{Introduction}
Anisotropic diffusion occurs in many physical applications in which the rate of diffusion in a certain direction can be orders of magnitude higher than the other. Thermal conductivity in fusion plasmas \cite{braginskii1965transport}, image processing \cite{perona1990scale}, biological process, and medical imaging \cite{basser2002diffusion} are some of the examples. Diffusion tensors can be extremely anisotropic in magnetized and high-temperature plasma process which poses a challenging problem for computational simulations. Due to extreme anisotropy, the diffusion phenomenon is effectively aligned with the magnetic field lines. Such alignment may lead to the parallel diffusion coefficient being orders of magnitude, up to $10^{9}$ in fusion plasmas, larger than perpendicular diffusion coefficient.

These extreme anisotropies can have strict requirements on the numerical methods used to model the anisotropic diffusion and magnetohydrodynamics equations since any misalignment of the grid can lead to significant numerical diffusion in the perpendicular direction to the magnetic field line. One approach is to solve the equations on a computational grid that is aligned with the applied magnetic field which can automatically take care of the directionality of the diffusion coefficients. This approach, known as the magnetic field-aligned mesh (MFAM), has been successfully used in modeling plasma propulsion devices and fusion plasmas \cite{marchand1996carre,degtyarev1986methods}. This approach can have problems in the case of crossing field lines where local non-alignment unavoidable. Also, one has to reconstruct the mesh for the situations involving time-varying field lines, or magnetic field induce flow, and it can be a cumbersome task. It is beneficial to develop numerical methods that are applicable even on non-aligned grids with minimal numerical diffusion.

Gunter et al. \cite{Gunter} has developed ``Symmetric'' and ``asymmetric'' scheme which are simple and easy to implement with low perpendicular numerical pollution. Sovinec et al. \cite{sovinec2004nonlinear} have used {\color{black} a} higher-order finite element method in the direction of the larger diffusion coefficient to reduce the numerical diffusion. {\color{black} On the other hand, Degond et al. have developed Asymptotic preserving schemes (AP) in a series of papers \cite{degond2010asymptotic,degond2012asymptotic} for the strongly anisotropic diffusion equation in which the equation itself is split into two parts, a limit problem for infinite anisotropy and the original singular perturbation problem. Mentrelli and Negulescu \cite{mentrelli2012asymptotic} further extended the AP schemes to the generalized nonlinear cases, and Chacon et al. \cite{Chacon2014} applied a more generic semi-Lagrangian approach to unsteady anisotropic diffusion.} Van Es et al. \cite{van2014finite} has developed a second-order accurate aligned finite difference method for anisotropic diffusion problems. Another important aspect of the anisotropic diffusion equation is the positivity of the temperature or space potential.  A numerical scheme should be robust against the development of nonphysical negative temperatures during the simulation and must satisfy positivity and monotonicity. For example, negative temperatures, in Hall thruster modeling can lead to decreased electron currents and negative Joule power density near the cathode region. Positivity has been attained by nonlinear schemes proposed in the literature, imposed via limiters by Kuzmin et al.\cite{kuzmin} and Sharma and Hammett \cite{sharma2007preserving}, but limiters can lead to low-order spatial accuracy and are also limited to moderate anisotropies of the order $\approx 10^3$. 

Recently an upwind formulation for diffusion equation has been introduced by Nishikawa \cite{Nishikawa2007} based on the residual-distribution (RD) method and later with a finite-volume method \cite{nishikawa_onetwothree_diffusion:JCP2014}. The mathematical strategy of this approach is to split the second order partial differential equation into a set of first-order differential equations by adding new variables and pseudo-time advancement terms such that the diffusion equation is reformulated as a hyperbolic system. This radical approach has been shown to offer several advantages over conventional methods, such as accelerated convergence for steady state solution and higher order of accuracy for both primary and gradient variables, as demonstrated 
for diffusion \cite{nishikawa_onetwothree_diffusion:JCP2014}, the incompressible/compressible Navier-Stokes equations \cite{nishikawa_hyperbolic_ns:AIAA2014,NakashimaWatanabeNishikawa_AIAA2016-1101}, third-order dispersion equations \cite{jcp2016v321pp593-605}, an incompressible magnetohydrodynamics model \cite{BatyNishikawa:MNRAS2016},  
an elliptic distance-function model \cite{WatsonTrojakTucker_aiaa2018-4261}, and so on. The original approach of Nishikawa has been further extended to a constant diffusion tensor by Lou et al. \cite{lou2017reconstructed, Lou2019} discretized with the reconstructed discontinuous Galerkin scheme (rDG). The hyperbolic approach has also been implemented for an anisotropic diffusion equation, based on high-order finite-volume schemes in Ref.\cite{chamarthi2018high} (see Example 4.1.5 in the reference). Even though fifth order accuracy was demonstrated for the test cases considered, it has been known that the method yields significantly large errors for increasing degree of anisotropy. To address this issue, we will extend the analysis in Ref.\cite{nishikawa2018dimensional} and derive an optimal length scale that defines a relaxation time in a hyperbolic formulation of anisotropic diffusion equations. As we will demonstrate, the optimal
length scale renders the method independent of the degree of anisotropy, thus achieving the same level of errors for a wide range of anisotropy. The method is developed based on a local-preconditioning formulation and extended to variable-coefficient and nonlinear anisotropic diffusion equations.

The rest of the paper is organized as follows. Hyperbolic approach and derivation of the preconditioned formulation for the anisotropic diffusion are described in Section 2. The cell-centered explicit and compact upwind finite difference upwind schemes and the implementation of boundary conditions are presented in Section 3. Several test cases in one and two-dimensional problems for anisotropic diffusion equation are presented in Section 4. These numerical experiments validate our numerical scheme and corroborate the high-order accuracy and implementation of boundary conditions. Finally, Section 5 summarizes our conclusions.

\section{{Hyperbolic formulation for anisotropic diffusion equation}}
\begin{figure}[H]
\centering
{\includegraphics[width=0.46\textwidth]{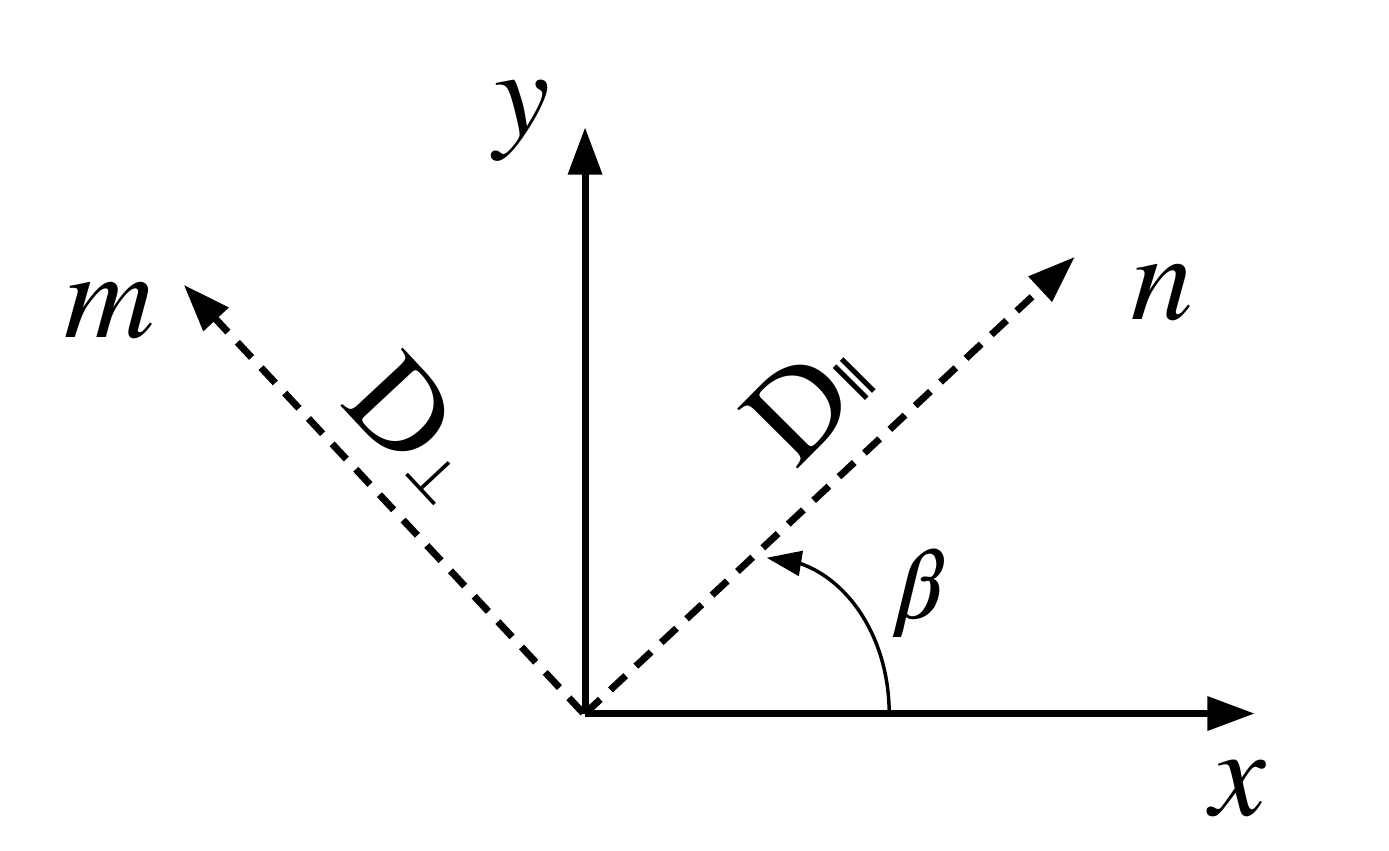}}
  \caption{Anisotropic diffusion with parallel and perpendicular diffusion coefficients on a non-aligned grid.}
  \label{fig:angle_mag}
\end{figure}

In this section, the hyperbolic approach for anisotropic diffusion equation is briefly explained. Anisotropic thermal diffusion equation can be described as
\begin{equation}
 \frac{\partial T}{\partial t}=-\nabla . \textbf{q}+ \textit{S}, \quad \textbf{q}=-(\textbf{D}.\nabla T),
 \label{eqn:aniso-original}
\end{equation}
where ${T}$, $\mathbf{q}$, $\mathbf{S}$ and $\mathbf{D}$ represent temperature, heat flux, source terms and the diffusion tensor respectively. For a two-dimensional problem the diffusion tensor is given by
\begin{align*}
\quad \mathbf{D} = \left[ \begin{array}{ccc}
D_{xx} & D_{xy} \\ \\
D_{yx} & D_{yy} \\
\end{array} \right]  = \left[\begin{matrix}D_{||} \cos^{2}{\left (\beta \right )} + D_{\perp}\sin^{2}{\left (\beta \right )} & \frac{1}{2} \left(D_{||} - D_{\perp}\right) \sin{\left (2 \beta \right )}\\ \\ \frac{1}{2} \left(D_{||} -D_{\perp}\right) \sin{\left (2 \beta \right )} & D_{||} \sin^{2}{\left (\beta \right )} + D_{\perp}\cos^{2}{\left (\beta \right )}\end{matrix}\right],
\end{align*}
where {$\mathbf {\beta}$} is the angle between the grid and the magnetic field lines and $D_{\perp}$ and $D_{||}$ are the perpendicular and parallel diffusion coefficients, shown in Fig. \ref{fig:angle_mag}, respectively. The diffusion tensor $\mathbf{D}$ is assumed to be positive definite: $ {\bf x}^t \mathbf{D} {\bf x} > 0 $ for an arbitrary non-zero vector ${\bf x}$.

%Anisotropic thermal diffusion equation can be described as
%\begin{equation}\label{eq:aniso-1}
% \frac{\partial T}{\partial t}=-\nabla . \textbf{q}+ \textit{S}, \quad \textbf{q}=-(\textbf{D}.\nabla T).
%\end{equation}
%where $\mathbf{T}$ represents temperature, $\mathbf{S}$  source term and $\mathbf{D}$ the diffusion tensor. For a two-dimensional problem the diffusion tensor is given by
%\begin{equation}
% \mathbf{D} = \left[ \begin{array}{ccc}
%D_{xx} & D_{xy} \\
%D_{yx} & D_{yy} \\
%\end{array} \right] = \left[\begin{matrix}D_{||} \cos^{2}{\left (\beta \right )} + D_{\perp}\sin^{2}{\left (\beta \right )} & \frac{1}{2} \left(D_{||} - D_{\perp}\right) \sin{\left (2 \beta \right )}\\\frac{1}{2} \left(D_{||} -D_{\perp}\right) \sin{\left (2 \beta \right )} & D_{||} \sin^{2}{\left (\beta \right )} + D_{\perp}\cos^{2}{\left (\beta \right )}\end{matrix}\right],
%\label{eq:tensor}
%\end{equation}
%where $D_{\perp}$ and $D_{||}$ represent the perpendicular and the parallel diffusion coefficients respectively, described in chapter-1 Fig. \ref{fig:angle_mag}. 
By dropping the time derivative and substituting \textbf{q} in the equation, we get the following steady state anisotropic diffusion equation
\begin{equation}
\begin{aligned}
0&=\nabla .(\textbf{D}\nabla T) + S,\\
&=\frac{\partial}{\partial x}\left(D_{xx}\frac{\partial T}{\partial x}+D_{xy}\frac{\partial T}{\partial y}\right)+\frac{\partial}{\partial y}\left(D_{yx}\frac{\partial T}{\partial x}+D_{yy}\frac{\partial T}{\partial y}\right) + S.
\end{aligned}
\label{eq:steady-aniso-1}
\end{equation}
Note that the physical time derivative can be retained for unsteady simulations. This paper focuses on 
steady problems and the developed steady solver can be employed to solve unsteady residual equations for unsteady simulations.
See Refs.\cite{LouLiLuoNishikawa_aiaa2018-4270,nishikawa_nakashima:jcp2018,liu_nishikawa_aiaa2017-0738} for the hyperbolic method applied to unsteady problems.

%Consider the following diffusion equation in two dimensions 
%\begin{equation}
%\begin{aligned}
%\frac{\partial T}{\partial t}&=\nabla .(\textbf{D}\nabla T) + S,\\
%&=\frac{\partial}{\partial x}\left(D_{xx}\frac{\partial T}{\partial x}+D_{xy}\frac{\partial T}{\partial y}\right)+\frac{\partial}{\partial y}\left(D_{yx}\frac{\partial T}{\partial x}+D_{yy}\frac{\partial T}{\partial y}\right) + S.
%\end{aligned}
%\end{equation}

%We introduce new variables $g$ and $h$, where $g=\frac{\partial T}{\partial x}$ and $h=\frac{\partial T}{\partial y}$, and the system is made hyperbolic by adding pseudo-time derivatives with respect to all variables as follows:
%\begin{equation}
%\begin{aligned}
%\frac{\partial T}{\partial t}-   \frac{ \partial}{\partial x}\left(D_{xx} g+D_{xy} h \right) - \frac{ \partial}{\partial y}\left( D_{xy} g+D_{yy} h\right) &= f ,\\
%\frac{\partial g}{\partial t} - \frac{\partial T}{\partial x}&=  -g ,\\
%\frac{\partial h}{\partial t} - \frac{\partial T}{\partial y}&=  -h.
%\label{eqn:hyper_1}
%\end{aligned}
%\end{equation}
%\begin{equation}\label{eqn:conc-aniso}
%\textbf{q}=-(\textbf{D}.\nabla T), \quad \frac{\partial T}{\partial t}=-\nabla . \textbf{q}+ \textit{f}
%\end{equation}

%\begin{equation}
%\frac{\partial T}{\partial t}=\frac{\partial}{\partial x}(D_{xx}\frac{\partial T}{\partial x}+D_{xy}\frac{\partial T}{\partial y})+\frac{\partial}{\partial y}(D_{yx}\frac{\partial T}{\partial x}+D_{yy}\frac{\partial T}{\partial y}) + S.
%\end{equation}
By introducing new variables $g$ and $h$, where $g=\frac{\partial T}{\partial x}$ and $h=\frac{\partial T}{\partial y}$, and the system is made hyperbolic by adding pseudo-time derivatives with respect to all variables as follows:
\begin{equation}
\begin{aligned}
\frac{\partial T}{\partial \tau}-   \frac{ \partial}{\partial x}\left(D_{xx} g+D_{xy} h\right) - \frac{ \partial}{\partial y}\left(D_{xy} g+D_{yy} h\right) &= S ,\\
\frac{\partial g}{\partial \tau} - \frac{\partial T}{\partial x}&=  -g ,\\
\frac{\partial h}{\partial \tau} - \frac{\partial T}{\partial y}&=  -h.
\label{eqn:hyper_form:2}
\end{aligned}
\end{equation}
%In this formulation implementing Dirichlet boundary conditions is straightforward. As \textbf{q} is not a variable in this formulation, flux boundary conditions, for example \textbf{q = 10}, should be set as as function of $g$ and $h$. 
This formulation is proposed by Lou et al. \cite{lou2017reconstructed,Lou2019}. The Eq.(\ref{eqn:hyper_form:2}) can be written as a preconditioned system as follows,
 \begin{equation}
           \mathbf{P}^{-1}\frac{\mathbf{\partial Q}}{\partial {\tau}}
                  +\frac{\partial \mathbf{E}_x}{\partial {x}}
                  +\frac{\partial \mathbf{E}_y}{\partial {y}}=\mathbf{S},
      \label{eqn:non_precon}
   \end{equation} 
   where
\begin{equation}\label{variables-h}
\begin{aligned}
&\mathbf{Q}= \left[ \begin{array}{ccc}T  \\ g \\ h \end{array} \right], 
\mathbf{E}_x = \left[ \begin{array}{ccc} -D_{xx} g -D_{xy} h  \\  -T \\ 0 \end{array} \right], 
&\mathbf{E}_y = \left[ \begin{array}{ccc} - D_{xy} g - D_{yy} h  \\  0 \\ -T \end{array} \right],
\mathbf{S} = \left[ \begin{array}{ccc} -S  \\  -g \\ -h \end{array} \right],
%      \label{vector:non_precon}
\end{aligned}
   \end{equation} 
    and the preconditioning matrix is given by,
 \begin{equation}   
    \mathbf{P}^{-1} = \left[\begin{matrix}1 & 0 & 0\\0 & T_r & 0 \\ 0 & 0 & T_r \end{matrix}\right],
    \end{equation}
where $T_r$ is the relaxation time which will be explained later. 
For an arbitrary $T_r$, this system is equivalent to the anisotropic diffusion equation (\ref{eq:steady-aniso-1}) in the steady state or as soon as the pseudo time
terms are dropped. The preconditioned hyperbolic formulation was first introduced in Ref.\cite{nishikawa_hyperbolic_ns:AIAA2011-3043}  to extend the hyperbolic method to nonlinear equations and simplify the analysis of the eigenstructure of the Navier-Stokes equations by eliminating the need to differentiate the viscosity given as a function of a solution. 
 It is also deliberately constructed as a preconditioned conservative form, so that any discretization method designed for a conservation law can be 
directly applied \cite{nishikawa_hyperbolic_ns:AIAA2011-3043}.
For the anisotropic diffusion equation, a similar formulation requires us to define four extra variables
corresponding to the four diffusive flux components: $D_{xx} g$, $D_{xy} h$, $D_{xy} g$ and $D_{yy} h$. 
To minimize the number of equations, we introduce, instead, the derivatives of
the temperature as extra variables. As a result, the diffusion coefficients remain in the flux, and thus they need to be differentiated in deriving the flux Jacobian. 
However, as pointed out in Ref.\cite{nishikawa_diff_discon:jcp2018}, to construct a numerical dissipation matrix, the flux Jacobian does not have to be exactly derived, and the coefficients can be frozen and thus not needed to be differentiated. This type of formulation is very useful for developing high-order methods \cite{LouLiLuoNishikawa_aiaa2018-2094, LiLouLuoNishikawa_aiaa2018-4160}. Note that the preconditioned formulation
is necessary, for whatever variables are chosen, to apply the hyperbolic method to variable-coefficient and nonlinear equations because $T_r$ is not a constant and thus cannot be included in the flux vectors \cite{nishikawa_diff_discon:jcp2018}. To construct an upwind flux, 
consider the preconditioned Jacobian of the flux projected along an arbitrary vector, $n=(n_x, n_y)$:
\begin{equation}
\mathbf{PA}_n
=\mathbf{P}\frac{\partial(\mathbf{E}_x n_x+\mathbf{E}_y n_y)}{\partial Q} 
= \left[\begin{matrix}0 & - D_{xx} n_{x} - D_{xy} n_{y} & - D_{xy} n_{x} - D_{yy} n_{y}\\
- \frac{n_{x}}{T_{r}} & 0 & 0\\- \frac{n_{y}}{T_{r}} & 0 & 0\end{matrix}\right],
\end{equation}
where the diffusion coefficients have been assumed to be constant for the sake of derivation (but they can be functions of 
space or solutions in the final form).
This system is hyperbolic since $\mathbf{PA}_n$ has real eigenvalues, which are given by
\begin{equation}
\lambda_1 = 0, \quad  \lambda_{2,3} = \pm \sqrt{\frac{1}{T_{r}} \left(D_{xx} n_{x}^{2} + 2 D_{xy} n_{x} n_{y} + D_{yy} n_{y}^{2}\right)},
\label{eig_values}
\end{equation}
 and linearly independent right-eigenvectors:
\begin{equation}
\mathbf{R}_n= \left[\begin{matrix}\frac{D_{xx} n_{x}^{2} + 2 D_{xy} n_{x} n_{y} + D_{yy} n_{y}^{2}}{n_{y} \sqrt{\frac{1}{T_{r}} \left(D_{xx} n_{x}^{2} + 2 D_{xy} n_{x} n_{y} + D_{yy} n_{y}^{2}\right)}} & 0 & - \frac{D_{xx} n_{x}^{2} + 2 D_{xy} n_{x} n_{y} + D_{yy} n_{y}^{2}}{n_{y} \sqrt{\frac{1}{T_{r}} \left(D_{xx} n_{x}^{2} + 2 D_{xy} n_{x} n_{y} + D_{yy} n_{y}^{2}\right)}}\\\frac{n_{x}}{n_{y}} & - \frac{D_{xy} n_{x} + D_{yy} n_{y}}{D_{xx} n_{x} + D_{xy} n_{y}} & \frac{n_{x}}{n_{y}}\\1 & 1 & 1\end{matrix}\right].
\end{equation}
The absolute Jacobian $|\mathbf{PA}_n|$, {\color{black}which is required for current upwind flux approach}, is constructed by right-eigenvector matrix $\mathbf{R}_n$ and the diagonal eigenvalue-matrix, $\mathbf{\Lambda_n}$,
\begin{equation}
\mathbf{\Lambda_n} = \sqrt{\frac{1}{T_{r}} \left(D_{xx} n_{x}^{2} + 2 D_{xy} n_{x} n_{y} + D_{yy} n_{y}^{2}\right)} \left[\begin{matrix}-1 & 0 & 0\\0 & 0 & 0\\0 & 0 & 1\end{matrix}\right]
\end{equation}
as follows:
\begin{equation}
|\mathbf{PA}_n|=\mathbf{R}_n|\mathbf{\Lambda}_n| \mathbf{R}^{-1}_n= \lambda \left[\begin{matrix}1 & 0 & 0\\0 & \frac{n_{x} \left(D_{xx} n_{x} + D_{xy} n_{y}\right)}{D_{xx} n_{x}^{2} + 2 D_{xy} n_{x} n_{y} + D_{yy} n_{y}^{2}} & \frac{n_{x} \left(D_{xy} n_{x} + D_{yy} n_{y}\right)}{D_{xx} n_{x}^{2} + 2 D_{xy} n_{x} n_{y} + D_{yy} n_{y}^{2}}\\0 & \frac{n_{y} \left(D_{xx} n_{x} + D_{xy} n_{y}\right)}{D_{xx} n_{x}^{2} + 2 D_{xy} n_{x} n_{y} + D_{yy} n_{y}^{2}} & \frac{n_{y} \left(D_{xy} n_{x} + D_{yy} n_{y}\right)}{D_{xx} n_{x}^{2} + 2 D_{xy} n_{x} n_{y} + D_{yy} n_{y}^{2}}\end{matrix}\right],
\label{eqn:flux-jacob}
\end{equation}
where $\lambda$=${\sqrt{\frac{1}{T_{r}} \left(D_{xx} n_{x}^{2} + 2 D_{xy} n_{x} n_{y} + D_{yy} n_{y}^{2}\right)}}$.

\subsection{Fourier analysis for optimum relaxation time and length scale}
{\color{black}Due to the significant difference between the eigenvalues, $10^9$ and 1, the condition number (ratio of the maximum eigenvalue to the minimum eigenvalue) of the system increases and degrades the convergence performance and also the accuracy of the numerical result. The optimal choice of relaxation time improves the condition number of the problem and thereby improves the accuracy such that the results are independent of the degree of anisotropy.} In this subsection, we derive optimal length scale and relaxation time such that the error remains the same regardless of the degree of anisotropy. The importance of length scale and relaxation time for the isotropic diffusion are discussed in Ref. \cite{Nishikawa2014b}. Optimal values for $T_r$ is derived by following the procedure briefly described in \cite{Nishikawa2007, Nishikawa2014b} for a two-dimensional hyperbolic diffusion system. In short, we substitute a Fourier mode into the first-order version of the finite-difference or finite-volume scheme in the semi-discrete form (on a regular grid) to derive a pseudo-time evolution of the amplitude of the Fourier mode, and derive the optimal values by requiring that the eigenvalues of the evolution matrix (the RHS of the ODE for the Fourier mode amplitude) become complex, so that the Fourier mode will propagate rather than purely get damped. If the Fourier mode propagates, the scheme has a property of removing errors by propagation in addition to error damping, which is much faster than pure damping. The slow convergence of relaxation schemes for elliptic problems comes from the pure damping, and the hyperbolic method improves the convergence by introducing error propagation into the scheme.

Consider the preconditioned form of the Eq.(\ref{eqn:non_precon}). To derive an optimal $T_r$, we ignore the physical time derivative $\frac{\partial Q}{\partial t}$ and focus on the pseudo-time system. Our goal is to define $T_r$ such that all possible Fourier modes will propagate by the system in the pseudo time. Define
the Fourier mode as

\begin{equation}
\mathbf{Q^{\beta}} = \mathbf{Q}_0 \exp (i(\beta_x x +\beta_y y)/h),
\end{equation}
on a uniform grid of spacing h, where $i$ = $\sqrt-1$ and $Q_0 = ({\color{black}T_0}, g_0, h_0)$ is the vector of amplitudes. Then, substitute it into the first-order system to obtain
\begin{equation}
\frac{d\mathbf{Q_0}}{d\tau} = \mathbf{M}Q_0,
\label{fourier_transformed}
\end{equation}
where 
 \begin{equation}   
    \mathbf{M} = \left[\begin{matrix} 0 & \frac{i(D_{xx}\beta_x+{\color{black}D_{xy}}\beta_y)}{h} & \frac{i(D_{xy}\beta_x+{\color{black}D_{yy}}\beta_y)}{h} \\ \\ \frac{i\beta_x}{hT_r} & -1/T_r & 0 \\ \\ \frac{i\beta_y}{hT_r} & 0 & -1/T_r \end{matrix}\right].
    \label{M_eigenvalues}
    \end{equation}
The eigenvalues of $\mathbf{M}$ are given by

\begin{equation}
\begin{aligned}
\lambda_{1,2} &= {\color{black}  - \frac{ h \pm \sqrt{z}  }{2 h T_r} , \quad z = h^2-4{{T_{r}} \left(D_{xx} \beta_{x}^{2} + 2 D_{xy} \beta_{x} \beta_{y} + D_{yy} \beta_{y}^{2}\right)}, }  \\
\lambda_{3}&=  - \frac{1}{T_r}.
\label{eqn:eigen}
\end{aligned}
\end{equation}
The third eigenvalue corresponds to the inconsistency-damping mode associated with the source terms \cite{Nishikawa2007}. We focus on the remaining pair and determine $T_r$ to make them complex conjugate, so that the Fourier mode propagates. That is, we require z $<$ 0 for all possible frequencies:

\begin{equation}
h^2-4{{T_{r}} \left(D_{xx} \beta_{x}^{2} + 2 D_{xy} \beta_{x} \beta_{y} + D_{yy} \beta_{y}^{2}\right)} < 0,
\end{equation}
which, since $\mathbf{D}$ is positive definite, leads to

\begin{equation}
{T_{r}} > \frac{h^2} {4 \left(D_{xx} \beta_{x}^{2} + 2 D_{xy} \beta_{x} \beta_{y} + D_{yy} \beta_{y}^{2}\right)}.
\end{equation}
The lower bound reaches the maximum for the smoothest discrete mode: $\beta_x$ = $\beta_y$ = $\pi$h, and therefore the inequality is satisfied for all possible frequencies if we set
{\color{black}
\begin{equation}
{T_{r}} = \frac{1} {(2\pi)^2\left(D_{xx} + 2 D_{xy} + D_{yy}\right)}.
\end{equation}
This implies that we define $T_r$ as
 \begin{equation}
 T_r=\frac{{L}_r^2}{\nu_{opt}}, \quad \textrm{where} \  \nu_{opt} = {D_{xx}+2D_{xy}+D_{yy}},
 \label{optimal_Tr}
 \end{equation}
where ${L}_r = 1/ (2  \pi)$ is a relaxation length}. In the hyperbolic approach proposed by Lou et al. \cite{lou2017reconstructed}, based on reconstructed Discontinuous Galerkin scheme (rDG), $\nu_{opt}$ is taken as 1. It will be shown here that such a hyperbolic diffusion scheme can be high order accurate, but it is not independent of the diffusion coefficient and loses accuracy with an increasing degree of anisotropy. It has been shown that the relaxation length associated with the hyperbolic diffusion system is a dimensional quantity and must be properly scaled by a reference length for a given problem to guarantee dimensional-invariance of numerical solutions \cite{nishikawa2018dimensional}. For each reference length chosen, an optimal value of the corresponding non-dimensionalized relaxation length needs to be found to achieve optimal performance of a hyperbolic solver, called an optimal reference length. The optimal value is given by $\frac{1}{2\pi}$ for a unit square domain, but it is not necessarily optimal for a general domain. A practical approach for obtaining a reference length is derived in Ref. \cite{nishikawa2018dimensional}. For a rectangular domain, (x, y) $\in$ $[0,L_x] \times [0,L_y ]$, the following equation can be used,  
 \begin{equation}
 L_r=\frac{2h}{\pi/\tilde{N}(\pi/\tilde{N}+4)}, \textrm{where} \ \tilde{N}=\frac{1}{h\sqrt{(\frac{1}{L^2_x}+\frac{1}{L^2_y})}},
 \label{optimal_Lr}
 \end{equation} 
{\color{black} where $h$ is grid spacing for a given uniform grid but has a very minor impact on $L_r$ when $h \ll1$.}
This approach is considered in this current paper and is found important for magnetized electrons test case.

\section{Review of numerical schemes and boundary conditions}
\begin{figure}[H]
\centering
\includegraphics[width=0.94\textwidth]{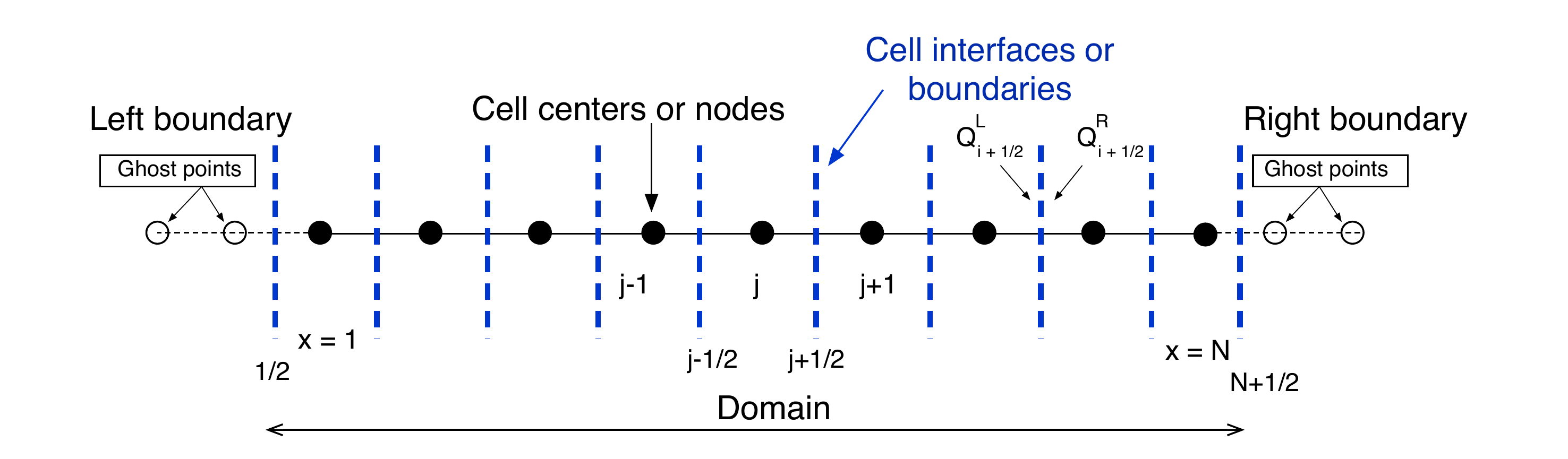}
  \caption{Grid and the domain boundary for cell-centered formulation}
  \label{fig:ghost_cell}
\end{figure}

In this section the procedures of cell centered upwind finite difference schemes \cite{Deng2000,Nonomura2013} and boundary conditions \cite{chamarthi2018high} are briefly explained. The domain is discretized uniformly into a Cartesian grid of size  $N_x \times N_y $, where $N_x$ and $N_y$ are the numbers of cells in $x$-and $y$-coordinate directions, and the domain is defined as $I_{j,k}=\left[x_{j-\frac{1}{2}},x_{j+\frac{1}{2}}\right]\times\left[y_{k-\frac{1}{2}},y_{k+\frac{1}{2}}\right]$ for $1 \leq j \leq N_x$, $1 \leq k \leq N_y$, where
\begin{equation}
  x_{j+\frac{1}{2}} =  x_a + j \Delta x, \quad
  y_{k+\frac{1}{2}} = y_a + k \Delta y
\end{equation}
\noindent and
\begin{equation}
  \Delta x = \frac{x_b - x_a}{N_x}, \quad \Delta y = \frac{y_b - y_a}{N_y}
\end{equation}
\noindent The Eq.(\ref{eqn:non_precon}) in semi-discrete form by finite difference scheme can be written as
\begin{equation}
  \left. \frac{\partial Q}{\partial t} \right|_{j,k} + \left. \frac{\partial E_x}{\partial x} \right|_{j,k} + \left. \frac{\partial E_y}{\partial y} \right|_{j,k}  = \left. S \right|_{j,k}
  \label{eqn:semi}
\end{equation}
\noindent where $\frac{\partial E_x}{\partial x} \big| _{j,k}$, $\frac{\partial E_y}{\partial y} \big| _{j,k}$, and $S |_{j,k}$ are the approximations of the flux derivatives in the $x$ and $y$ directions, and source term at cell centers $(x_j, y_k)$, shown in Fig. \ref{fig:ghost_cell}, respectively.
%where
%\begin{equation}
%  x_i = \frac{x_{i-\frac{1}{2}} + x_{i+\frac{1}{2}}}{2}, \quad
%    y_j = \frac{y_{j-\frac{1}{2}} + y_{j+\frac{1}{2}}}{2}
%\end{equation}

\subsection{Overview of upwind finite difference schemes}
\label{upwind_fd}
For the evaluation of the flux derivatives, Lele has developed a family of compact schemes in \cite{lele1992compact}. Through a detailed Fourier analysis, it has been shown that compact schemes on a cell-centered mesh, staggered, provide better resolution in comparison with the node-centered, collocated, mesh. The cell-centered method involves interpolation of variables to cell interfaces, which can be carried out either by a centered stencil or an upwind biased stencil. Nagarajan et al. \cite{Nagarajan2003a} and Boersma \cite{Boersma2011a} used cell-centered compact schemes with a centered stencil to simulate compressible flows, and their numerical tests indicate that their methods are quite robust. Unfortunately, central interpolations are known to cause oscillations in the presence of discontinuities and are not suitable for the current hyperbolic approach. On the other hand, Deng and Zhang \cite{Deng2000} developed a class of weighted nonlinear compact schemes (WCNS) based on the idea of  Weighted Essentially Non-oscillatory approach (WENO). In this approach, the interpolation of variables is carried out through a nonlinear combination of several lower-order stencils, which are upwind biased, similar to the methodology of the WENO scheme \cite{jiang1995}. These schemes capture the discontinuities without spurious oscillations and recover formal order of accuracy in smooth regions. Deng et al. \cite{deng2015family} has also developed a family of hybrid cell-interface and cell-node compact linear schemes. {\color{black}In this paper, both linear and nonlinear weighted compact schemes are used for spatial discretization. It should be noted that the finite-difference schemes considered here are more efficient and cheaper for the nonlinear problems than the high-order finite volume methods (FVM) which require additional flux computations at quadrature points \cite{Titarev2004}. To achieve higher order accuracy, third or more, in FVM, the integrals of the fluxes are discretized utilizing a high-order Gaussian quadrature with suitable Gaussian integration points over the faces of the control volume. Due to these difficulties, the higher order finite volume approach is at least 3 to 12 times expensive compared to the finite difference methods in multidimensional problems. Nonomura et al. have discussed the advantages of these schemes over high-order finite volume schemes for multi-component flows in detail(readers can refer \cite{Nonomura2012} for further information).} The WCNS consists of three parts: high-order flux difference for flux derivatives, high-order interpolation for cell-interface values, and upwind flux evaluation at the cell interface.

In the first part, a high-order flux difference, cell-interface-to-node, scheme is adapted to calculate the flux derivatives, which are

\begin{equation} \label{eqn-differencing}
\begin{aligned}
     \rm{4^{th}\ order:}; \Bigl|\frac{\partial E_x}{\partial x}\Bigr|_{j,k} &= \frac{1}{\Delta x} \Bigl[\frac{9}{8}\left({E}_{j + \frac{1}{2},k} - {E}_{j - \frac{1}{2},k}\right) -\frac{1}{24}\left({E}_{j + \frac{3}{2},k} - {E}_{j - \frac{3}{2},k}\right) \Bigr]\\     
          \rm{6^{th}\ order:};  \Bigl|\frac{\partial E_x}{\partial x}\Bigr|_{j,k} &= \frac{1}{\Delta x}  \left[
      \frac{75}{64} \left({E}_{j+\frac{1}{2},k} - {E}_{j-\frac{1}{2},k} \right)
      - \frac{25}{384} \left( {E}_{j+\frac{3}{2},k} -  {E}_{j-\frac{3}{2},k} \right) \right. \\
     & \left. + \frac{3}{640} \left({E}_{j+\frac{5}{2},k} - {E}_{j-\frac{5}{2},k} \right)
    \right],
        \end{aligned}
\end{equation}
where $E_{j+\frac{1}{2},k}$ are fluxes approximated at cell interfaces by linear or nonlinear interpolations. The derivatives in the $y$ direction are computed similarly. The flux derivatives can also be computed compact difference schemes, but Deng and Zhang \cite{Deng2000} report that the resolution of the explicit difference scheme is almost same as that of the compact difference scheme and also the explicit differencing is computationally more efficient than the compact scheme. Nonomura et al.\cite{Nonomura2013} proposed a family of cell interface-and-node-to-node differencing schemes for the computation of the first derivatives which are robust than Eq.(\ref{eqn-differencing}) but in our computations, we found no difference between the results.

%During the weighted interpolation for the cell-centered value, we use the upwind biased stencils to get the approximation.
In the second part, high-order linear or nonlinear interpolations are used for the computation of $E_{j+\frac{1}{2}}$ in Eq.(\ref{eqn-differencing}) which requires knowledge of the solutions $Q^R$ and $Q^L$, shown in Fig. \ref{fig:ghost_cell}. The solutions of $Q^R$ and $Q^L$ are obtained by upwind interpolation.   The ${Q}_{j+\frac{1}{2}}^{(L)}$ is biased to left and similarly ${Q}_{j+\frac{1}{2}}^{(R)}$ is biased to the right for correct upwinding. The interpolation can be either explicit or implicit in space, also known as compact schemes. Details of the implementation of nonlinear interpolation schemes (WCNS) \cite{Deng2000, Wong2017} are given in Appendix A. Compact linear interpolation polynomials, denoted as U-5C, for the left and right interface are as follows,
\begin{equation}\label{eqn-compact}
\rm{5^{th} \ order \ compact}
\begin{cases}
\frac{1}{2} Q^{L}_{j-\frac{1}{2}}+ Q^{L}_{j+\frac{1}{2}} + \frac{1}{10} Q^{L}_{j+\frac{3}{2}}= \frac{1}{10} Q_{j-1}+  Q_{j} + \frac{1}{2} Q_{j+1} \\
\frac{1}{10} Q^{R}_{j-\frac{1}{2}}+ Q^{R}_{j+\frac{1}{2}} + \frac{1}{2} Q^{R}_{j+\frac{3}{2}}= \frac{1}{2} Q_{j}+  Q_{j+1} + \frac{1}{10} Q_{j+2}.
\end{cases}
\end{equation}
Polynomials of accuracy up to $9^{th}$ order are considered for the computations. Explicit third and fifth order linear interpolation formulas, denoted as U-3E and U-5E, are given below 

\begin{equation}\label{eqn-explicit}
\begin{aligned}
   \rm{3^{rd} \ order}
&\begin{cases}
{Q}_{j+\frac{1}{2}}^{(L)} = \frac{1}{8}\left(-Q_{j-1} + 6Q_{j} + 3Q_{j+1} \right) \\
{Q}_{j+\frac{1}{2}}^{(R)} = \frac{1}{8}\left(3Q_{j} + 6Q_{j+1} - Q_{j+2} \right),
\end{cases}\\
   \rm{5^{th} \ order}
&\begin{cases}
{Q}_{j+\frac{1}{2}}^{(L)}= \frac{1}{128} \left(3Q_{j-2} -20Q_{j-1} +90Q_{j}+60Q_{j+1}-5Q_{j+2}  \right) \\
{Q}_{j+\frac{1}{2}}^{(R)}= \frac{1}{128} \left(-5Q_{j-1} +60Q_{j} +90Q_{j+1} -20Q_{j+2} + 3Q_{j+3} \right).
\end{cases}
\end{aligned}
\end{equation}

Finally, in the third part, the upwind flux will be computed by an appropriate Riemann solver. One of the advantages of WCNS is that the interpolation procedure is independent of the numerical flux calculations, unlike the WENO interpolation. The numerical flux can be evaluated by both flux vector splitting and flux difference-splitting schemes (see \cite{Deng2000} for details). We evaluate the numerical flux by the following upwind flux:
\begin{equation}
\mathbf{E}_{j+\frac{1}{2},k} =  \frac{1}{2}(\mathbf{E}^L + \mathbf{E}^R) -
  \frac{1}{2} \mathbf{P}^{-1}|\mathbf{ P}\mathbf {A}|(\mathbf{Q}^R-\mathbf{Q}^L),
\label{eqn:precon-aniso}
\end{equation}
where $|\mathbf {A}|$ is the flux Jacobian, given in Eq.(\ref{eqn:flux-jacob}). In the local preconditioning method \cite{nishikawa2003general}, the numerical flux
  $\mathbf{E}_{j+\frac{1}{2},k} $ is constructed based on the preconditioned Jacobian. $\mathbf{P}^{-1}$  is multiplied in the dissipation term in the above expression to cancel the effect of $\mathbf{P}$ that will be multiplied in the final residual which is a standard procedure in the construction of upwind fluxes for a preconditioned PDE. The complete procedure can be summarized as follows : 

\begin{enumerate}
\item Interpolate the conservative variables from cell-node to cell interface, Eq.(\ref{eqn-compact}) or (\ref{eqn-explicit}).
\item compute the upwind flux evaluation at the cell interface, Eq.(\ref{eqn:precon-aniso}), and
\item evaluate the derivative by cell-interface-to-node differencing, Eq.(\ref{eqn-differencing}). Fourth order differencing formula is used for third order interpolation, and sixth order differencing formula is used for fifth order explicit and implicit interpolation formulae.
\end{enumerate}

\subsection{Time discretization}
After discretizing the spatial derivative, we form a set of ordinary differential equations: e.g., at a cell $(j,k)$, 
\begin{equation}
%\mathbf{Q_t} = \mathbf{Res(Q)}, 
\mathbf{P}^{-1}_{j,k}  \frac{ d \mathbf{Q}_{j,k} }{ d{\tau} } = \mathbf{Res}_{j,k},
\end{equation}
where
 \begin{equation}
 \mathbf{Res}_{j,k} =  - \left. \frac{\partial E_x}{\partial x} \right|_{j,k} - \left. \frac{\partial E_y}{\partial y} \right|_{j,k}  + \left. S \right|_{j,k},
\end{equation}
and integrate it in the pseudo time by the following third order TVD Runge-Kutta method, to drive the solution to the steady state:
\begin{eqnarray}  \label{rk}
\mathbf{Q}^{(1)} & = & \mathbf{Q}^n + \Delta \tau \mathbf{Res}(\mathbf{Q}^n),  \nonumber \\
\mathbf{Q}^{(2)} & = & \frac{3}{4} \mathbf{Q}^n + \frac{1}{4} \mathbf{Q}^{(1)} + \frac{1}{4}\Delta \tau \mathbf{Res}(\mathbf{Q}^{(1)}) ,\\
\mathbf{Q}^{n+1}& = & \frac{1}{3} \mathbf{Q}^n + \frac{2}{3} \mathbf{Q}^{(2)} + \frac{2}{3}\Delta \tau \mathbf{Res}(\mathbf{Q}^{(2)}),  \nonumber
\end{eqnarray}
where $\Delta \tau$ is the pseudo time step given by the following equation

\begin{equation}
\Delta \tau = \textrm{CFL} \ {\textrm{min}\left(\frac{x_{i+1}-x_{i}}{|\sqrt{D_{xx}/T_r}|},\frac{y_{i+1}-y_{i}}{|\sqrt{D_{yy}/T_r}|}\right)}.
\end{equation}

Time integration is performed with a CFL $=$ 0.2 for all the schemes until the residual is reduced by ten orders of magnitude in the L1 norm.

\subsection{Boundary conditions}
In this section, the implementation of boundary conditions is described. In the hyperbolic approach the Neumann boundary conditions are also implemented as Dirichlet boundary condition through the gradient variables, and therefore we describe only Dirichlet boundary conditions, more details are given in Ref. \cite{chamarthi2018high}
\begin{enumerate}
\item \textbf{Primary variable $T$}\\
The physical domain is extended through ghost cells and the Dirichlet boundary condition is employed at the cell interface, say $T_{\frac{1}{2}}$. The $3^{rd}$ and $5^{th}$   order accurate polynomials for the left and right boundaries are given by Eqs. (\ref{ghost_3-2}) and (\ref{ghost_5-right}).
%Lagrange extrapolation formula given by Eq.(\ref{lagrange-2}) is used to compute the values in ghost cells:

%%
%\begin{equation}\label{lagrange-2}
%u^{(r)}_{\frac{1}{2}}(x)=\sum _{j=0}^{k} u_{i-r+j}C_{rj}(x), \quad C_{rj}(x)=\prod^{k} _{\begin{smallmatrix} l= 0\\l\neq j\end{smallmatrix}}{\frac {x-x_{i-r+l}}{x_{i-r+j}-x_{i-r+l}}},
%\end{equation}
%%
%where r is the order of the Lagrange polynomial.
\begin{equation}
\begin{aligned}
   \rm{Left \ boundary}
&\begin{cases}
T_0 &= \frac{1}{3}\left(8T_\frac{1}{2}-6T_{1} + T_{2} \right),\\
T_{0} &=\frac{128}{35}\left(T_\frac{1}{2} -\frac{35}{32} T_{1} +\frac{35}{64} T_{2}-\frac{7}{32} T_{3}+\frac{5}{128}T_{4} \right).\label{ghost_3-2}
\end{cases}\\
\end{aligned}
\end{equation}

\begin{equation}
\begin{aligned}
   \rm{Right \ boundary}
&\begin{cases}
T_{N+1} &= \frac{1}{3}\left(8 T_{N+\frac{1}{2}} + T_{N-1}-6T_{N} \right),\\
 T_{N+1} &=\frac{128}{35}\left(T_{N+\frac{1}{2}} +\frac{5}{128} T_{N-3} -\frac{7}{32} T_{N-2}-\frac{35}{64} T_{N-1}-\frac{35}{32}T_{N} \right).\label{ghost_5-right}
 \end{cases}\\
\end{aligned}
\end{equation}
 
\item \textbf{Gradient variables $g$ and $h$}\\ 
In isotropic diffusion problems boundary conditions are required only for one of the gradient variables, i.e. $g$ in x-direction and $h$ in y-direction, but in anisotropic diffusion boundary values should be specified for both the gradient variables, $g$ and $h$. The $3^{rd}$ and $5^{th}$ order cell-center to cell-center extrapolation formula are given by Eq.(\ref{extra-left}) and \ref{extra-right},  
\begin{equation}
\begin{aligned}
   \rm{Left \ boundary}\label{extra-left}
&\begin{cases}
g_0&= 3g_{1} - 3g_{2} +  g_{3},\\
g_0&= 5g_{1}-10g_{2}+10g_{3}-5g_{4}+g_{5}.
\end{cases}\\
\end{aligned}
\end{equation}

\begin{equation}\label{extra-right}
\begin{aligned}
   \rm{Right \ boundary}
&\begin{cases}
g_{N+1}& = 3g_{N} - 3g_{N-1}+g_{N-2},\\
g_{N+1}  &= g_{N-4} - 5g_{N-3} + 10g_{N-2} -10g_{N-1}+5g_{N}.
\end{cases}\\
\end{aligned}
\end{equation}%
\end{enumerate}

%Lagrange extrapolation can be approximated by a (s-k)th order Taylor expansion and a general formula is given by Eqn. \ref{extra_g-2} 
%\begin{equation} 
%\sum_{k=0}^{r}  \frac{r!}{k!(r-k)!} (-1)^k p_{j-k}=0, \quad j = N+1,.....,N+3
%\label{extra_g-2} 
%\end{equation} 
%
%
%
%
%
%
%
%
%
%
\section{Numerical tests}
In this section, various test cases for anisotropic diffusion equation are solved by the upwind schemes and the boundary conditions described earlier. All the simulations are carried until the residuals are dropped below $10^{-10}$ in $L_1$ norm by the TVD-RK time integration. Uniform cartesian meshes are used for all the problems. For problems with constant diffusion tensor, $D_{\perp}$ is taken as 1.

\subsection{Grid aligned test case}
\begin{figure}[H]
\centering
\begin{onehalfspacing}
\subfigure[]{\includegraphics[width=0.46\textwidth]{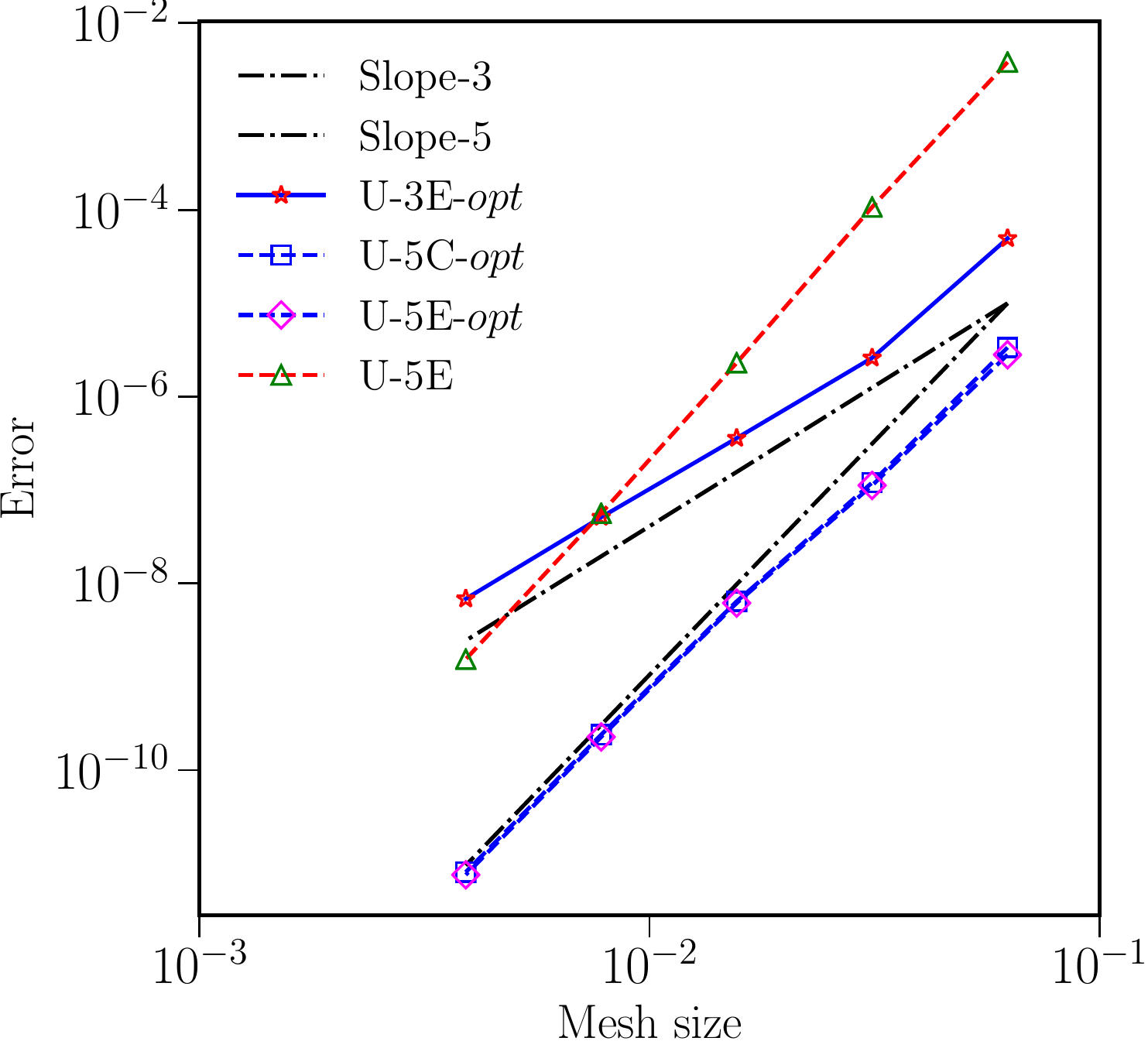}
\label{fig:kuz_max_error}}
\subfigure[]
{\includegraphics[width=0.44\textwidth]{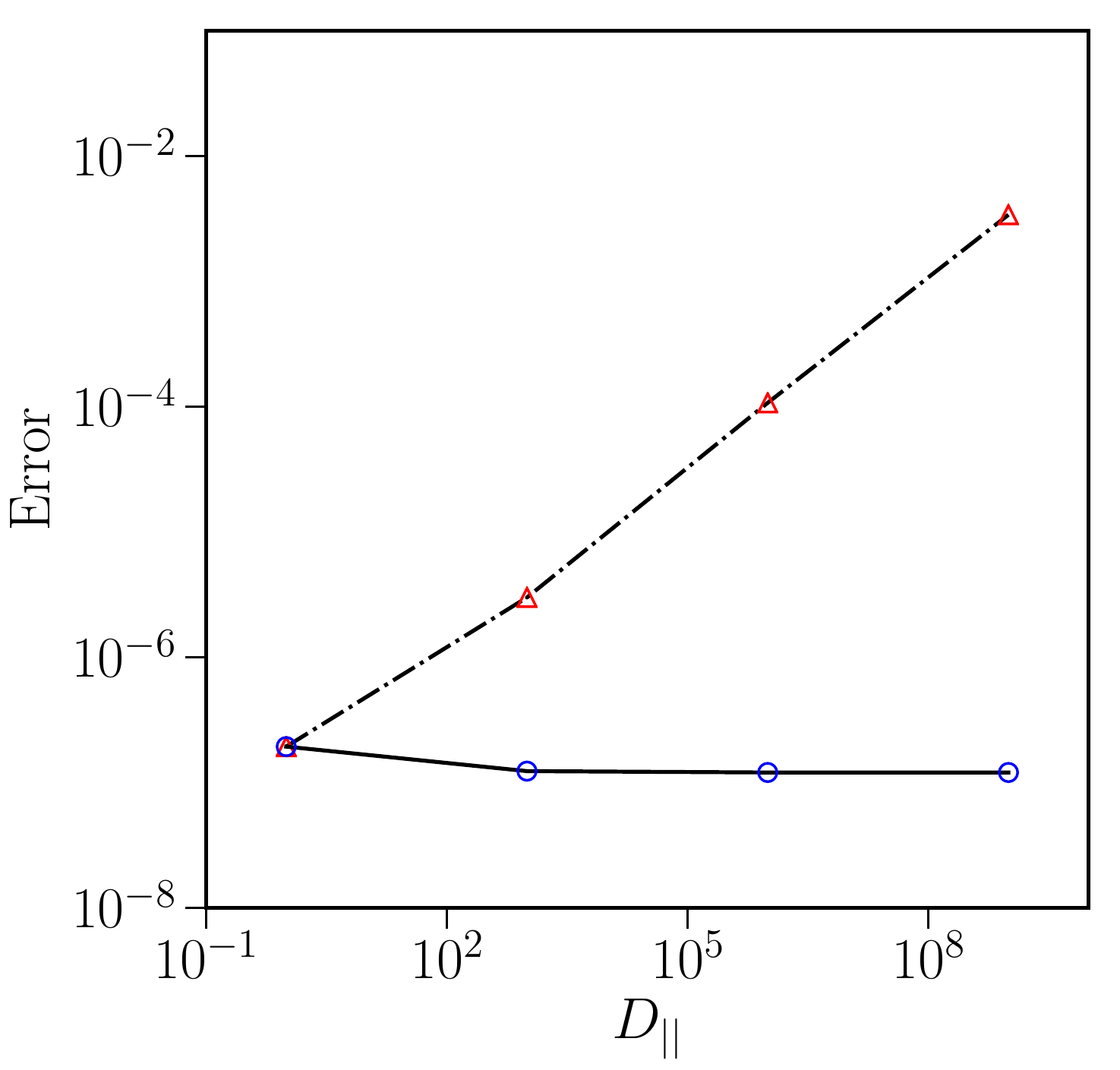}
\label{fig:2d-aniso-error}}
\caption{(a) $L_2$ convergence errors for various schemes. Blue lines: $\nu_{opt}$; red lines: $\nu =1$ and (b) $L_2$ error for increasing degree of anisotropy for U-5E. Blue circles: $\nu_{opt}$; red triangles: $\nu =1$ for \hyperref[sec:4.1]{Example 4.1}.}
\end{onehalfspacing}
\end{figure}

\begin{figure}[H]
\centering
\begin{onehalfspacing}
\subfigure[]{%
\includegraphics[width=0.44\textwidth]{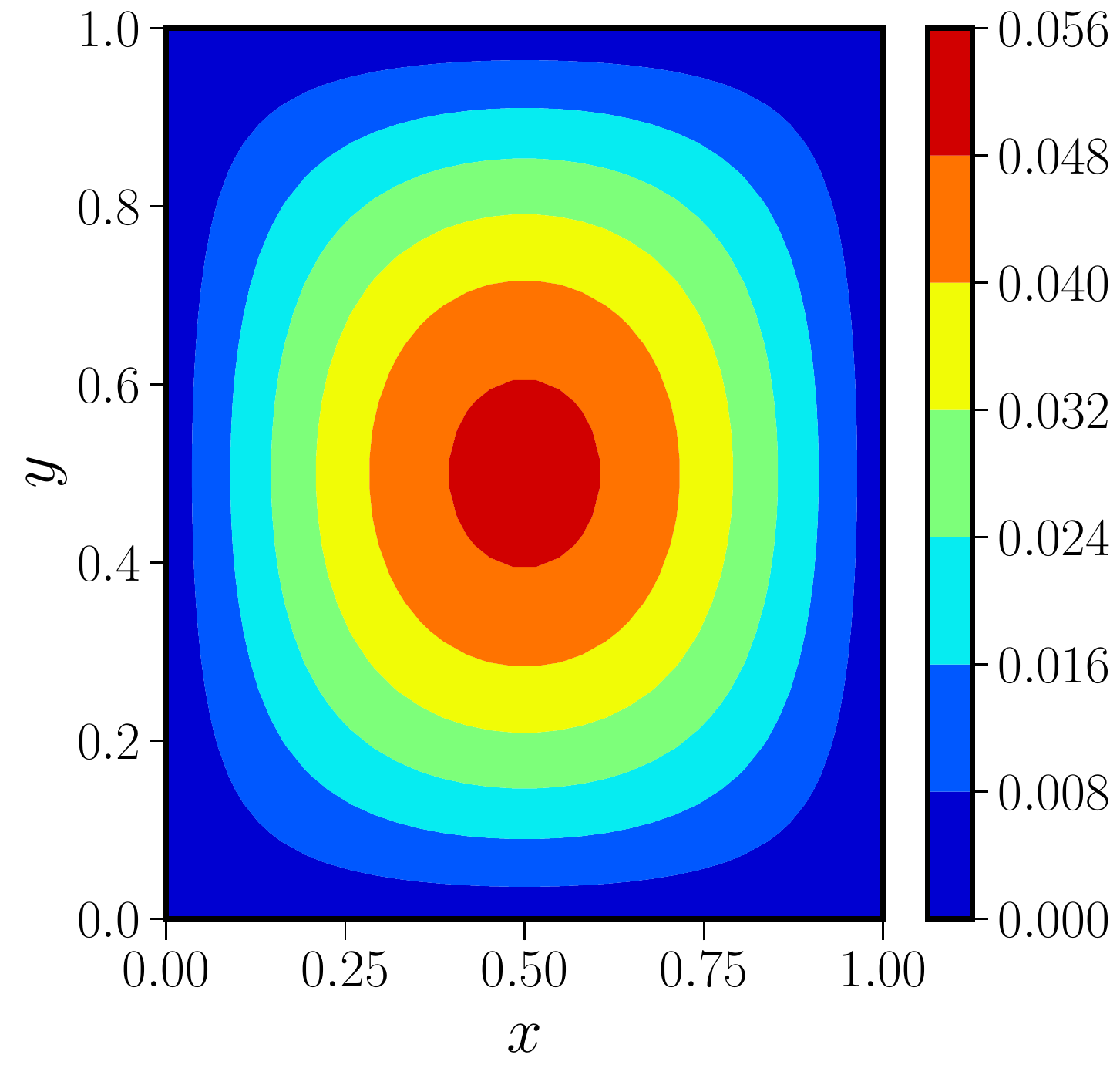}
\label{fig:3e-aniso-source}}
\subfigure[]
{\includegraphics[width=0.44\textwidth]{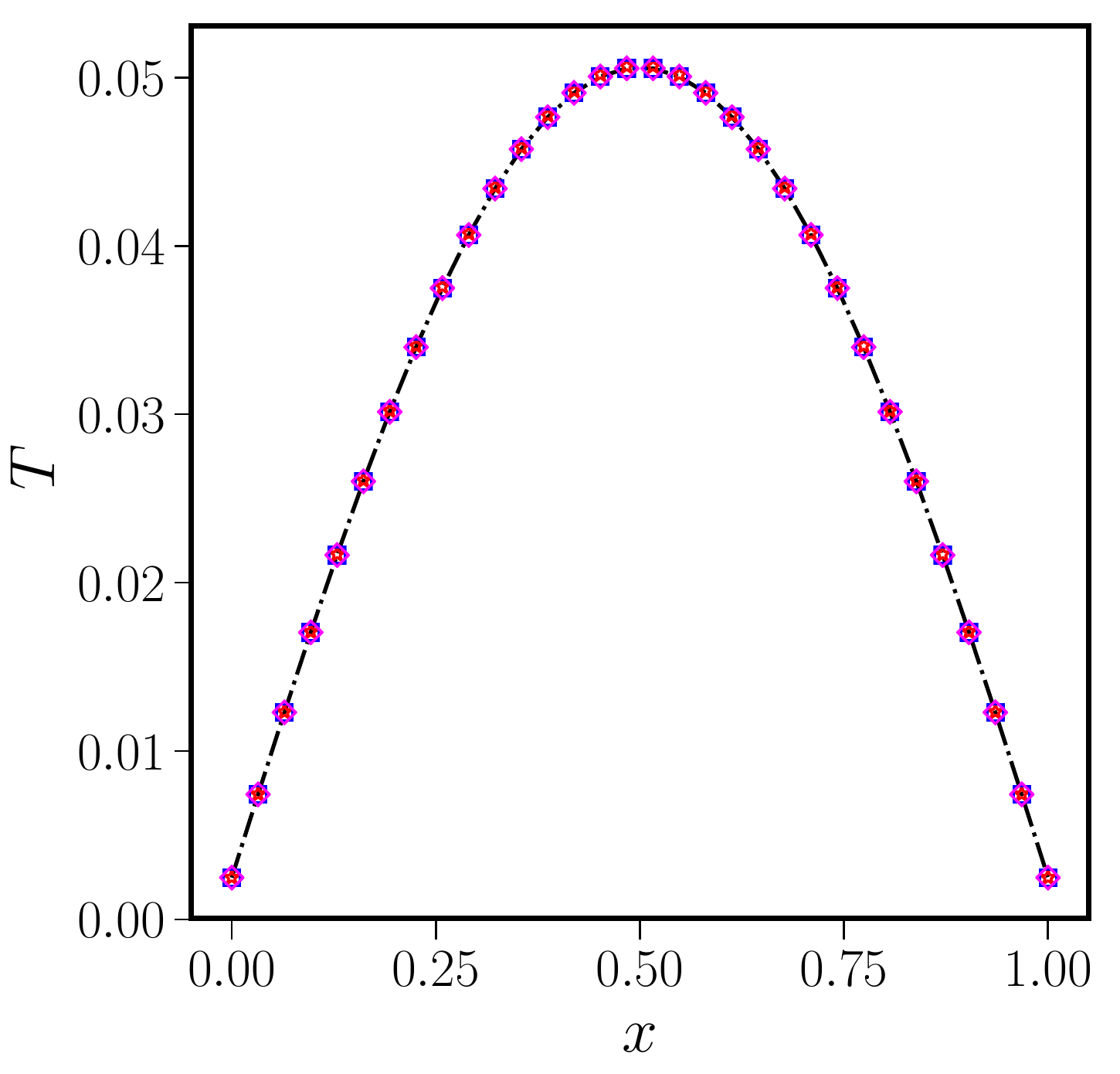}
\label{fig:s2d-aniso-center}}
\caption{(a) Solution contours for anisotropic diffusion by upwind scheme U-5E and (b) computed values at geometric center by $\nu_{opt}$; Dashed line: analytical; red stars: U-3E; blue squares: U-5C; magenta diamonds: U-5E.}
\end{onehalfspacing}
\end{figure}

\noindent \textbf{Example 4.1.}\label{sec:4.1} The following anisotropic diffusion equation identified in Ref. \cite{kuzmin}, with a spatial domain $(x,y) \in [0,1] \times [0,1]$,  has been considered. 
\begin{equation}\label{eq:diffu-ani}
\begin{aligned}
\frac{\partial T}{\partial t} = D_{xx} \frac{\partial ^2 T}{\partial x^2} + D_{yy} \frac{\partial ^2 T}{\partial y^2} + S,
\end{aligned}
\end{equation}
where the diffusion tensor and source term are 
 \begin{equation}\label{eq:tensor_1}
 \mathbf{D} = \left[ \begin{array}{ccc}
D_{xx} & 0 \\
0 & D_{yy} \\
\end{array} \right] = \left[ \begin{array}{ccc}
D_{||} & 0 \\
0 & D_{\perp} \\
\end{array} \right] =  \left[ \begin{array}{ccc}
10^{\gamma} & 0 \\
0 & 1 \\
\end{array} \right], \quad  S = 50.5 \sin(\pi x)\sin(\pi y),
\end{equation}
where $\gamma$ is varied from 0 to 9 and the exact solution for this test case is given by, 
 \begin{equation}\label{eq:kuz}
 T_{exact}(x,y) =\frac {1}{2\pi^2}  \sin(\pi x)\sin(\pi y), x,y \in [0,1],
\end{equation}
which is also imposed as Dirichlet boundary conditions.  In this test case, the diffusion coefficients $D_{xy}$ and $D_{yx}$ are zero. Therefore this test case may be considered as a grid-aligned situation where angle $\beta = 0^{\circ}$, shown in Fig. \ref{fig:angle_mag}. Grid refinement study is carried out for all the upwind schemes for two different relaxation times $\nu=1$ and $\nu_{opt}$. It can be seen from Fig. \ref{fig:kuz_max_error} and Table \ref{tab:aniso_tr} that the design order of accuracy is obtained for both the relaxation times for the primary variable, $T$, for all the schemes, but the $L_2$ error obtained by $\nu_{opt}$ is much lower than $\nu = 1$. Fig. \ref{fig:2d-aniso-error} shows that the error remains the same for increasing value of $D_{||}$, which is varied from $10^0$ to $10^9$, for $\nu_{opt}$ whereas it increases for $\nu =1$. It is also important to point out that the convergence speed of the computation also improves significantly and is independent of the anisotropy by including the diffusion coefficient in $T_r$. It will be shown in further numerical tests that the choice of $\nu_{opt}$ can make the scheme independent of anisotropy and also the angle of misalignment. The solution contours obtained by U-5E on a 32 $\times$ 32 grid and computed values at the geometric center for various schemes are shown in Figs. \ref{fig:3e-aniso-source} and \ref{fig:s2d-aniso-center} respectively.

\begin{table}[H]
  \centering
\footnotesize
  \caption{$L_2$ errors and order of convergence for $\nu_{opt}$ and $\nu = 1$ by 3rd order explicit, 5th order explicit and compact schemes for anisotropic diffusion problem given in \hyperref[sec:4.1]{Example 4.1}.}
    \begin{tabular}{|c|c|c|c|c|c|c|c|c|}
    \hline
          & \multicolumn{6}{c|}{$\nu_{opt}$}                          & \multicolumn{2}{c|}{$\nu=1$} \\
    \hline
    Number  & Upwind-3E &       & Upwind-5E &       & Upwind-5C &       & Upwind-5C &  \\
\hline
    of points & Error & Order & Error & Order & Error & Order & Error & Order \\
\hline
        $8^2$     & 4.95E-05 &       & 2.81E-06 &       & 3.36E-06 &       & 3.82E-03 &  \\
    \hline
    $16^2$    & 2.60E-06 & 4.25  & 1.12E-07 & 4.65  & 1.20E-07 & 4.80  & 1.07E-04 & 5.16 \\
        \hline
    $32^2$    & 3.59E-07 & 2.86  & 6.15E-09 & 4.19  & 6.41E-09 & 4.23  & 2.30E-06 & 5.54 \\
            \hline
    $64^2$    & 5.12E-08 & 2.81  & 2.27E-10 & 4.76  & 2.40E-10 & 4.74  & 5.64E-08 & 5.35 \\
                \hline
    $128^2$   & 6.84E-09 & 2.90  & 7.57E-12 & 4.91  & 8.10E-12 & 4.89  & 1.54E-09 & 5.20 \\
                    \hline
    \end{tabular}%
  \label{tab:aniso_tr}%
\end{table}%

\subsection{Constant angle of misalignment }
In the following test cases the applicability of the approach to full diffusion tensor, with a constant angle of misalignment $\beta$ shown in Eq.(\ref{eqn:aniso-original}), is analyzed. For all problems in this subsection, we set 
$D_{\perp}=1$ and $D_{||} = 10^\gamma$, where $\gamma$ varies again 0 to 9.

\begin{figure}[H]
\begin{onehalfspacing}
\centering
\subfigure[]{\includegraphics[width=0.41\textwidth]{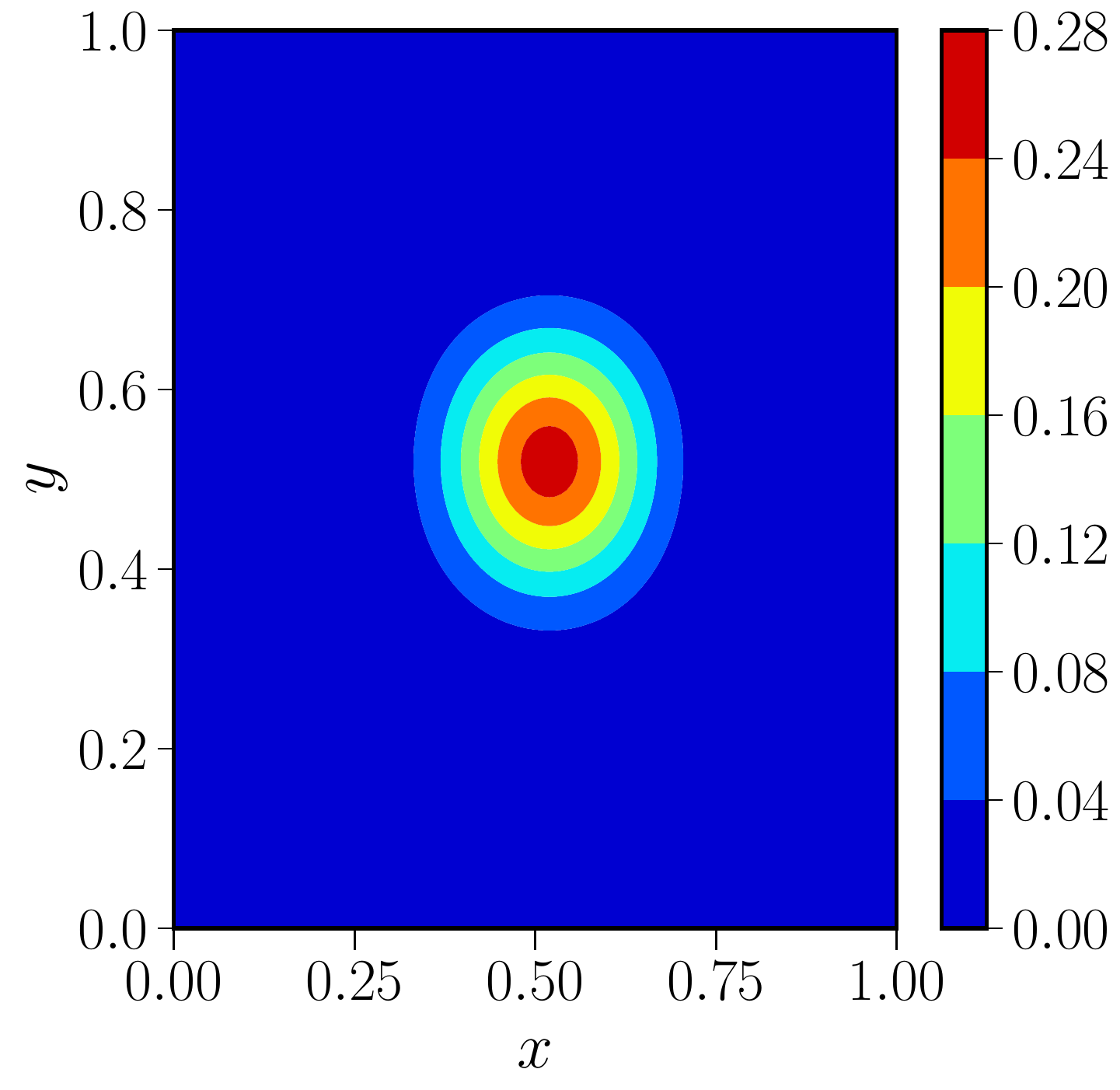}
\label{fig:van_cont}}
\subfigure[]{\includegraphics[width=0.41\textwidth]{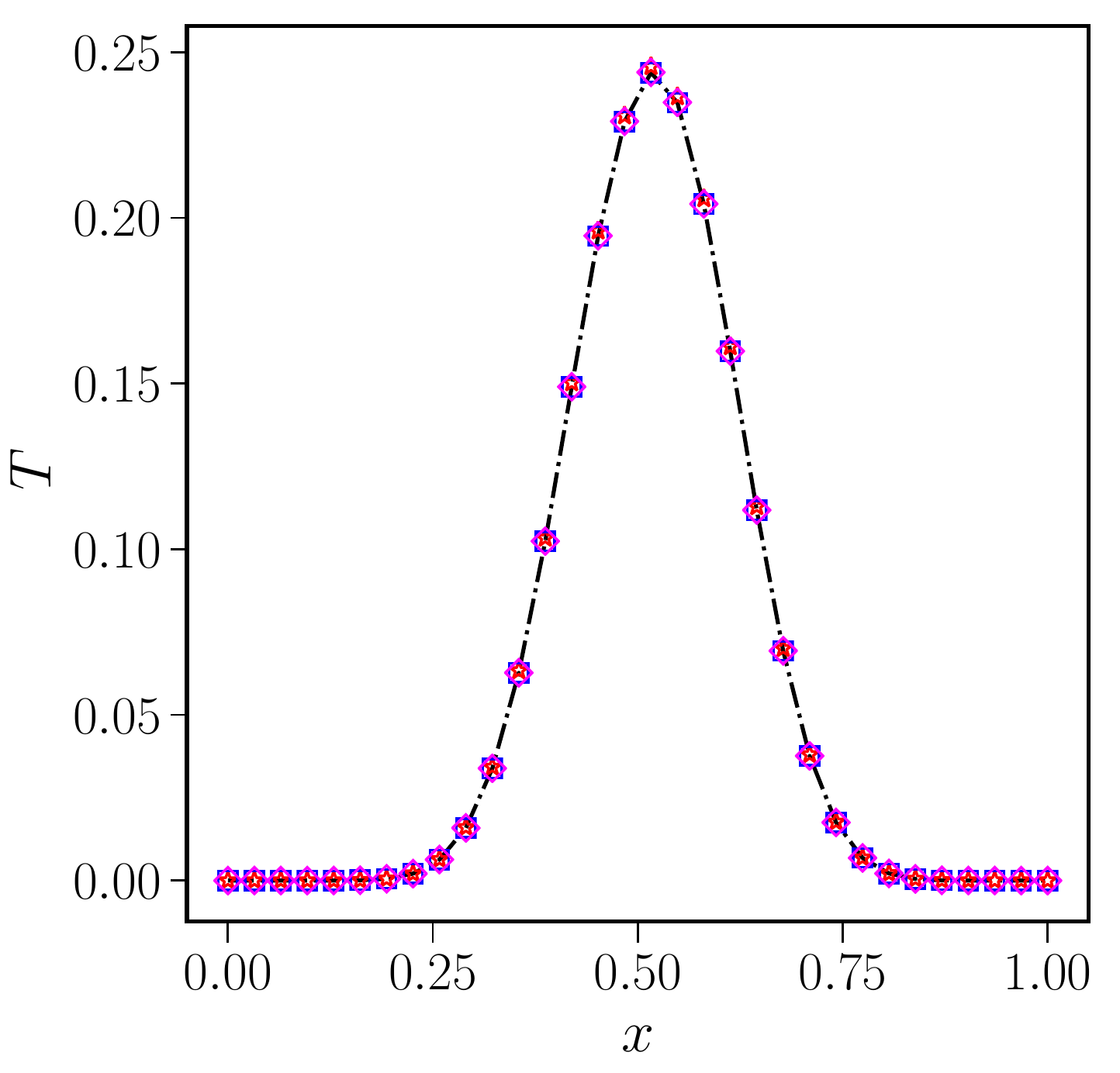}
\label{fig:van_center}}
\subfigure[]{\includegraphics[width=0.44\textwidth]{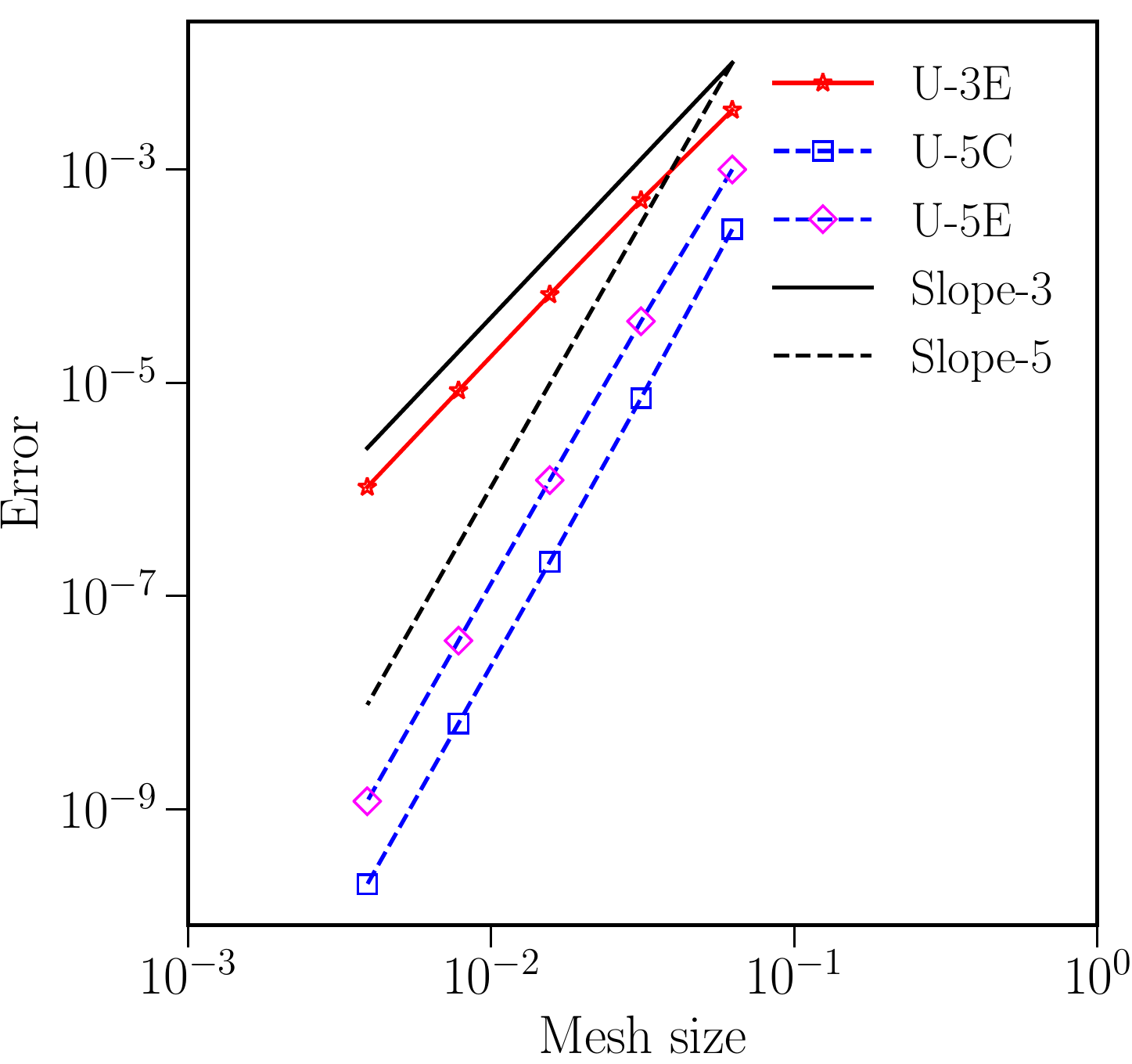}
\label{fig:5C_error_2}}
  \caption{(a) Solution contour for U-5E on a grid size of 96 by 96, (b) computed values at geometric center on 32 $\times$ 32 grid and (c)$L_2$ error for upwind schemes for $\beta$ =$30^0$ and anisotropy $10^9$. Dashed line: analytical; red stars: U-3E; blue squares: U-5C; magenta diamonds: U-5E. \hyperref[sec:4.2a]{Example 4.2a}.}
    \label{fig:Van-sol}
    \end{onehalfspacing}
\end{figure}

\noindent \textbf{Example 4.2a.}\label{sec:4.2a} Second, we consider a simple steady diffusion problem given in Ref. \cite{van2014finite} with the following exact solution, which simulates a temperature peak, 
\begin{equation}
 T_{exact}(x,y) = xy(\sin(\pi x)sin(\pi y))^{\omega}, x,y \in [0,1], 
 \label{eqn1:source}
\end{equation}
where $\omega = 10$ and the angle of misalignment $\beta$ is set to a constant value of $30^{\circ}$. The source term is calculated by plugging Eq.(\ref{eqn1:source}) into Eq. (\ref{eqn:aniso-original}). The simulations are conducted with grid refinements from 8$\times$8 to 128$\times$128. Figure \ref{fig:5C_error_2} shows the results obtained with $\nu_{opt}$ that the design order accuracy is obtained for all the schemes and fifth order compact schemes are significantly more accurate in comparison with explicit schemes. As expected, the relaxation time with $\nu_{opt}$ makes the schemes independent of the anisotropy. 
%\begin{figure}[H]
%\begin{onehalfspacing}
%\centering
%
%  \caption{ for \hyperref[sec:4.2a]{Example 4.2a}.}
%    \label{fig:Van-c}
%    \end{onehalfspacing}
%\end{figure}
Figures \ref{fig:van_cont} and \ref{fig:van_center} show the solution contours and order of accuracy for an anisotropy of $10^9$ at a constant $\beta$ of $30^{\circ}$ respectively. All the schemes can capture the temperature peak without any numerical diffusion.
\begin{figure}[H]
\centering
\begin{onehalfspacing}
\subfigure[U-3E]{\includegraphics[width=0.44\textwidth]{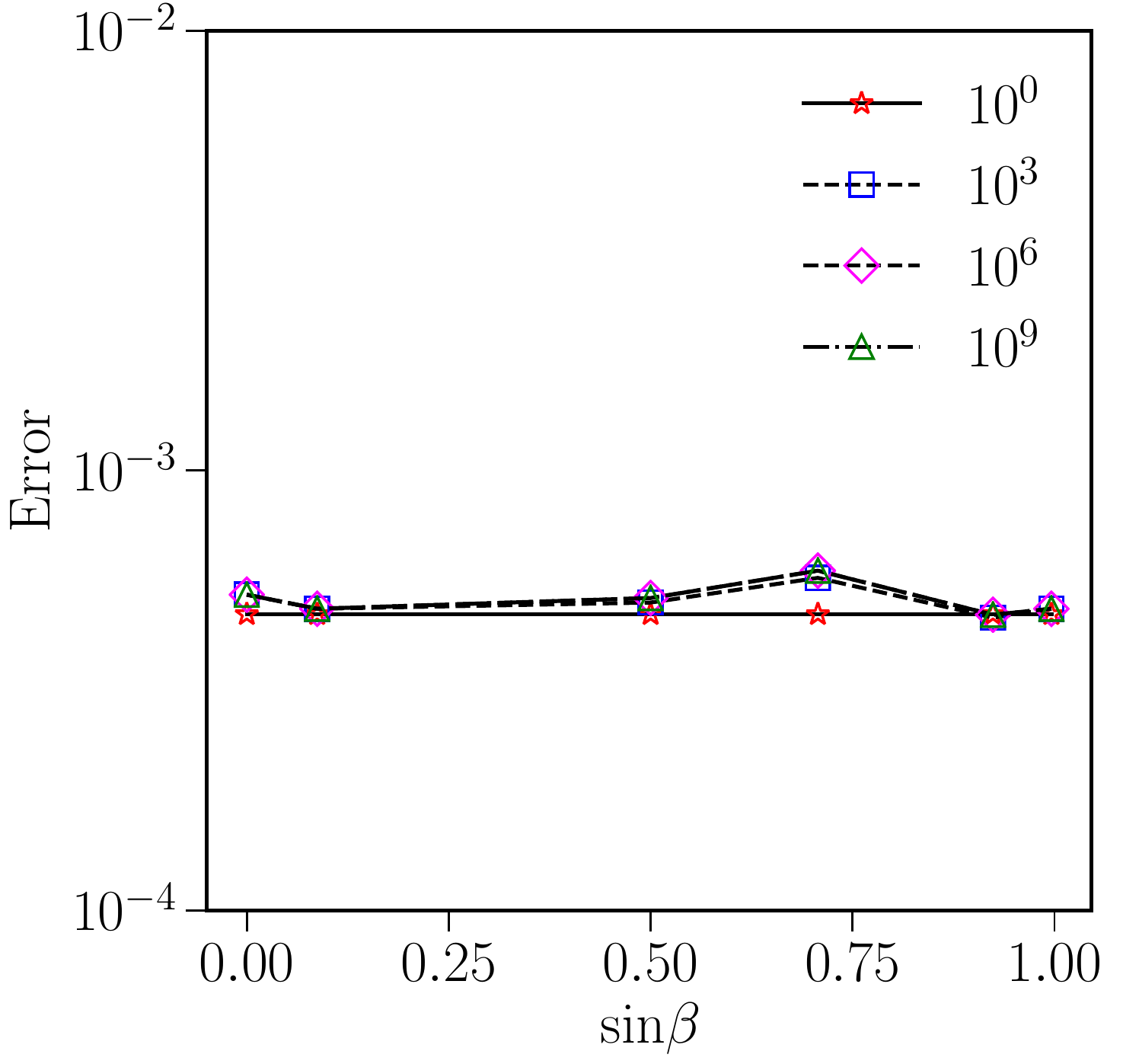}
\label{fig:3E_error_1}}
\subfigure[U-5E]{\includegraphics[width=0.44\textwidth]{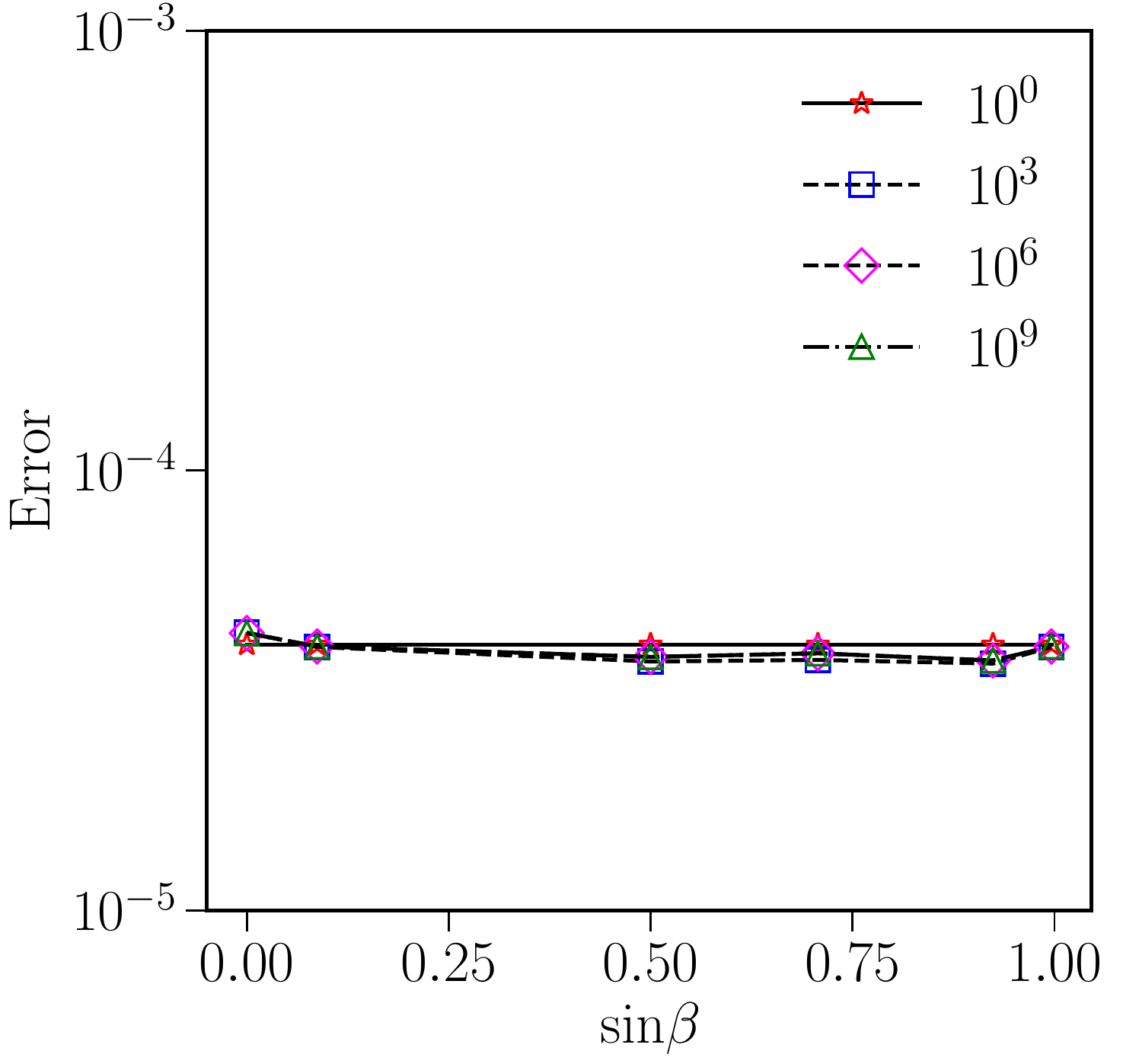}
\label{fig:5E_error_1}}
\subfigure[U-5C]{\includegraphics[width=0.44\textwidth]{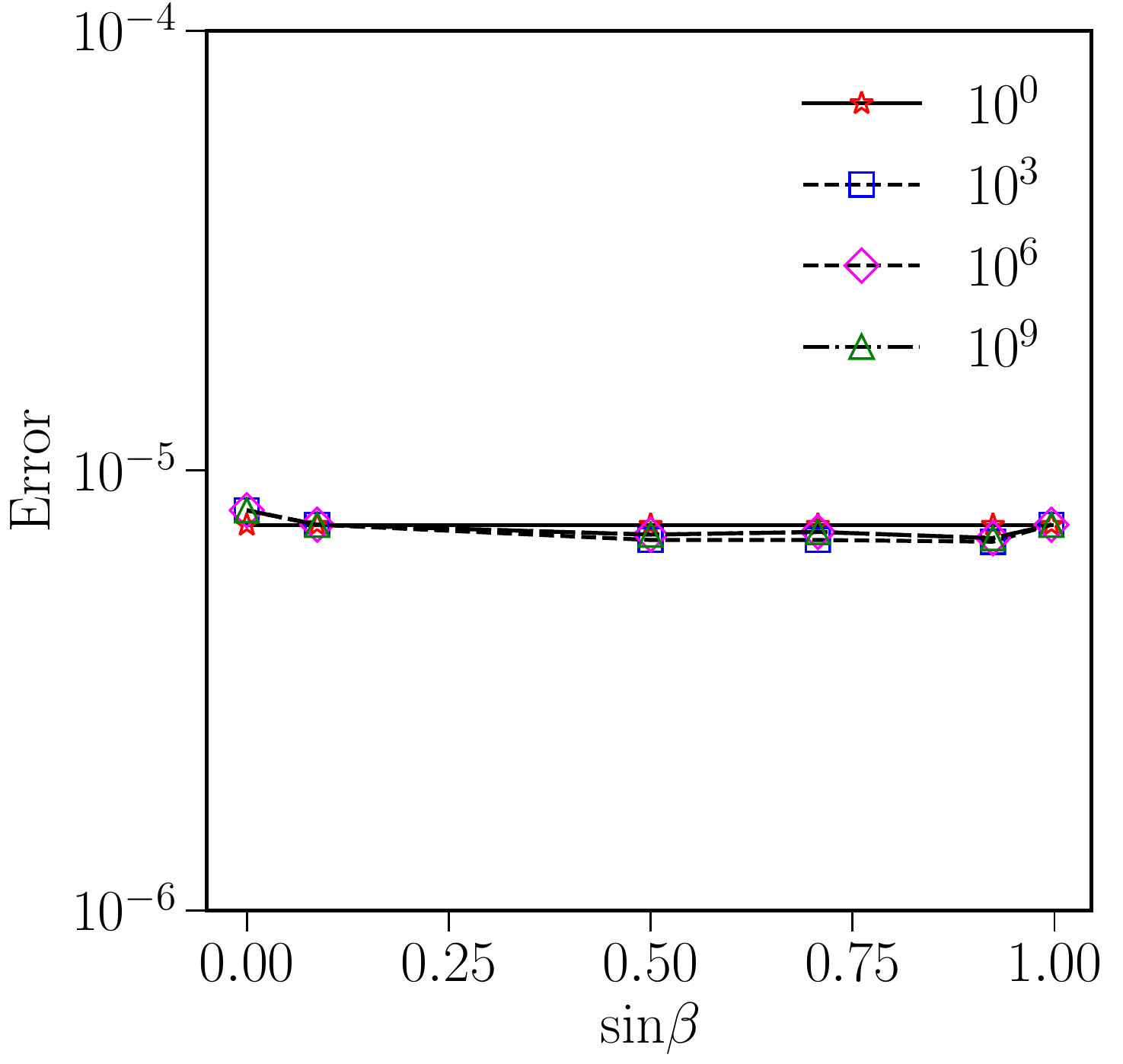}
\label{fig:5C_error_1}}
\subfigure[]{\includegraphics[width=0.44\textwidth]{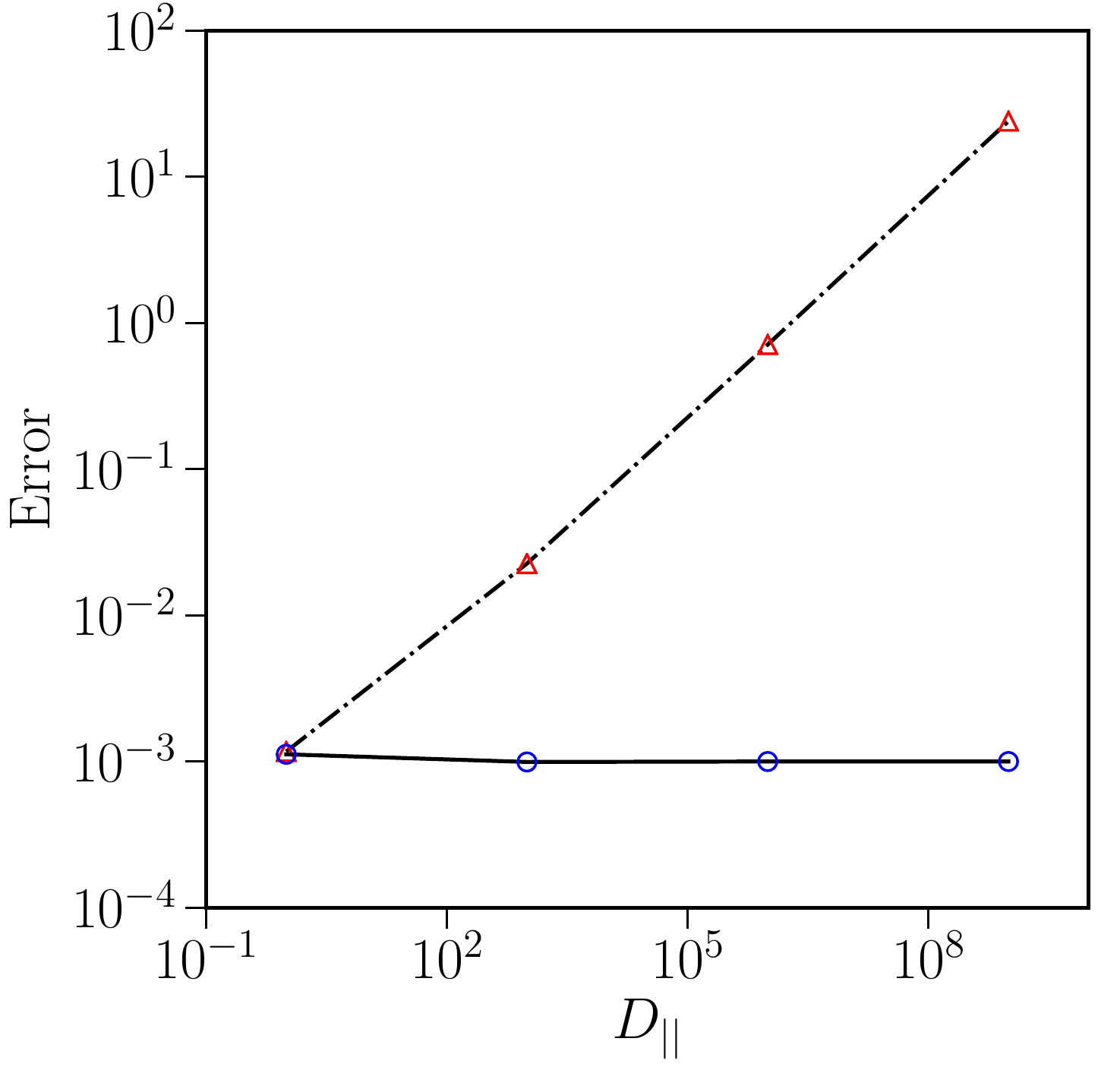}
\label{fig:error_Van}}
  \caption{Figures (a),(b) and (c) show the accuracy of U-3E, U-5E and U-5C schemes for different angles and increasing levels of anisotropy for $\nu_{opt}$ on a grid size of 32 $\times$ 32 (d) Accuracy plot for increasing degree of anisotropy on a grid size of 16 $\times$ 16 for U-5E. Blue circles: $\nu_{opt}$; red triangles: $\nu =1$ for \hyperref[sec:4.2a]{Example 4.2a}.}
  \label{fig:Tr-1r}
  \end{onehalfspacing}
\end{figure}
It can be easily confirmed from the Figs. \ref{fig:3E_error_1}, \ref{fig:5E_error_1}, and \ref{fig:5C_error_1} that the schemes are independent of both angle and degree of anisotropy even for an extreme an anisotropy of $10^{9}$ as the angle of misalignment, $\beta$, is varied from ${0^{\circ}}$ to ${90^{\circ}}$. It can be seen from Fig. \ref{fig:error_Van} that the results obtained by $\nu_{opt}$ are independent of anisotropy whereas the error increases with anisotropy for $\nu=1$.
\begin{table}[H]
  \centering
\footnotesize
  \caption{$L_2$ errors and order of convergence for $\nu_{opt}$ by 3rd order explicit, 5th order explicit and compact schemes for anisotropic diffusion problem given in \hyperref[sec:4.2a]{Example 4.2a}.}
    \begin{tabular}{| c | c | c | c | c | c | c | c | c|}
\hline
\hline
    Number & \multicolumn{2}{c|}{Upwind-3E} & \multicolumn{2}{c|}{Upwind-5C} & \multicolumn{2}{c|}{Upwind-5E}  \\
    \cline{2-7}
    of points& error & order & error & order & error & order    \\
    \cline{1-7}
    \hline
    $16^2$    & 3.61E-03 &   -    & 2.75E-04 &  -     & 1.00E-03 & - \\
    \hline
    $32^2$    & 5.13E-04 & 2.82  & 7.15E-06 & 5.26  & 3.77E-05 & 4.73 \\
    \hline
    $64^2$    & 6.71E-05 & 2.93  & 2.09E-07 & 5.09  & 1.22E-06 & 4.95 \\
    \hline 
    $128^2$   & 8.44E-06 & 2.99  & 6.39E-09 & 5.04  & 3.81E-08 & 5.00 \\
    \hline
    $256^2$   & 1.05E-06 & 3.00  & 1.98E-10 & 5.01  & 1.19E-09 & 5.01 \\
    \hline 
    \end{tabular}%
  \label{tab:4.2a}%
\end{table}%

\begin{figure}[H]
\centering
\begin{onehalfspacing}
\subfigure[]{\includegraphics[width=0.46\textwidth]{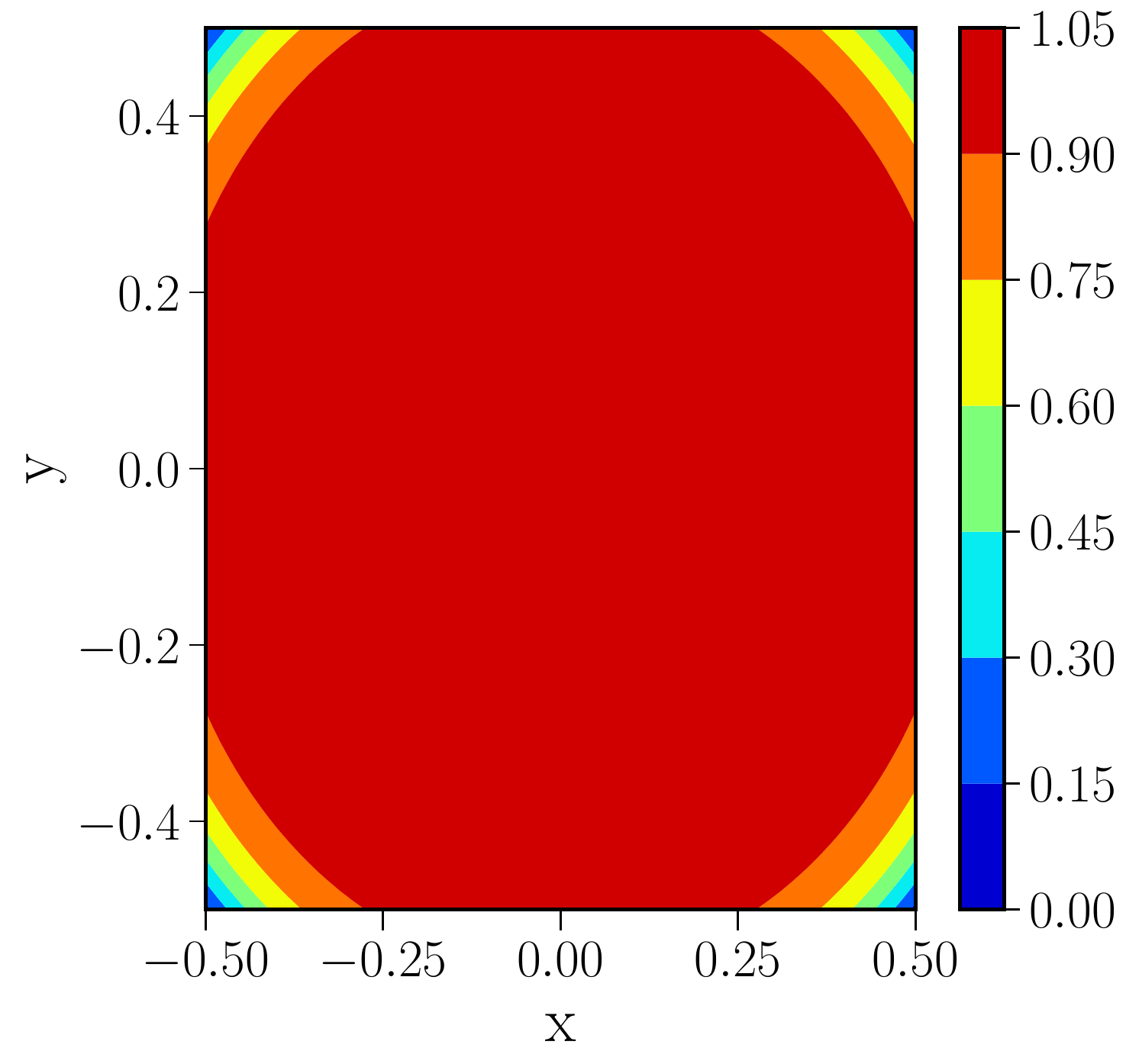}
\label{fig:5C_soln}}
\subfigure[]{\includegraphics[width=0.44\textwidth]{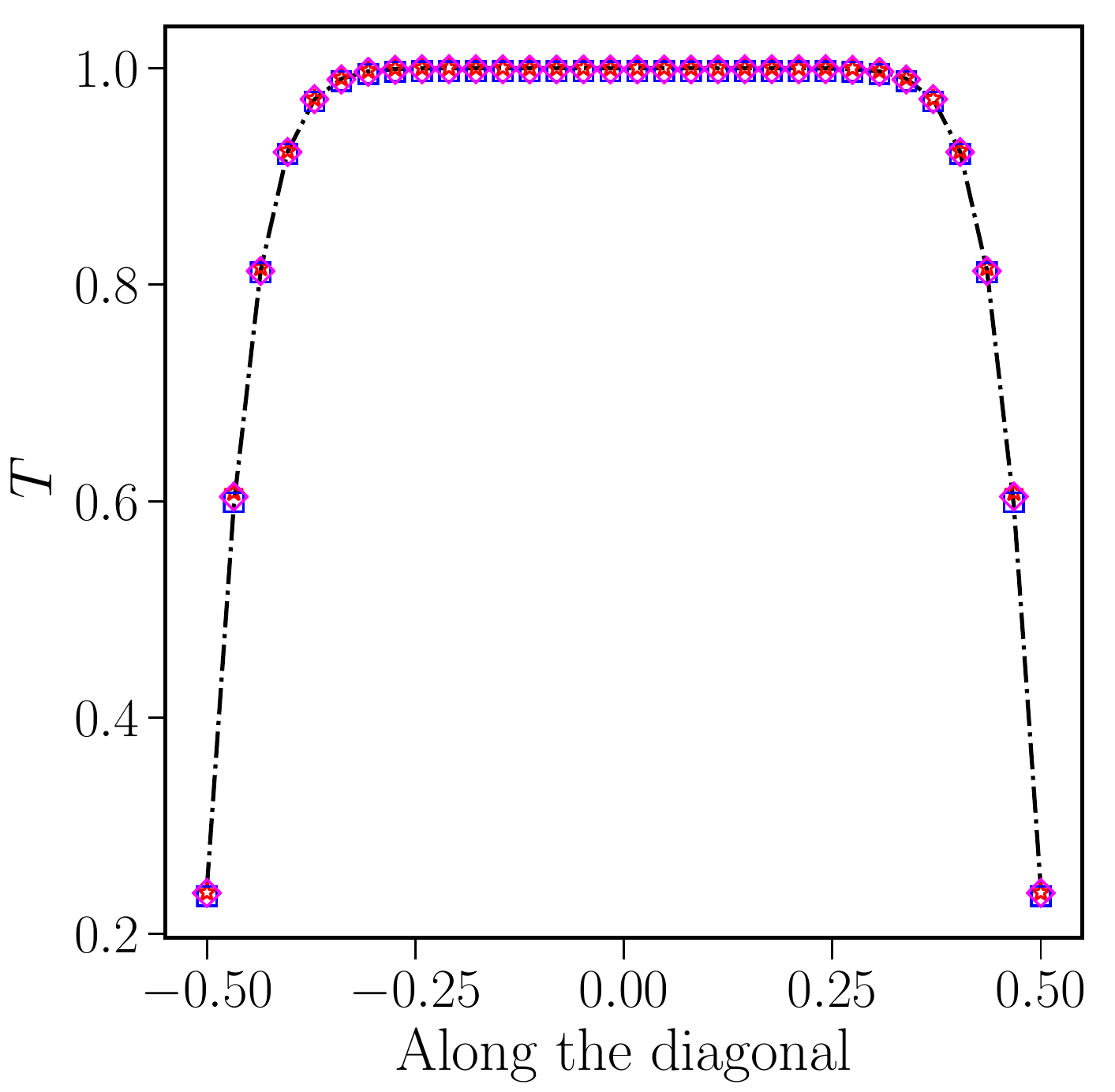}
\label{fig:error_high}}
\subfigure[]{\includegraphics[width=0.44\textwidth]{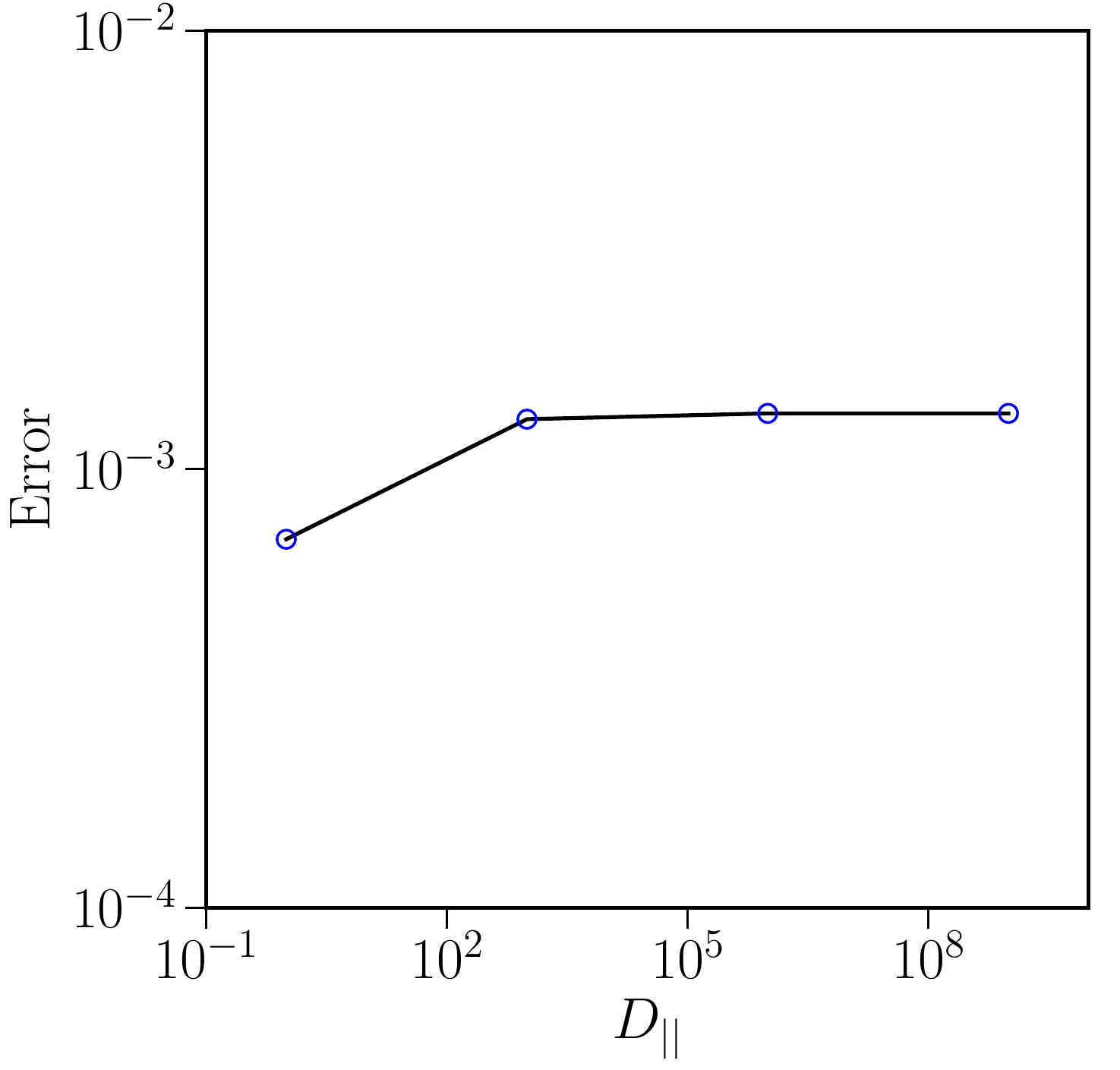}
\label{fig:error_opt}}
  \caption{(a) Solution contours by U-5E on a grid size of 96 $\times$ 96, (b) comparison of numerical solution obtained by various schemes along the diagonal on a grid size of 32 $\times$ 32 and (c) Accuracy plot for increasing degree of anisotropy on a grid size of 16 $\times$ 16 for U-5E. Blue circles: $\nu_{opt}$ for \hyperref[sec:4.2b]{Example 4.2b}. Dashed line: analytical; red stars: U-3E; blue squares: U-5C; magenta diamonds: U-5E. }
    \label{fig:gradient-c}
    \end{onehalfspacing}
\end{figure}

\noindent \textbf{Example 4.2b.}\label{sec:4.2b} In this test case, the following temperature distribution from Ref. \cite{van2016finite} is considered to test the effect of steep temperature gradients,
\begin{equation}
 T_{exact}(x, y) = \frac{1}{1 - e^{- 2^{\frac{1}{2 \omega}}}} \left({e^{- 2^{\omega} \left(x^{2} + y^{2}\right)^{\omega}} - e^{- 2^{\frac{1}{2 \omega}}}}\right), x,y \in [-0.5, 0.5],
\end{equation}
where {\color{black} $D_{||} = 1, 10^3, 10^6$, and $10^9$} and $\omega$ is taken as 6, and the angle of misalignment is set to a constant value of $45^{\circ}$ for the anisotropy of $10^9$. Finite-volume schemes proposed in Ref. \cite{van2016finite} were unable to capture the sharp gradients in temperature and were not independent of anisotropy. Only third-order boundary conditions are considered for all the schemes due to steep gradients close to the boundary. It can be seen from Fig. \ref{fig:error_high} that the hyperbolic scheme can also handle situations with severe anisotropy as well as sharp gradients without losing accuracy. Table \ref{tab:aniso_gradient} show that the third order of accuracy is obtained for the primary variable, $T$, for all the schemes. 

\begin{table}[H]
  \centering
\begin{onehalfspacing}
\footnotesize
  \caption{$L_2$ errors and order of convergence for $D_{||}=10^9$, \hyperref[sec:4.2b]{Example 4.2b}.}
    \begin{tabular}{| c | c | c | c | c | c | c | c | c|}
\hline
\hline
    Number & \multicolumn{2}{c|}{Upwind-3E} & \multicolumn{2}{c|}{Upwind-5E} & \multicolumn{2}{c|}{Upwind-5C}  \\
    \cline{2-7}
    of points& error & order & error & order & error & order    \\
    \cline{1-7}
    $8^2$     & 5.81E-02 &       & 3.03E-02 &       & 3.18E-02 &  \\
        \hline
    $16^2$    & 6.90E-03 & 2.13  & 6.03E-03 & 2.33  & 5.16E-03 & 2.62 \\
            \hline
    $32^2$    & 1.78E-03 & 1.95  & 1.34E-03 & 2.17  & 1.24E-03 & 2.06 \\
                \hline
    $64^2$    & 3.04E-04 & 2.55  & 2.10E-04 & 2.68  & 1.92E-04 & 2.69 \\
                    \hline
    $128^2$   & 4.11E-05 & 2.89  & 2.64E-05 & 2.99  & 2.06E-05 & 3.22 \\
    \hline
    \end{tabular}%
  \label{tab:aniso_gradient}%
  \end{onehalfspacing}
\end{table}%

\subsection{Variable diffusion tensor}\label{sec:4.3}
In the earlier test cases the angle of misalignment is constant, and thereby the diffusion tensor is constant in space. It is found from the earlier test cases the $T_r$ with $\nu_{opt}$ is required to make the schemes independent of the diffusion tensor. In the following test cases, we further test the approach where diffusion coefficients are space varying which can also be thought of as varying angle of misalignment. The physical domain is [0, 1] $\times$ [0, 1] for both the test cases. The preconditioned formulation {\color{black} employed in this study 
is necessary for variable diffusion tensor problems \cite{nishikawa_diff_discon:jcp2018}, where $\nu_{opt}$ is a function of space. Again, we set $D_{\perp}=1$ for all problems considered below. }

\begin{figure}[H]
\centering
\begin{onehalfspacing}
\subfigure[Solution contours]{\includegraphics[width=0.44\textwidth]{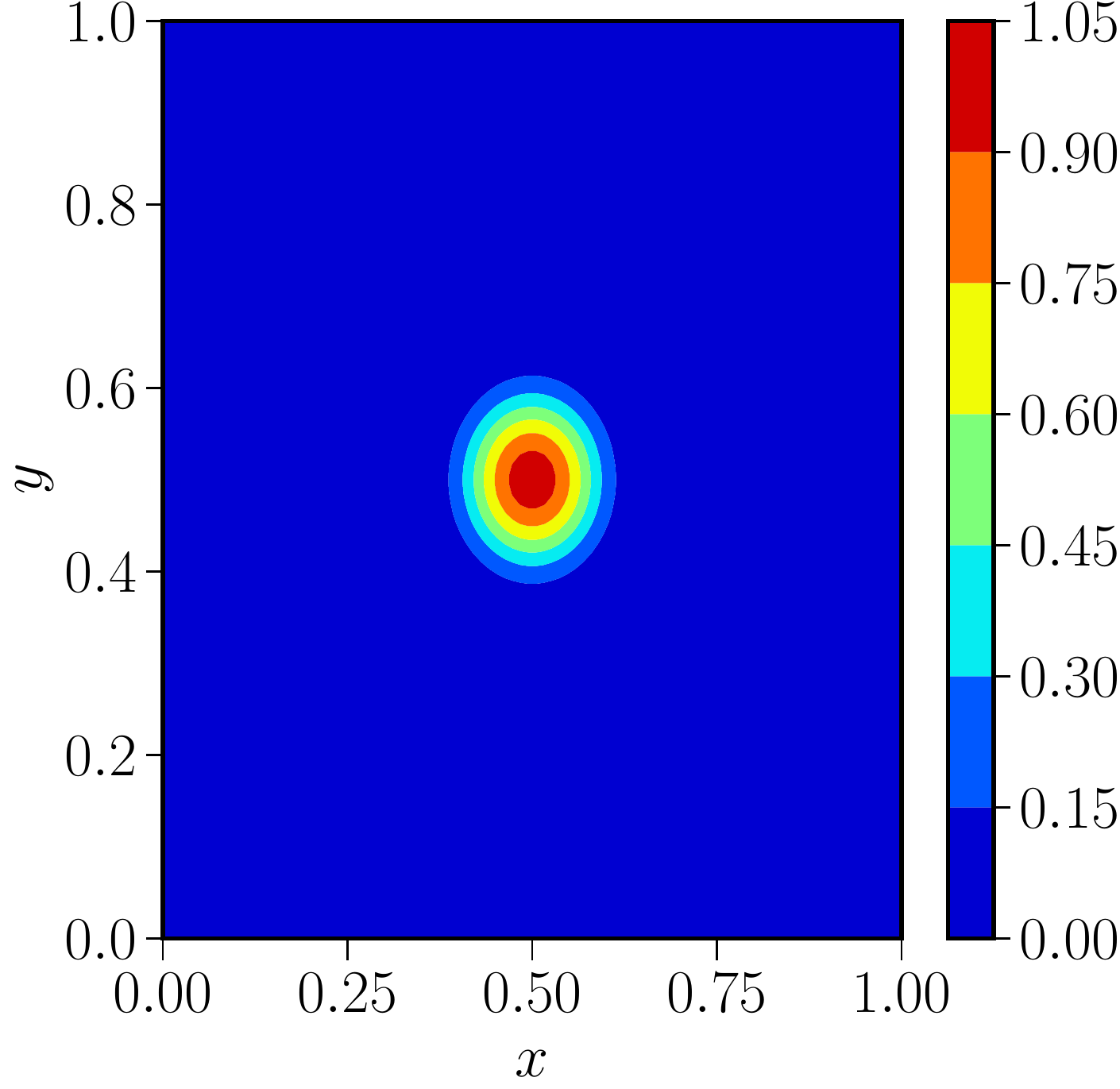}
\label{fig:lou_cont}}
\subfigure[Order of accuracy]{\includegraphics[width=0.44\textwidth]{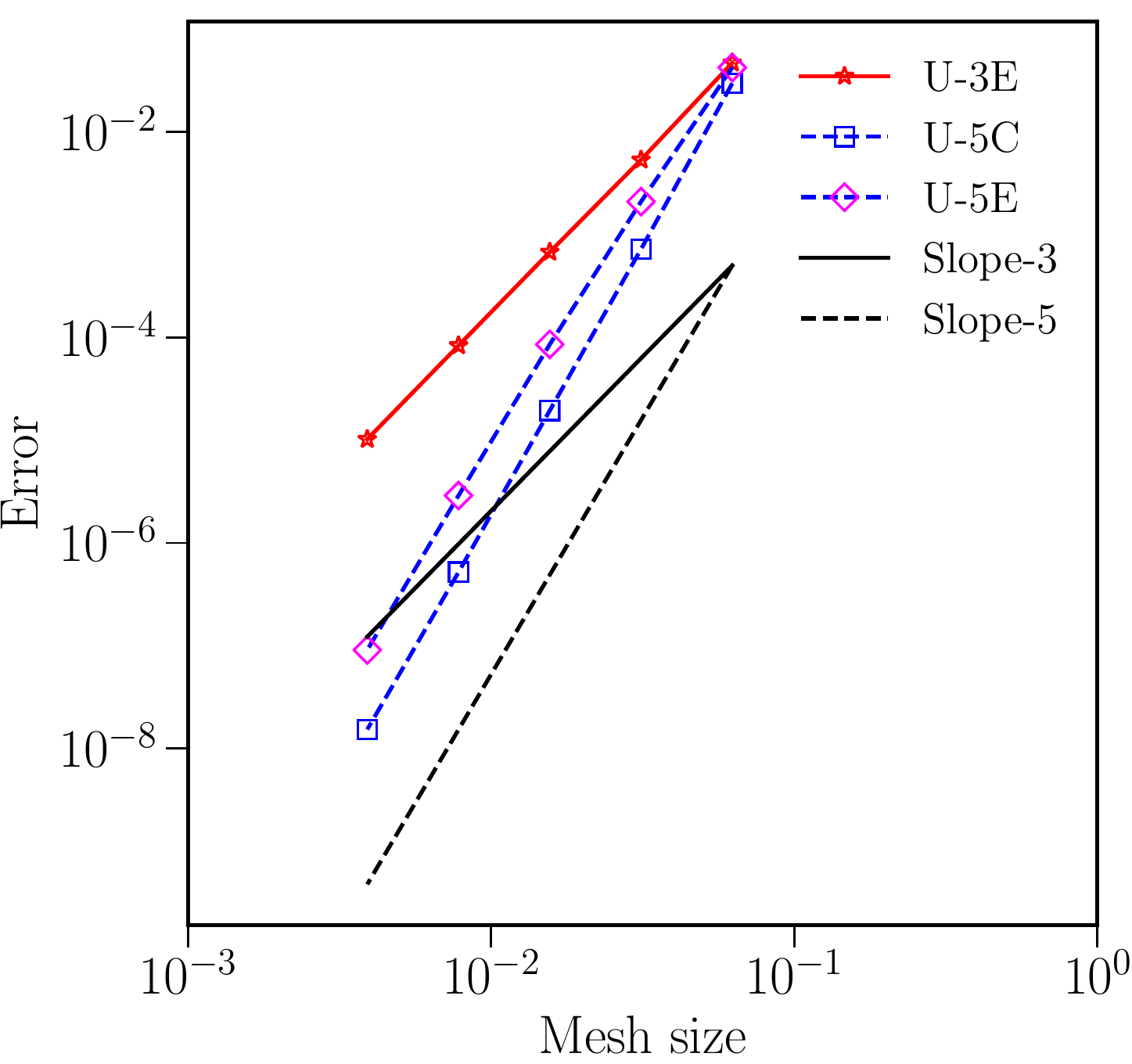}
\label{fig:lou_error}}
  \caption{Solution contours and  order of accuracy obtained by various schemes for \hyperref[sec:4.3a]{Example 4.3a}. }
  \end{onehalfspacing}
\end{figure}

\noindent \textbf{Example 4.3a}\label{sec:4.3a} For the first test case, the exact solution is given by $T(x,y)  = 1-\tanh \left(\frac{(x-0.5)^2+(y-0.5)^2}{0.01}\right)$  and the diffusion tensor given by Lou et al. \cite{lou2017reconstructed,Lou2019} is modified by including extreme anisotropy through $D_{||}$,
\begin{equation}
D = \left[\begin{matrix} D_{||}(y^{2} + \left(x + 1\right)^{2}) & - x y\\ \\ - x y & \left(y + 1\right)^{2}\end{matrix}\right] = \left[\begin{matrix} 10^{\gamma}(y^{2} + \left(x + 1\right)^{2}) & - x y\\ \\ - x y & \left(y + 1\right)^{2}\end{matrix}\right],
\end{equation}
%\begin{align}
%F &= - \left(\tanh^{2}{\left (100.0 \left(x - 0.5\right)^{2} + 100.0 \left(y - 0.5\right)^{2} \right )} - 1\right) \notag\\ 
% & \left(- x \left(200.0 x - 100.0\right) + 200.0 y^{2} - y \left(200.0 y - 100.0\right) + 200.0 \left(x + 1\right)^{2} + 2 \left(x + 1\right) \left(200.0 x - 100.0\right) 
%  + 2 \left(200.0 x - 100.0\right) \left(x y \left(200.0 y - 100.0\right) - \left(200.0 x - 100.0\right) \left(y^{2} + \left(x + 1\right)^{2}\right)\right)
%  \tanh{\left (100.0 \left(x - 0.5\right)^{2} + 100.0 \left(y - 0.5\right)^{2} \right )} + 200.0 \left(y + 1\right)^{2} + 2 \left(y + 1\right)
%   \left(200.0 y - 100.0\right) + 2 \left(200.0 y - 100.0\right) \left(x y \left(200.0 x - 100.0\right) - \left(y + 1\right)^{2} \left(200.0 y - 100.0\right)\right)
%    \tanh{\left (100.0 \left(x - 0.5\right)^{2} + 100.0 \left(y - 0.5\right)^{2} \right )}\right)
%\end{align}
where $\gamma =  9$. Again, the grid refinement study is carried out for all three upwind schemes, and the design order of accuracy is achieved for all them, as shown in Fig. \ref{fig:lou_error}. Once again, the compact scheme is more accurate than explicit interpolation. Solution contours computed on a grid size of 96$\times$96 is shown in Fig. \ref{fig:lou_cont}. 

\begin{figure}[H]
\centering
\subfigure[]{\includegraphics[width=0.44\textwidth]{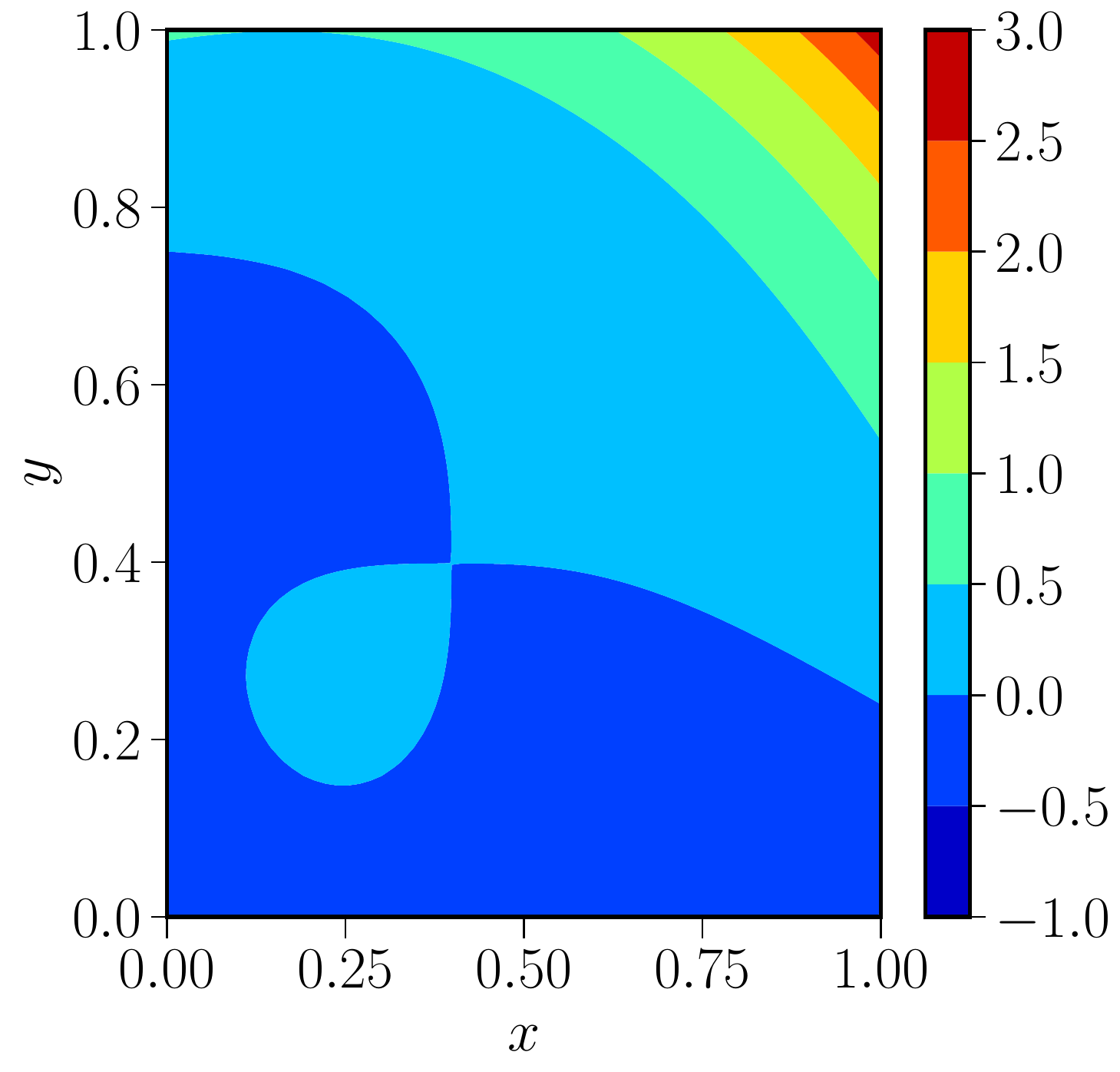}
\label{fig:tilt_cont}}
\subfigure[]{\includegraphics[width=0.44\textwidth]{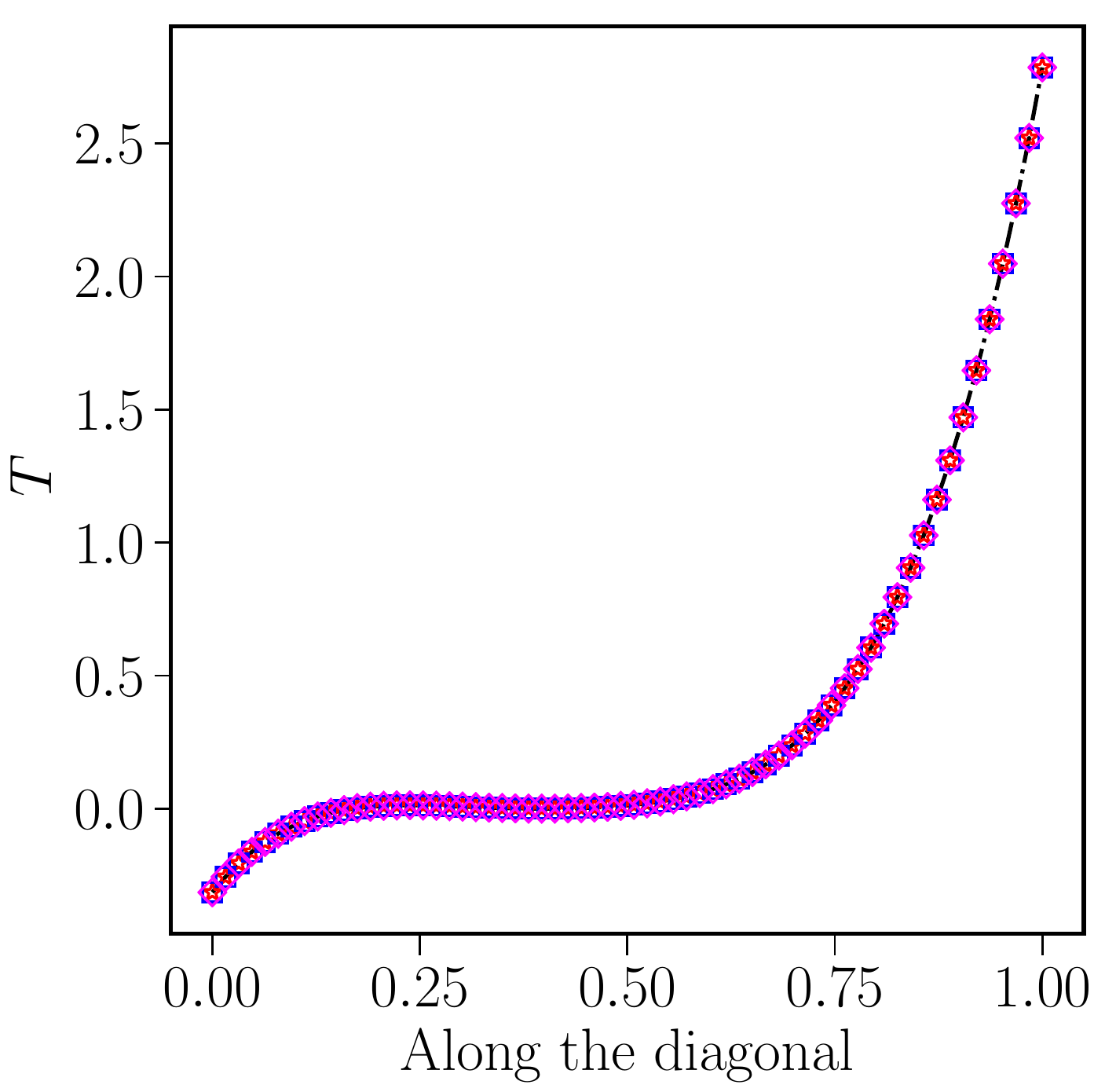}
\label{fig:tilt_center}}
\subfigure[]{\includegraphics[width=0.44\textwidth]{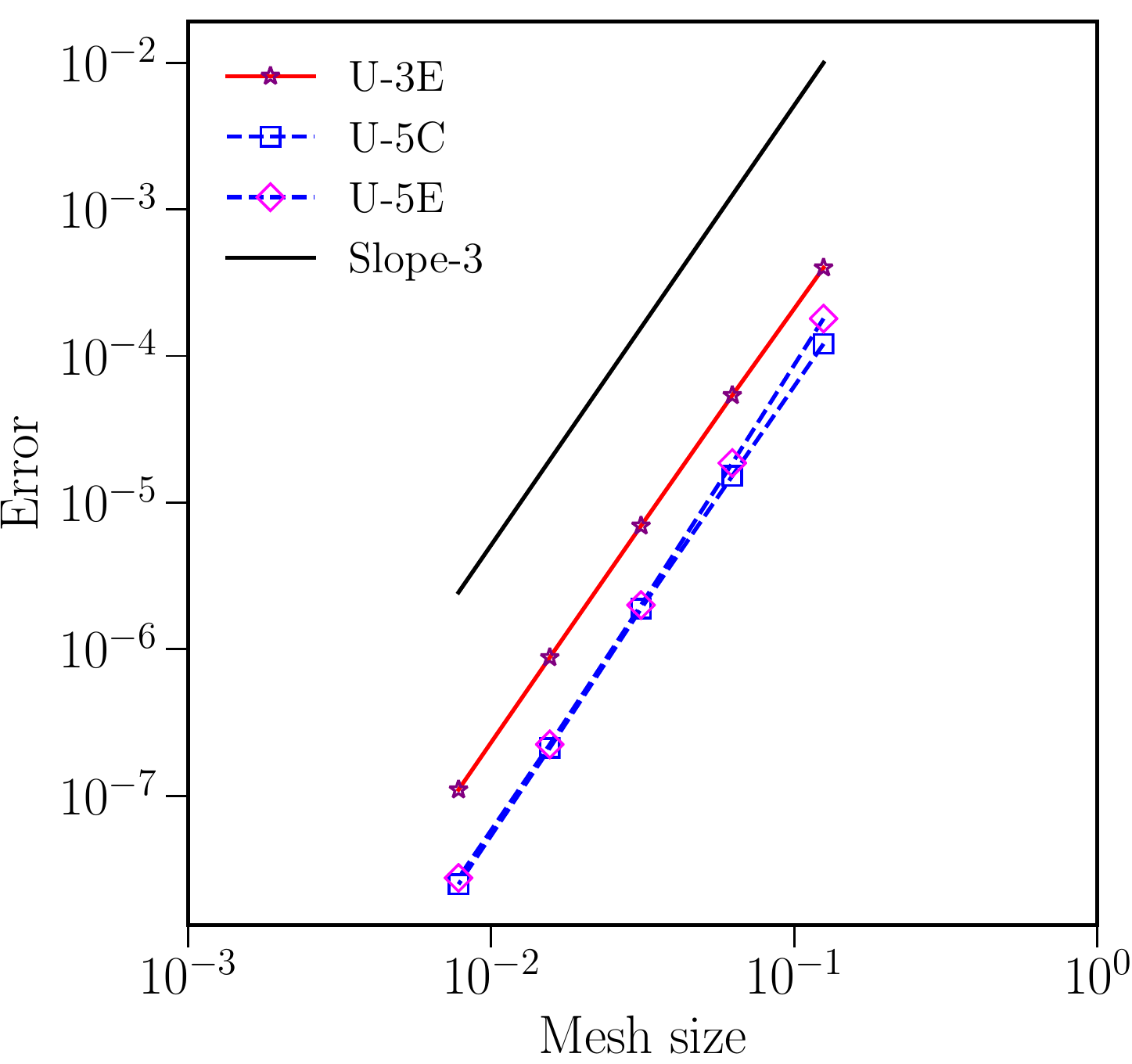}
\label{fig:tilt_order}}
  \caption{(a) Solution contour for U-5E on a grid size of 64 $\times$ 64, (b) computed values at geometric center on 64 $\times$ 64 grid and (c)$L_2$ error for upwind schemes for an anisotropy of $10^9$. Dashed line: analytical; red stars: U-3E; blue squares: U-5C; magenta diamonds: U-5E. \hyperref[sec:4.3b]{Example 4.3b}.}
    \label{fig:Van-sol-2}
\end{figure}

\noindent \textbf{Example 4.3b}\label{sec:4.3b}
In the second test case, the exact solution is 
\begin{equation}
T(x,y) = (xy+(ax+by)(x^2+y^2)^\frac{3}{2}),
\end{equation}
where a = 2 and b =5. The diffusion is tensor is varied by the angle $\beta$ as follows,
\begin{align*}
\quad D = \left[\begin{matrix}D_{||} \cos^{2}{\left (\beta \right )} + D_{\perp}\sin^{2}{\left (\beta \right )} & \frac{1}{2} \left(D_{||} - D_{\perp}\right) \sin{\left (2 \beta \right )}\\ \\ \frac{1}{2} \left(D_{||} -D_{\perp}\right) \sin{\left (2 \beta \right )} & D_{||} \sin^{2}{\left (\beta \right )} + D_{\perp}\cos^{2}{\left (\beta \right )}\end{matrix}\right],
\end{align*}
where {\color{black} $D_{||} =10^9$} and $\mathbf{\beta} = \arctan\left(x+y\right)$.
Again, grid refinement study is carried out for all three upwind schemes and the design order of accuracy is achieved for all them, as shown in Fig. \ref{fig:Van-sol-2}. All the schemes are 3rd order accurate because of the boundary conditions as shown in Table \ref{tab:aniso_angle}. Solution contours computed on a grid size of 64$\times$64 is shown in Fig. \ref{fig:tilt_cont}.

\begin{table}[H]
  \centering
  \caption{$L_2$ errors and order of convergence for $D_{||}=10^9$, \hyperref[sec:4.3b]{Example 4.3b}.}
  \begin{onehalfspacing}
  \footnotesize
    \begin{tabular}{| c | c | c | c | c | c | c | c | c|}
\hline
\hline
    Number & \multicolumn{2}{c|}{Upwind-3E} & \multicolumn{2}{c|}{Upwind-5C} & \multicolumn{2}{c|}{Upwind-5E}  \\
    \cline{2-7}
    of points& error & order & error & order & error & order    \\
    \cline{1-7}
    \hline
    $16^2$    & 4.00E-04 &       & 1.21E-04 &       & 1.80E-04 &  \\
    \hline 
    $32^2$    & 5.39E-05 & 2.89  & 1.52E-05 & 3.00  & 1.86E-05 & 3.27 \\
    \hline 
    $64^2$    & 6.94E-06 & 2.96  & 1.89E-06 & 3.01  & 2.00E-06 & 3.22 \\
    \hline
    $128^2$   & 8.78E-07 & 2.98  & 2.12E-07 & 3.15  & 2.24E-07 & 3.15 \\
    \hline
    $256^2$   & 1.10E-07 & 2.99  & 2.49E-08 & 3.09  & 2.75E-08 & 3.03 \\
    \hline 
    \hline
    \end{tabular}%
  \label{tab:aniso_angle}%
  \end{onehalfspacing}
\end{table}%fig:Van-sol

\subsection{Nonlinear examples}
Until now we have considered only linear problems where the diffusion tensor is either constant or varying in space. In real-world applications of MHD or Hall thruster simulations, the diffusion coefficient can be a function of the temperature. In this section, we demonstrate the applicability of the approach for nonlinear anisotropic diffusion problems. 
{\color{black} Note that the preconditioned formulation needs to be employed for nonlinear problems \cite{nishikawa_diff_discon:jcp2018}, 
where $\nu_{opt}$ depends on the solution value.}

\noindent {\textbf{Example 4.4a}}\label{sec:4.4a}
\noindent First test case is a steady state nonlinear diffusion problem in a unit square, i.e., 
\begin{equation}
0 = {\partial_x (\mathbf{D} \partial_x T})+{\partial_y (\mathbf{D} \partial_y T}) - S,
\end{equation}
where the diffusion coefficient, the exact solution, and the source term given by
\begin{equation}\label{eq:aniso-non-1}
\begin{aligned}
& \mathbf{D} = \left[ \begin{array}{ccc}
D_{xx} & 0 \\
0 & D_{yy} \\
\end{array} \right] = \left[ \begin{array}{ccc}
D_{||}(1+ T^2) & 0 \\
0 &  1+T^2  \\
\end{array} \right] =  \left[ \begin{array}{ccc}
10^{\gamma}(1+T^2) & 0 \\
0 & 1+T^2 \\
\end{array} \right]\\
&T_{exact}= \sin(\pi x) \sin(\pi y), \\
&S = \pi^{2} \left(3 D_{||} \sin^{2}{\left (\pi x \right )} \sin^{2}{\left (\pi y \right )} - 2 D_{||} \sin^{2}{\left (\pi y \right )} + D_{||} + 3 \sin^{2}{\left (\pi x \right )} \sin^{2}{\left (\pi y \right )} - 2 \sin^{2}{\left (\pi x \right )} + 1\right) \\ & \sin{\left (\pi x \right )} \sin{\left (\pi y \right )}.
\end{aligned}
\end{equation}
The problem reduces to isotropic diffusion, which is considered in Ref. \cite{nishikawa2018dimensional}, for $D_{||} = 1$. Table \ref{tab:aniso_non} shows the $L_2$ errors for $D_{||}$ = $10^0$ and $10^9$. It can be seen that the scheme are independent of the degree of anisotropy and design order of accuracy is obtained for all them.
{\color{black}
\begin{table}[H]
  \centering
  \caption{$L_2$ errors and order of convergence for primary variable T by 3rd order explicit, 5th order explicit and compact schemes for anisotropic diffusion problem given in \hyperref[sec:4.4a]{Example 4.4a}.}
  \footnotesize
    \begin{tabular}{| c | c | c | c | c | c | c | c | c|}
\hline
\hline
    Number & \multicolumn{2}{c|}{Upwind-3E} & \multicolumn{2}{c|}{Upwind-5C} & \multicolumn{2}{c|}{Upwind-5E}  \\
    \cline{2-7}
    of points& error & order & error & order & error & order    \\
    \cline{1-7}
   \hline
   \hline
   \multicolumn{7}{c}{$D_{||}=10^0$} \\
   \hline
   \hline
   $16^2$    & 1.69E-04 &  -     & 4.30E-06 &    -   & 4.73E-06 & - \\
    \hline
   $32^2$    & 2.29E-05 & 2.89  & 1.50E-07 & 4.84  & 1.68E-07 & 4.81 \\
   \hline
   $64^2$    & 2.97E-06 & 2.95  & 4.73E-09 & 4.98  & 5.47E-09 & 4.94 \\
   \hline
   $128^2$   & 3.77E-07 & 2.98  & 1.47E-10 & 5.01  & 1.73E-10 & 4.98 \\
   \hline
   $256^2$   & 4.74E-08 & 2.99  & 4.56E-12 & 5.01  & 5.42E-12 & 4.99 \\
   \hline
   \hline
   \multicolumn{7}{c}{$D_{||}=10^9$} \\
    \hline
    \hline
    $16^2$    & 8.94E-05 &       & 3.32E-06 &       & 3.54E-06 &  \\
    \hline
    $32^2$    & 1.23E-05 & 2.87  & 1.27E-07 & 4.71  & 1.35E-07 & 4.71 \\
    \hline
    $64^2$    & 1.62E-06 & 2.92  & 4.16E-09 & 4.93  & 4.54E-09 & 4.90 \\
    \hline
    $128^2$   & 2.08E-07 & 2.96  & 1.31E-10 & 4.99  & 1.45E-10 & 4.97 \\
    \hline
    $256^2$   & 2.64E-08 & 2.98  &  4.10E-12     & 5.00 & 4.60E-12 & 4.98 \\
    \hline
    \end{tabular}%
  \label{tab:aniso_non}%
\end{table}}%

\begin{figure}[H]
\centering
\subfigure[]{\includegraphics[width=0.4\textwidth]{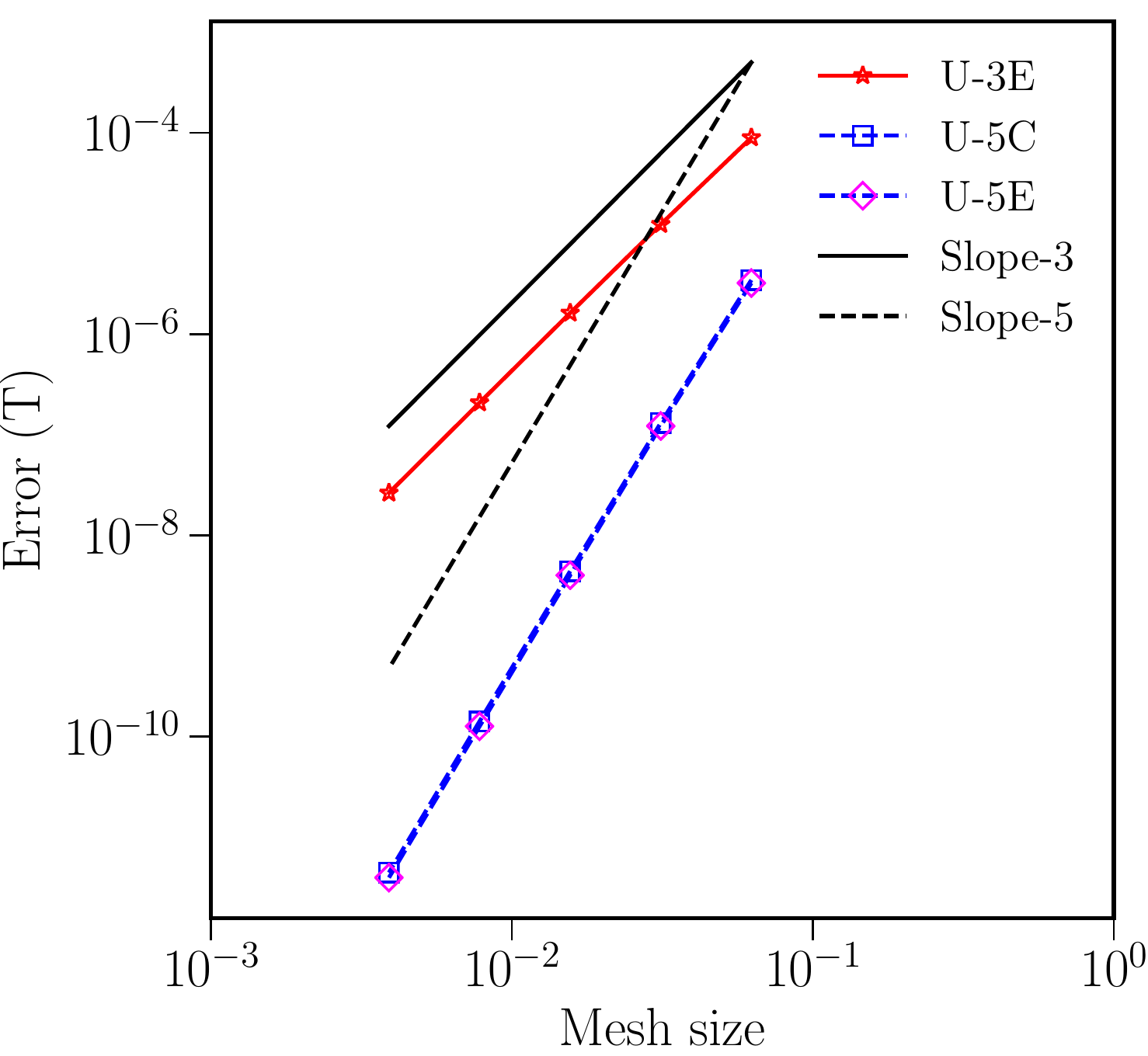}
\label{fig:t_1}}
\subfigure[]{\includegraphics[width=0.4\textwidth]{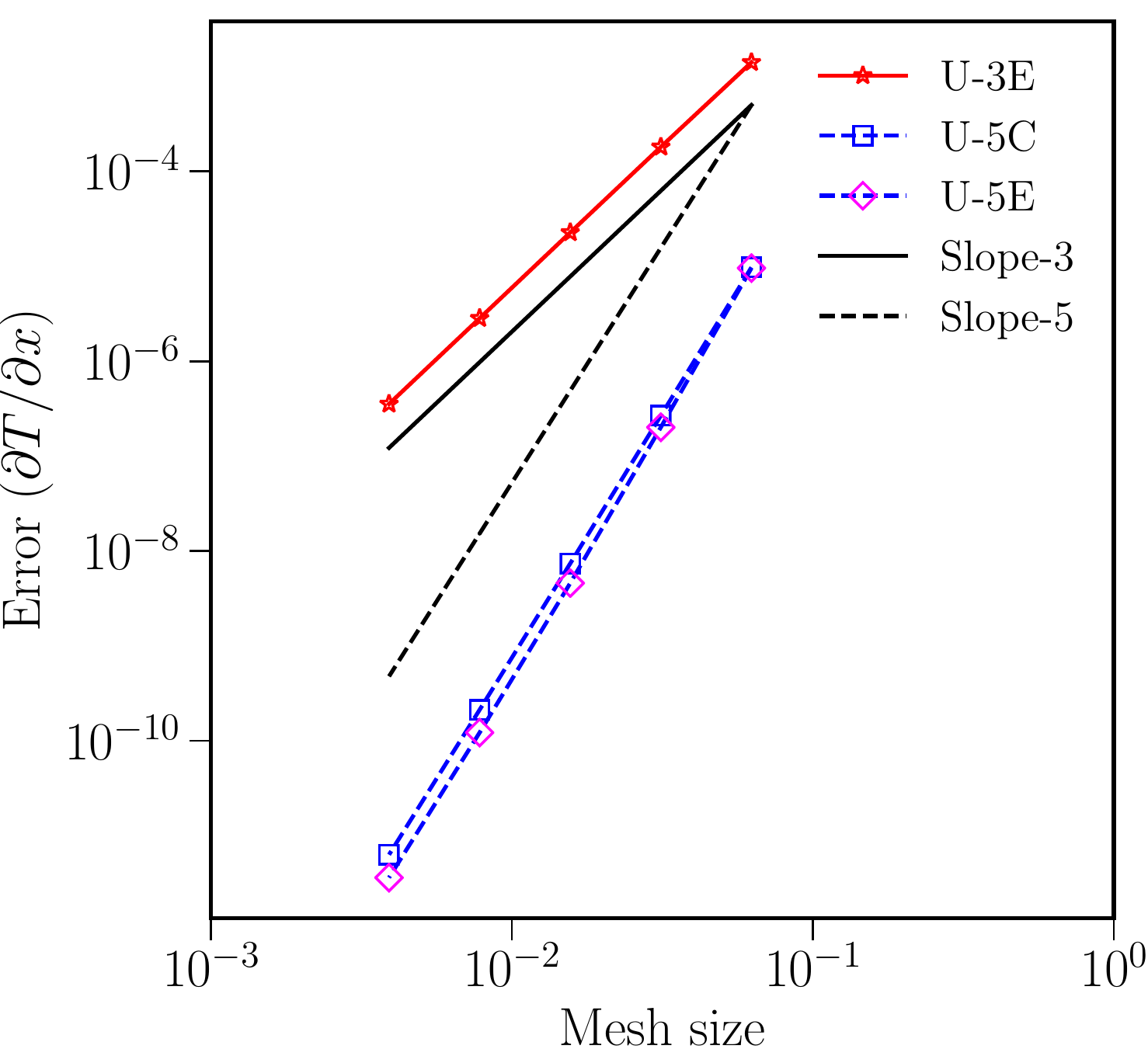}
\label{fig:g_1}}
\subfigure[]{\includegraphics[width=0.4\textwidth]{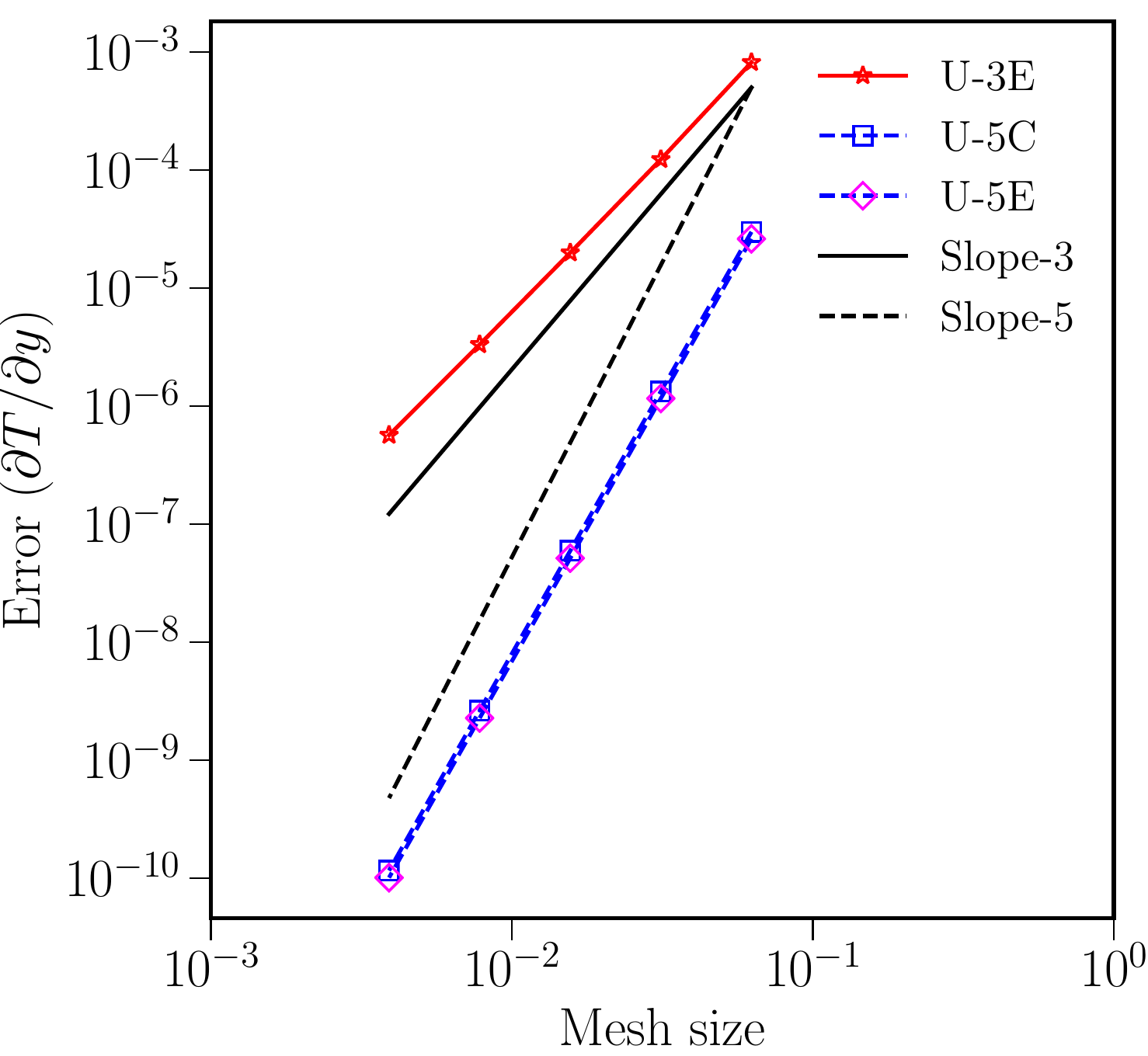}
\label{fig:h_1}}
\subfigure[]{\includegraphics[width=0.4\textwidth]{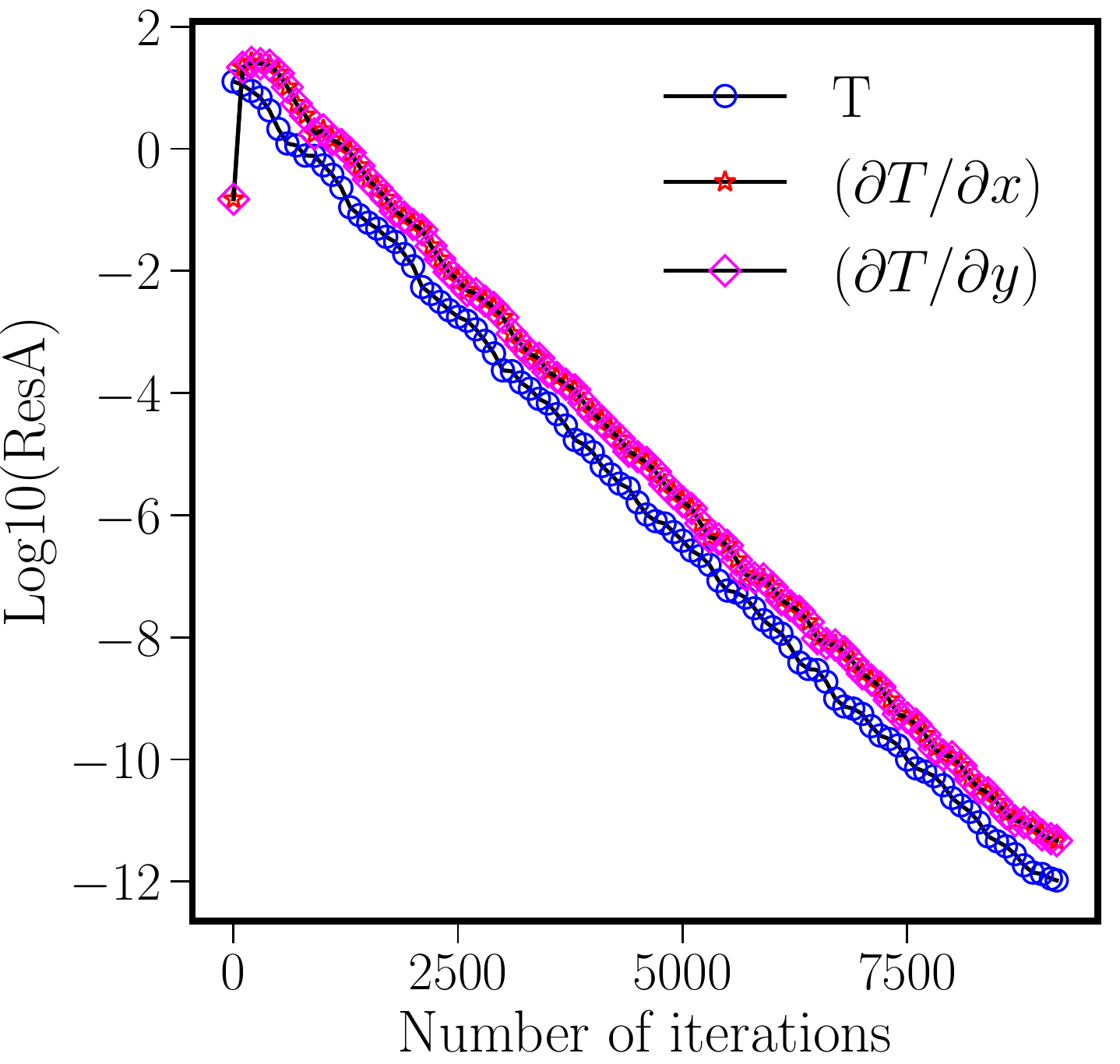}
\label{fig:r_1}}
  \caption{(a), (b) and (c) show $L_2$ errors of T, $\frac{\partial T}{\partial x}$, and $\frac{\partial T}{\partial y}$ respectively for upwind schemes for an anisotropy of $10^9$. Fig. (d) shows iterative convergence of U-5C for all the variables.  \hyperref[sec:4.4a]{Example 4.4a}.}
    \label{fig:non-1}
\end{figure}

\noindent {\textbf{Example 4.4b}}\label{sec:4.4b} In the final test case we consider the following full diffusion tensor, 

\begin{equation}\label{eq:aniso-non-2}
\begin{aligned}
\mathbf{D} &= \left[ \begin{array}{ccc}
D_{xx} & D_{xy} \\
D_{yx} & D_{yy} \\
\end{array} \right] = {\Theta^{-1}} \left[ \begin{array}{ccc}
 D_{||} & 0 \\
0 &  D_{\perp}(1+g+h)  \\
\end{array} \right]\   \Theta, \ \textrm{where} \  \Theta =  \left[
   \begin{array}{cccc}
    \cos\beta  & -\sin\beta \\
    \sin\beta & \cos\beta
   \end{array}
   \right]   \\ 
\\ \\
%&=\left[\begin{matrix}D_{||} \cos^{2}{\left (\beta \right )} + \left(g + h + 1\right) \sin^{2}{\left (\beta \right )} & \frac{1}{2} \left(D_{||} - g - h - 1\right) \sin{\left (2 \beta \right )}\\\frac{1}{2} \left(D_{||} - g - h - 1\right) \sin{\left (2 \beta \right )} & D_{||} \sin^{2}{\left (\beta \right )} + \left(g + h + 1\right) \cos^{2}{\left (\beta \right )}\end{matrix}\right].
\end{aligned}
\end{equation}
%where
%\begin{equation}
%
%   \end{equation}
and $g$ and $h$ are temperature gradients $\frac{\partial T}{\partial x}$ and $\frac{\partial T}{\partial y}$ respectively. Such temperature dependence is proposed by Lafleur et al. \cite{lafleur2016theory} for magnetized plasmas in Hall thrusters. The exact solution for this problem is the same as that of the \hyperref[sec:4.2a]{Example 4.2a}. Table \ref{tab:aniso_Hall} shows the $L_2$ errors and order of convergence, and we can observe that the design order of accuracy is obtained for all the numerical schemes. There is no qualitative difference between the nonlinear problem and the results obtained in \hyperref[sec:4.2a]{Example 4.2a}.
%\begin{equation}
%\left[\begin{matrix}D_{||} \cos^{2}{\left (\beta \right )} + \left(g + h + 1\right) \sin^{2}{\left (\beta \right )} & \frac{1}{2} \left(D_{||} - g - h - 1\right) \sin{\left (2 \beta \right )}\\\frac{1}{2} \left(D_{||} - g - h - 1\right) \sin{\left (2 \beta \right )} & D_{||} \sin^{2}{\left (\beta \right )} + \left(g + h + 1\right) \cos^{2}{\left (\beta \right )}\end{matrix}\right]
%\end{equation}
\begin{table}[H]
  \centering
  \caption{$L_2$ errors and order of convergence for $D_{||}=10^9$, \hyperref[sec:4.4b]{Example 4.4b}.}
  \begin{onehalfspacing}
  \footnotesize
    \begin{tabular}{| c | c | c | c | c | c | c | c | c|}
\hline
\hline
    Number & \multicolumn{2}{c|}{Upwind-3E} & \multicolumn{2}{c|}{Upwind-5C} & \multicolumn{2}{c|}{Upwind-5E}  \\
    \cline{2-7}
    of points& error & order & error & order & error & order    \\
    \cline{1-7}
    \hline
    $16^2$    & 3.07E-03 &  -     & 5.99E-04 &   -    & 8.87E-04 & -  \\
    \hline
    $32^2$    & 3.80E-04 & 3.01  & 1.18E-05 & 5.67  & 2.83E-05 & 4.97 \\
    \hline
    $64^2$    & 4.73E-05 & 3.00  & 2.47E-07 & 5.58  & 9.08E-07 & 4.96 \\
    \hline
    $128^2$   & 5.86E-06 & 3.01  & 5.89E-09 & 5.39  & 2.79E-08 & 5.02 \\
    \hline
    $256^2$   & 7.28E-07 & 3.01  & 1.65E-10 & 5.16  & 8.69E-10 & 5.00 \\
    \hline
    \end{tabular}%
  \label{tab:aniso_Hall}%
  \end{onehalfspacing}
\end{table}%

\begin{figure}[H]
\centering
\subfigure[]{\includegraphics[width=0.44\textwidth]{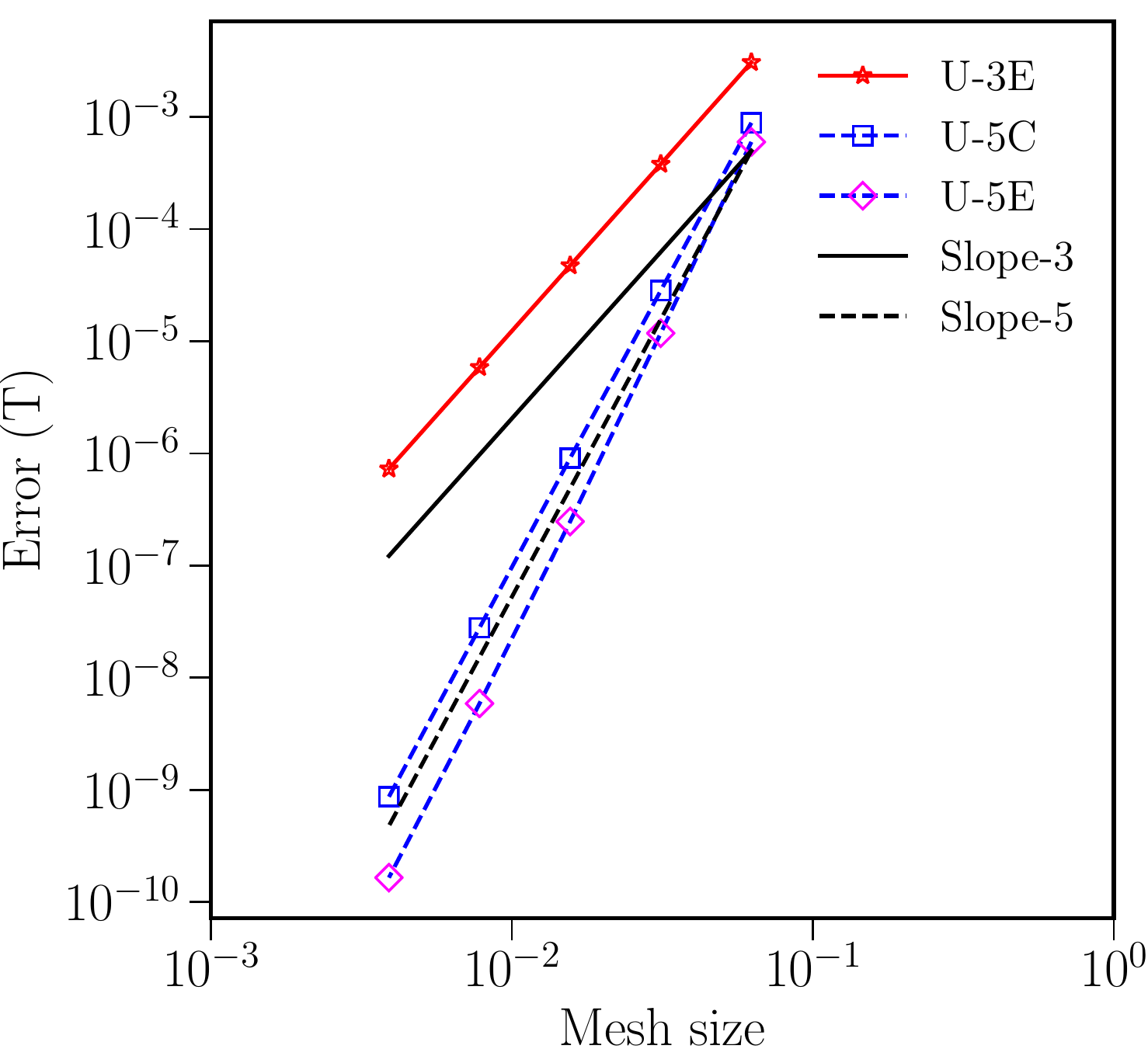}
\label{fig:t_2}}
\subfigure[]{\includegraphics[width=0.44\textwidth]{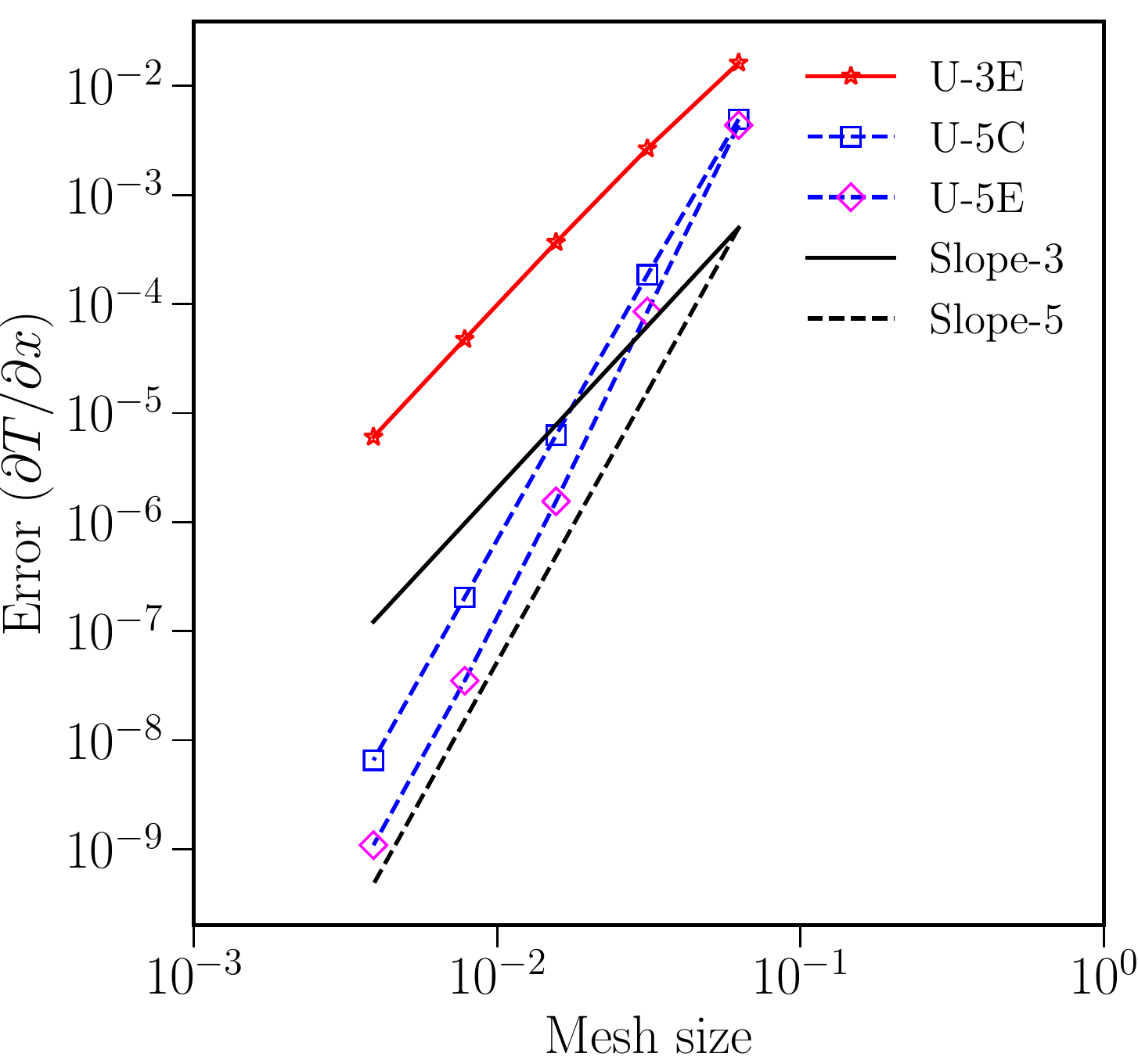}
\label{fig:g_2}}
\subfigure[]{\includegraphics[width=0.44\textwidth]{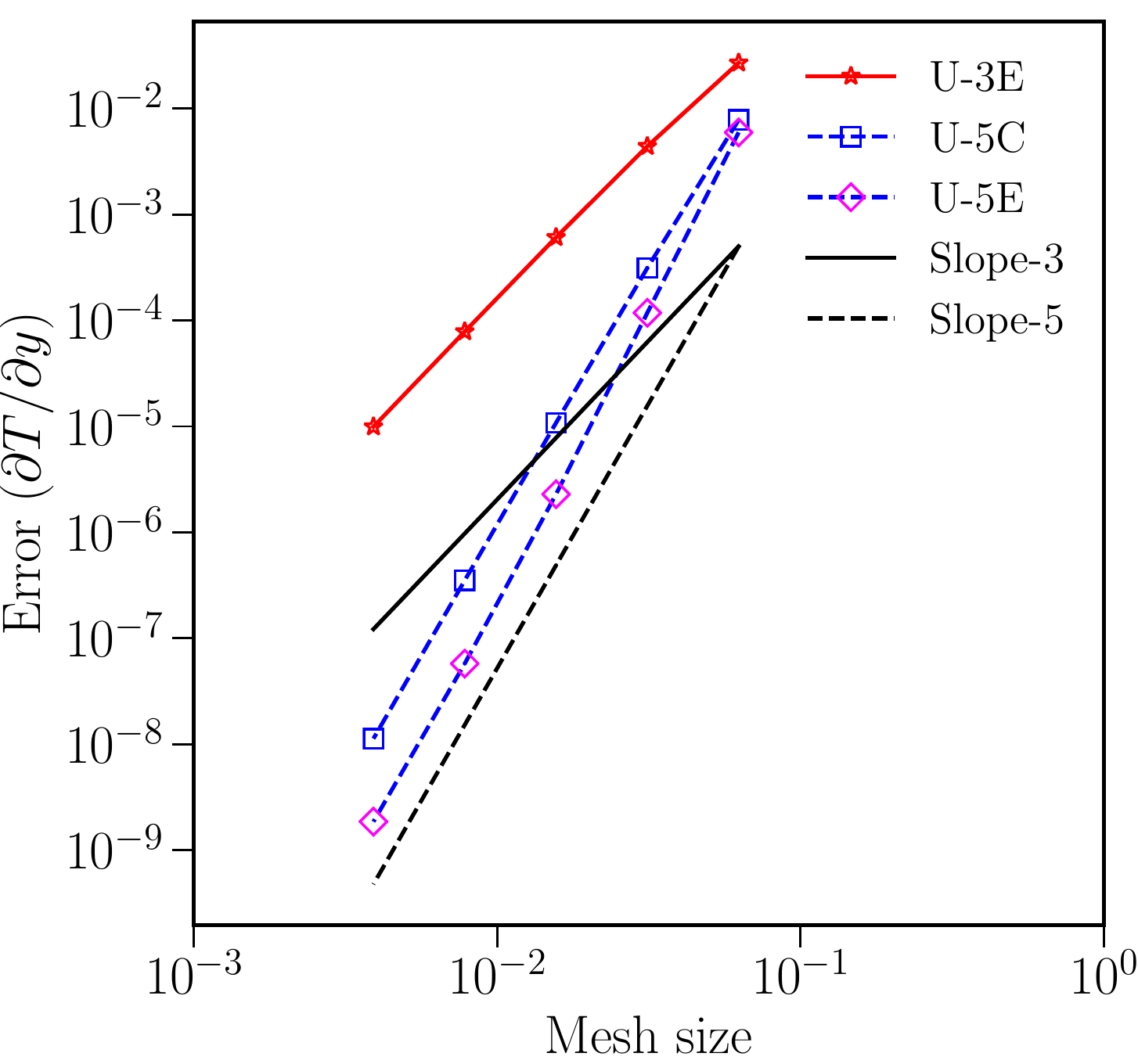}
\label{fig:h_2}}
\subfigure[]{\includegraphics[width=0.4\textwidth]{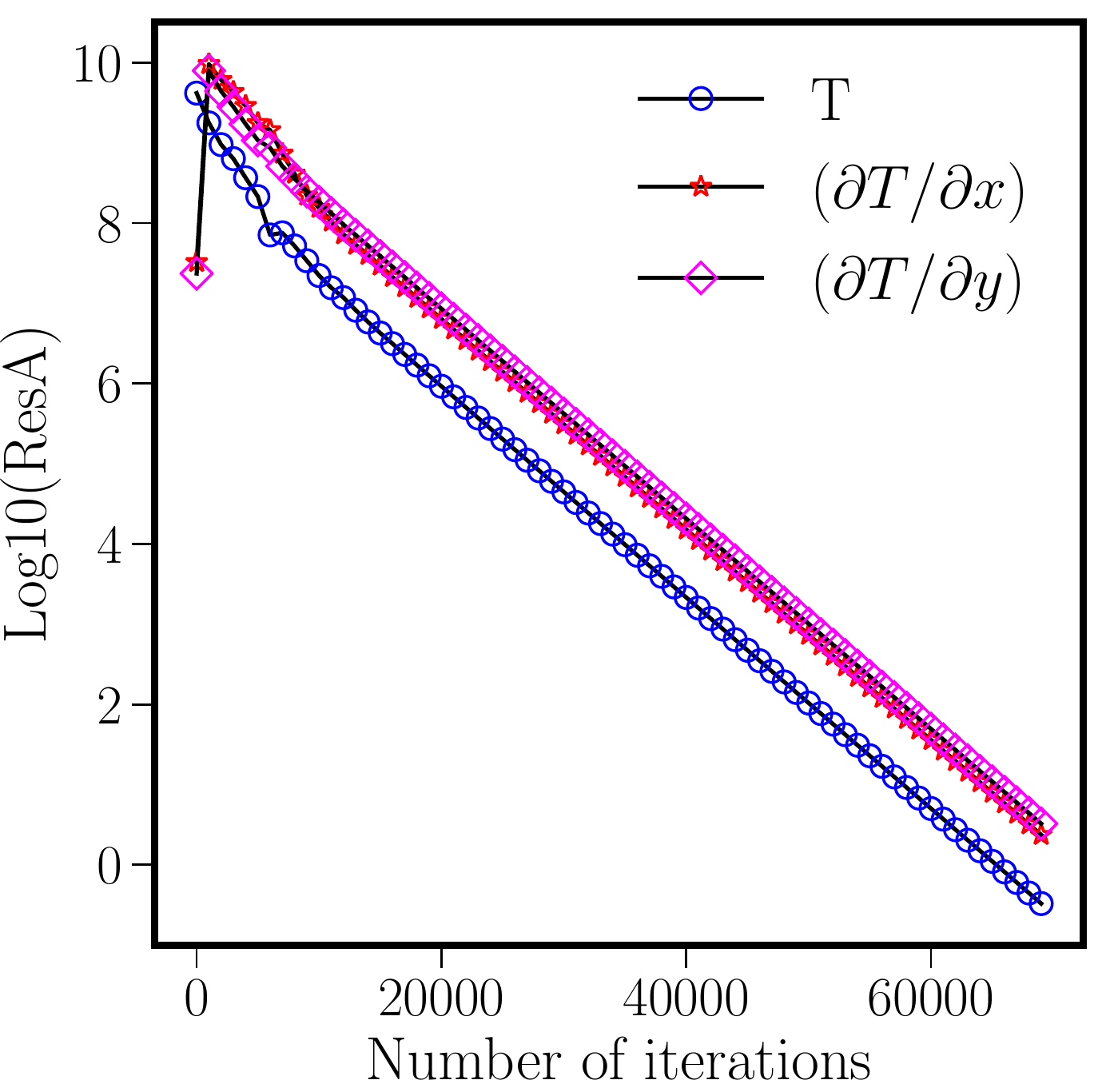}
\label{fig:r_2}}
  \caption{(a), (b) and (c) show $L_2$ errors of T, $\frac{\partial T}{\partial x}$, and $\frac{\partial T}{\partial y}$ respectively for upwind schemes for an anisotropy of $10^9$ and (d) shows iterative convergence of U-5C.   \hyperref[sec:4.4a]{Example 4.4b}.}
    \label{fig:non-2}
\end{figure}

%{\color{blue}
%The temperature dependence proposed by Lafleur et al. \cite{lafleur2016theory} is given by
%\begin{equation}
%\begin{aligned}
%D_{\perp eff} = \mu_{\perp eff} =& \frac{\frac{|q|}{m\nu_m}}{1+\frac{\omega^2}{\nu^2_m}}\left[1+\frac{\omega}{\nu_m}\frac{R_{ei}}{|q|n_eE_z}\right], \\
%\rm where \quad R_{ei} = & \frac{q}{c_s4\sqrt{6}}|\nabla.(n_eT_e)|
%\end{aligned}
%\end{equation}}
%
%\section{Conclusions}
%\begin{enumerate}
%\item Hyperbolic approach is extended to anisotropic diffusion equation and design order of accuracy is obtained for all the test cases.
%\item The objective is to simulate Hall thrusters by next year IEPC.
%\item Extend to Energy equation (2018 Dec) $\rightarrow$ Semi-Lagrangian method (WENO?)for Ions (2019 IEPC) $\rightarrow$ Self-consistent HET simulation (long term goal)
%\end{enumerate}
\subsection{ Magnetized electron test case:}
Final numerical problem is the magnetized electron test case, which has anisotropic diffusion, considered in \cite{Kawashima2015, chamarthi2018high}. 
The hyperbolized mass and momentum equations, readers can refer the aforementioned references for further details, are given by
   \begin{equation}
   \begin{aligned}
      \frac{n_{\rm e}}{T_{\rm e}}\frac{\partial \phi}{\partial t}-\nabla \cdot \left(n_{\rm e} \vec{u}_{\rm e}\right)&=-n_{\rm e}\nu_{\rm {ion}},\\
      \frac{1}{\nu_{\rm {col}}}
      \frac{\partial}{\partial t}\left(n_{\rm e}\vec{u}_{\rm e}\right)
      - n_{\rm e}\left[\mu \right]\nabla \phi+\left[\mu\right]\nabla\left(n_{\rm e}T_{\rm e}\right)&=-n_{\rm e}\vec{u}_{\rm e},
    \label{eq:mass-momentum}
  \end{aligned}
   \end{equation}
 where  \(n_{\rm e}\), \(\vec{u}_{\rm e} {\color{black} = (u_x, u_y) } \), \(\phi\), \(\nu_{\rm {ion}}\), and \(T_{\rm e}\), are the electron number density, electron velocity, space potential, {\color{black} ionization collision frequency, and electron temperature respectively}. The electron mobility tensor $\left[\mu\right]$, similar to $[D]$ in Eq.(\ref{eqn:aniso-original}), can be expressed as follows, 
   \begin{equation}
   \begin{aligned}
      \left[\mu\right]={\left[
   \begin{array}{cccc}
   \mu_{xx}  & \mu_{xy}    \\
   \mu_{xy}  & \mu_{yy}
   \end{array}
   \right]}&=\Theta^{-1}\left[
   \begin{array}{cccc}
   \mu_{||}  & \    \\
   \  & \mu_{\perp}
   \end{array}
   \right]\Theta,\hspace{20pt}
   %\end{equation}
   %\begin{equation}
   \Theta=\left[
   \begin{array}{cccc}
    \cos\theta  & -\sin\theta \\
    \sin\theta & \cos\theta
   \end{array}
   \right], \\ 
   \\
    \mu_{||}&= \frac{e}{m_{\rm e}\nu_{\rm {col}}}, \quad \mu_{\perp}=\frac{\mu_{||}}{1+\left(\mu_{||}B\right)^2}.
       \end{aligned}
          \label{eqn:mobility}
      \end{equation}        
For simplified analysis, the following values are assumed for all the test calculations,
   \begin{equation}
      {n}_{\rm e}=1,\hspace{20pt}{T}_{\rm e}=1,\hspace{20pt}{\nu}_{{\rm col}}=1,\hspace{20pt}{\nu}_{\rm {ion}}=0.
   \end{equation}

{\color{black} Then, the system (\ref{eq:mass-momentum}) can be written in the form (\ref{eqn:non_precon}) with
\begin{equation}
\begin{aligned}
&\mathbf{Q}= \left[ \begin{array}{ccc} \phi  \\ u_x \\ u_y \end{array} \right], \quad 
\mathbf{E}_x = \left[ \begin{array}{ccc} -u_x \\ - \phi \\ 0 \end{array} \right],  \quad
&\mathbf{E}_y = \left[ \begin{array}{ccc} -u_y  \\  0 \\ - \phi \end{array} \right], \quad
\mathbf{S} = \left[ \begin{array}{ccc} 0  \\  -\hat{u}_x \\ -\hat{u}_y \end{array} \right],
%      \label{vector:non_precon}
\end{aligned}
   \end{equation}
where $(\hat{u}_x,\hat{u}_y) =  \left[\mu\right]^{-1} \vec{u}_{\rm e}$, and 
   \begin{equation}
   \mathbf{ P}^{-1} = \Theta^{-1}P^{-1} \Theta, \quad
      P^{-1}=\left(
         \begin{array}{ccc}
         1 & 0 & 0   \\
         0 & \frac{T_r}{{\mu}_{||}} & 0 \\
         0 & 0 & \frac{T_r}{{\mu}_{\perp}}
         \end{array}
      \right),
      \quad
      \Theta=\left(
   		\begin{array}{ccc}
   		1 & 0 & 0   \\
   		0 & \cos\theta & \sin\theta \\
   		0 & -\sin\theta & \cos\theta
   		\end{array}
   	\right).
      \label{eqn:precon_rotation}
   \end{equation}
Note that the relaxation time $T_r$ has been introduced, which allows us to incorporate the dimensionally-consistent optimal length scale, i.e., 
Equation (\ref{optimal_Tr}) with Equation (\ref{optimal_Lr}), without changing steady solutions that we seek. In contrast, the formulation in the previous 
studies \cite{chamarthi2018high,Kawashima2015} corresponds to the following:
   \begin{equation}
   \mathbf{ P}^{-1} 
   = 
  \left[\mu\right]^{-1}
    \left(
         \begin{array}{ccc}
         1 & 0 & 0   \\
         0 & {\mu}_{xx} & 0 \\
         0 & 0 & {\mu}_{yy}
         \end{array}
      \right)
       ,
   \end{equation}
and thus no consideration was given on the optimal length scale nor the dimensional consistency. 
Even though the results obtained with the previous formulation are non-oscillatory  \cite{chamarthi2018high}, 
further studies indicated that the results are not independent of the scale.
As we demonstrate below, the optimal definition of $T_r$ is very important for achieving anisotropy- and scale-independent solutions.
Note also that the above formulation is in the preconditioned conservative form with fluxes $\mathbf{E}_x$ and $\mathbf{E}_y$, and 
therefore can be discretized straightforwardly by any method designed for conservative systems, including the method described in Section \ref{upwind_fd}.
The previous formulation, on the other hand, is in the quasi-linear form with coefficient matrices, and therefore requires the evaluation of the coefficient
matrices at each cell center. 

It should be noted that the above formulation is slightly different from Equation (\ref{eqn:non_precon}) because the diffusion tensor
is not included in the first equation: it gives
\begin{equation}\label{eqn-form-2}
\mathbf{PA}_n
=\mathbf{P}\frac{\partial(\mathbf{E}_x n_x+\mathbf{E}_y n_y)}{\partial Q} 
= \left[\begin{matrix}
0 & -n_x & -n_y\\
- \frac{\mu_{xx} n_x }{T_{r}} & - \frac{\mu_{xy} n_y }{T_{r}} & 0\\
- \frac{\mu_{xy} n_x }{T_{r}} & - \frac{\mu_{xyy} n_y }{T_{r}} & 0
\end{matrix}\right],
\end{equation}
where $\mathbf{P}$ is defined by Equation (\ref{eqn:precon_rotation}).
Fortunately, it leads to the same set of eigenvalues (\ref{M_eigenvalues}) for the matrix ${\bf M}$ in Equation (\ref{fourier_transformed}) when the Fourier mode is substituted
 (although the matrix itself is different).
Therefore, the relaxation time as derived in Equation (\ref{optimal_Tr}) is optimal for this system also.
The dissipation term is, however, slightly different from Equation (\ref{eqn:flux-jacob}) because, although the eigenvalues are the same,
\begin{equation}
\mathbf{\Lambda_n} = \lambda
\left[\begin{matrix}-1 & 0 & 0\\0 & 0 & 0\\0 & 0 & 1\end{matrix}\right], \quad 
\lambda =  \sqrt{\frac{1}{T_{r}} \left(\mu_{xx} n_{x}^{2} + 2 \mu_{xy} n_{x} n_{y} + \mu_{yy} n_{y}^{2}\right)} ,
\end{equation}
the right-eigenvectors are different:
\begin{equation}
\mathbf{R}_n= \left[\begin{matrix}
\frac{  \mu_{xx} n_{x}^{2} + 2 \mu_{xy} n_{x} n_{y} + \mu_{yy} n_{y}^{2}   }{ \lambda }  & 0 & -  \frac{  \mu_{xx} n_{x}^{2} + 2 \mu_{xy} n_{x} n_{y} + \mu_{yy} n_{y}^{2}   }{ \lambda } \\
 \mu_{xx} n_{x}  +  \mu_{xy}   n_{y}    &   -n_y & \mu_{xx} n_{x}  +  \mu_{xy}   n_{y}   \\
  \mu_{xy} n_x + \mu_{yy} n_ y & n_x &  \mu_{xy} n_x + \mu_{yy} n_ y 
 \end{matrix}\right],
\end{equation}
which results in the following dissipation matrix:
\begin{equation}
{\bf P}^{-1} |\mathbf{PA}_n|  
=  {\bf P}^{-1}  \mathbf{R}_n|\mathbf{\Lambda}_n| \mathbf{R}^{-1}_n
=
 \lambda \left[\begin{matrix}
 1 &   0 &   0 \\  
0 & \frac{ n_x^2}{  \lambda^2 } &    \frac{ n_x n_y}{  \lambda^2 } \\  
0 &   \frac{ n_x n_y }{ \lambda^2  } &     \frac{ n_y^2}{  \lambda^2  }
\end{matrix}\right],
\label{eqn:flux-jacob2}
\end{equation}
}

\begin{figure}[htbp]
\centering
{\includegraphics[width=0.6\textwidth]{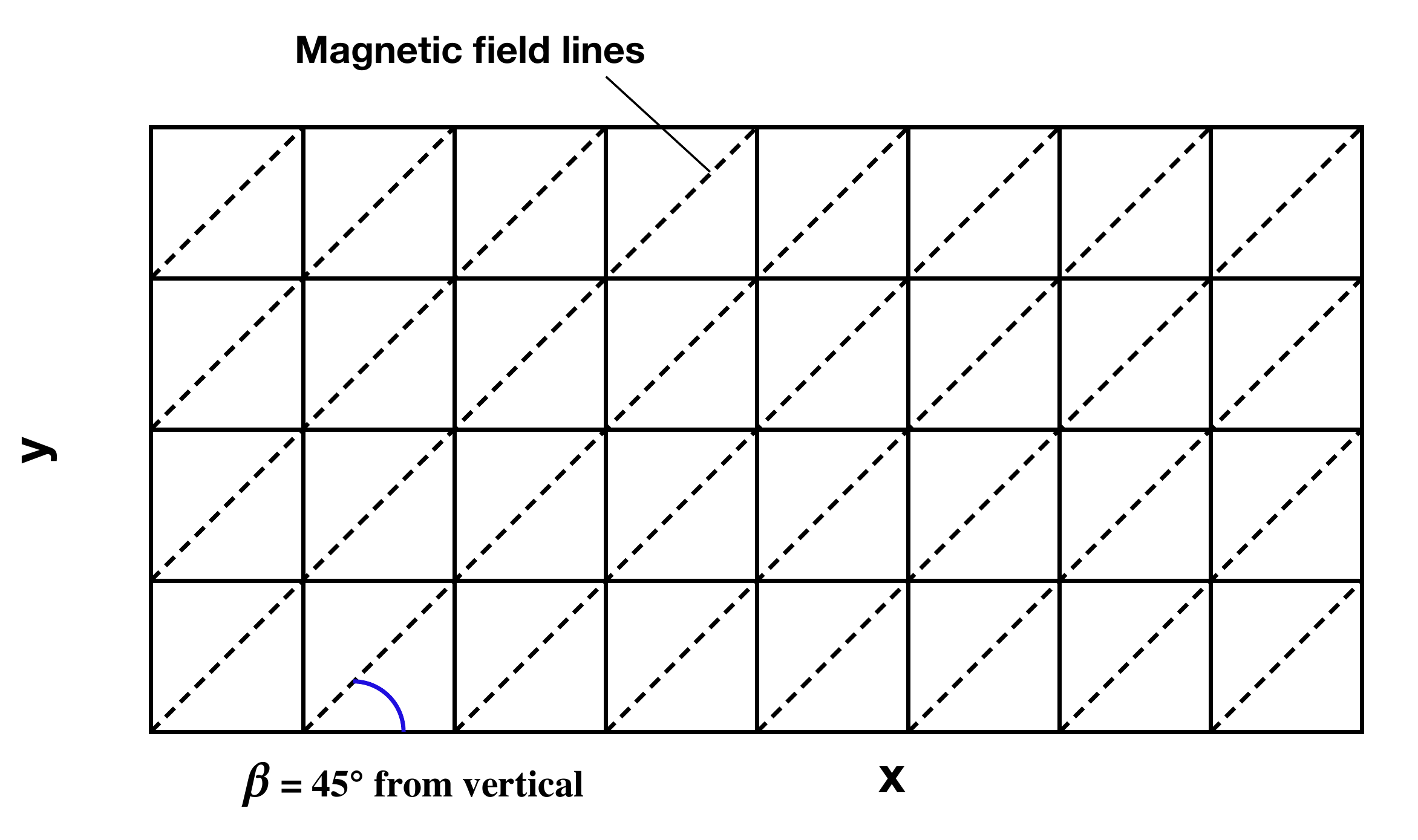}}
\caption{Sketch of the magnetic field lines for  $45^{\circ}$ angle.}
\label{fig:45}
\end{figure}

\begin{figure}[H]
\centering
\begin{onehalfspacing}
\subfigure[]{\includegraphics[width=0.48\textwidth]{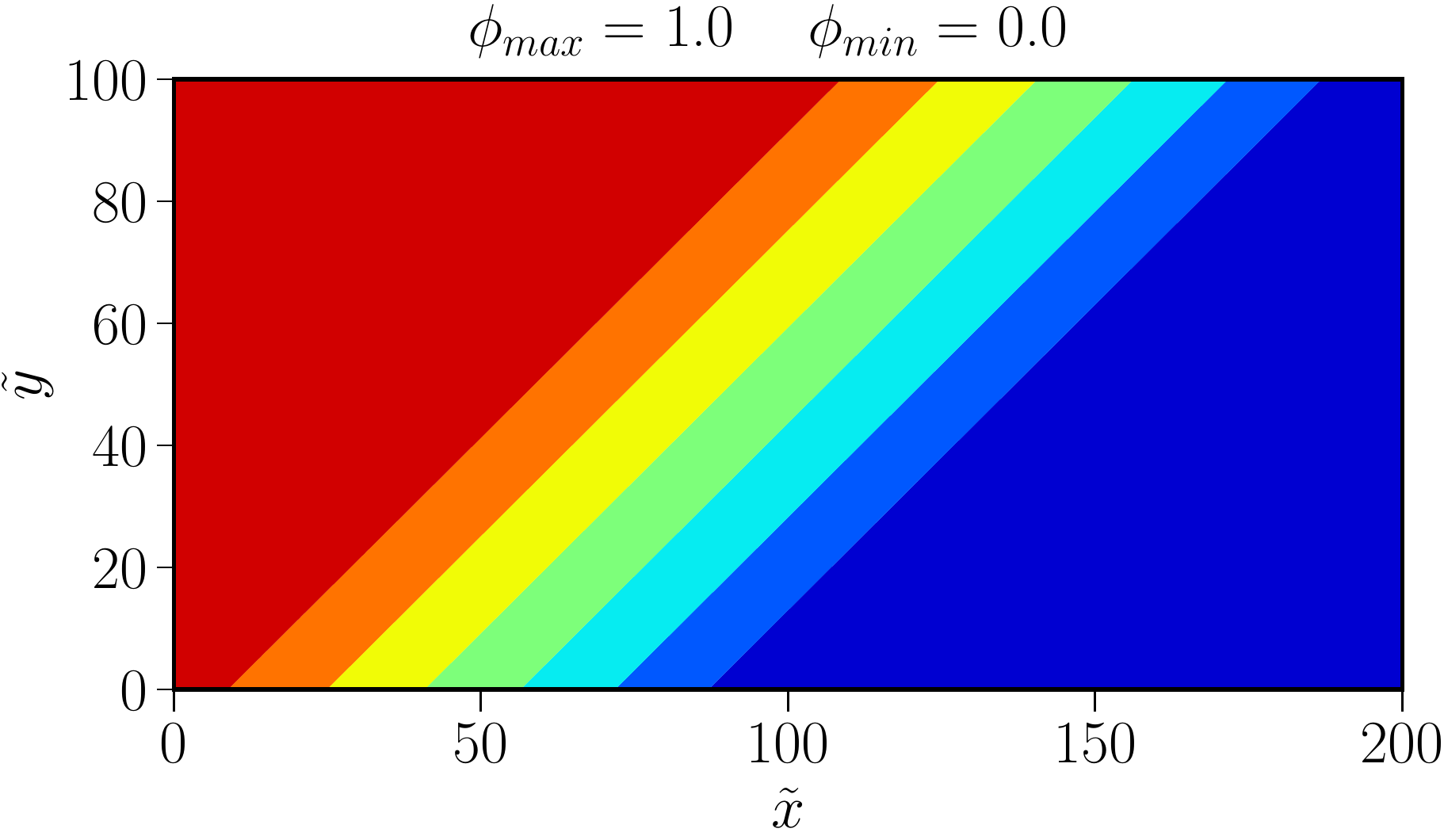}
\label{fig:space_large}}
\subfigure[]{\includegraphics[width=0.48\textwidth]{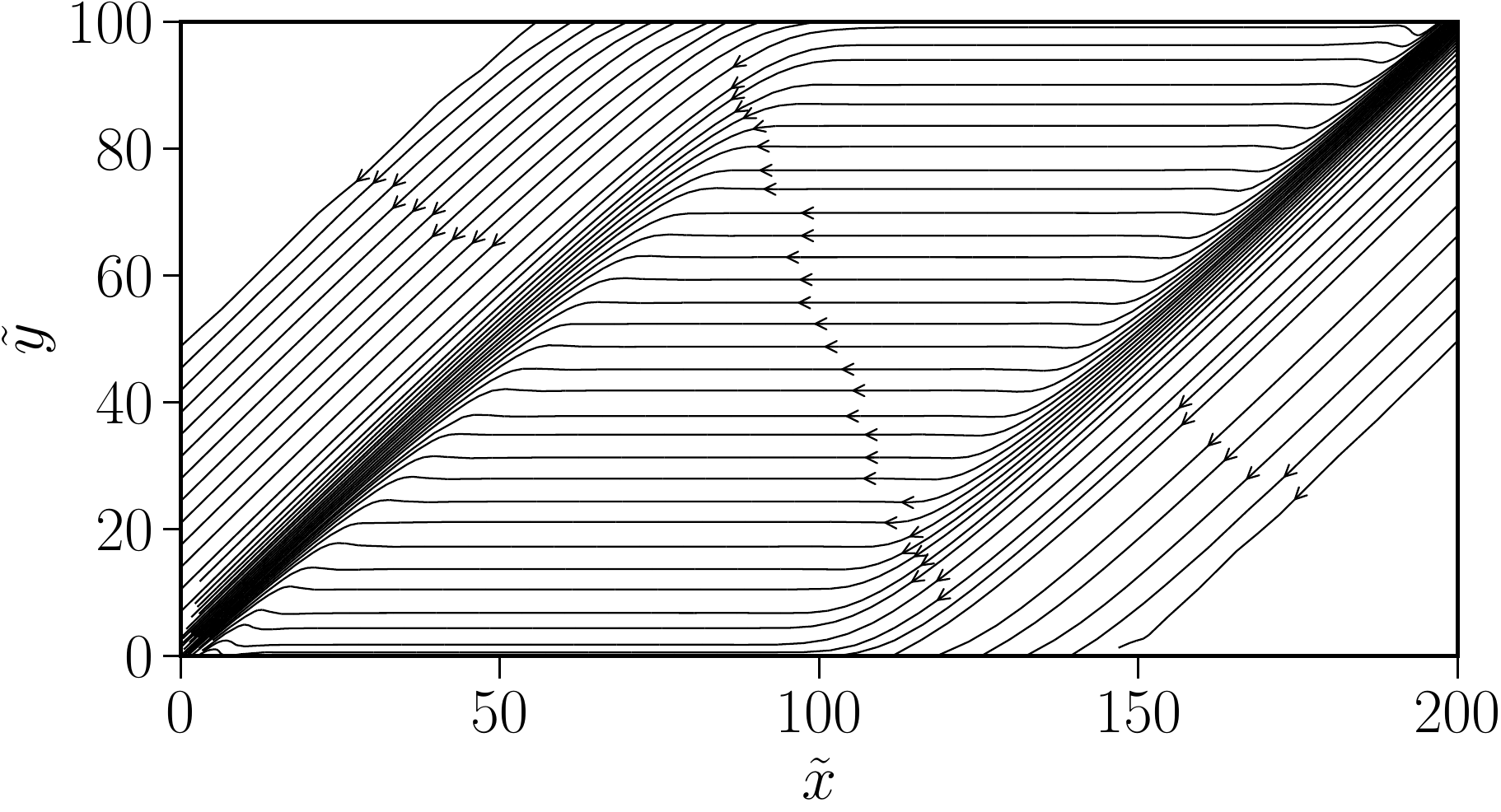}
\label{fig:ele_large}}
\subfigure[]{\includegraphics[width=0.48\textwidth]{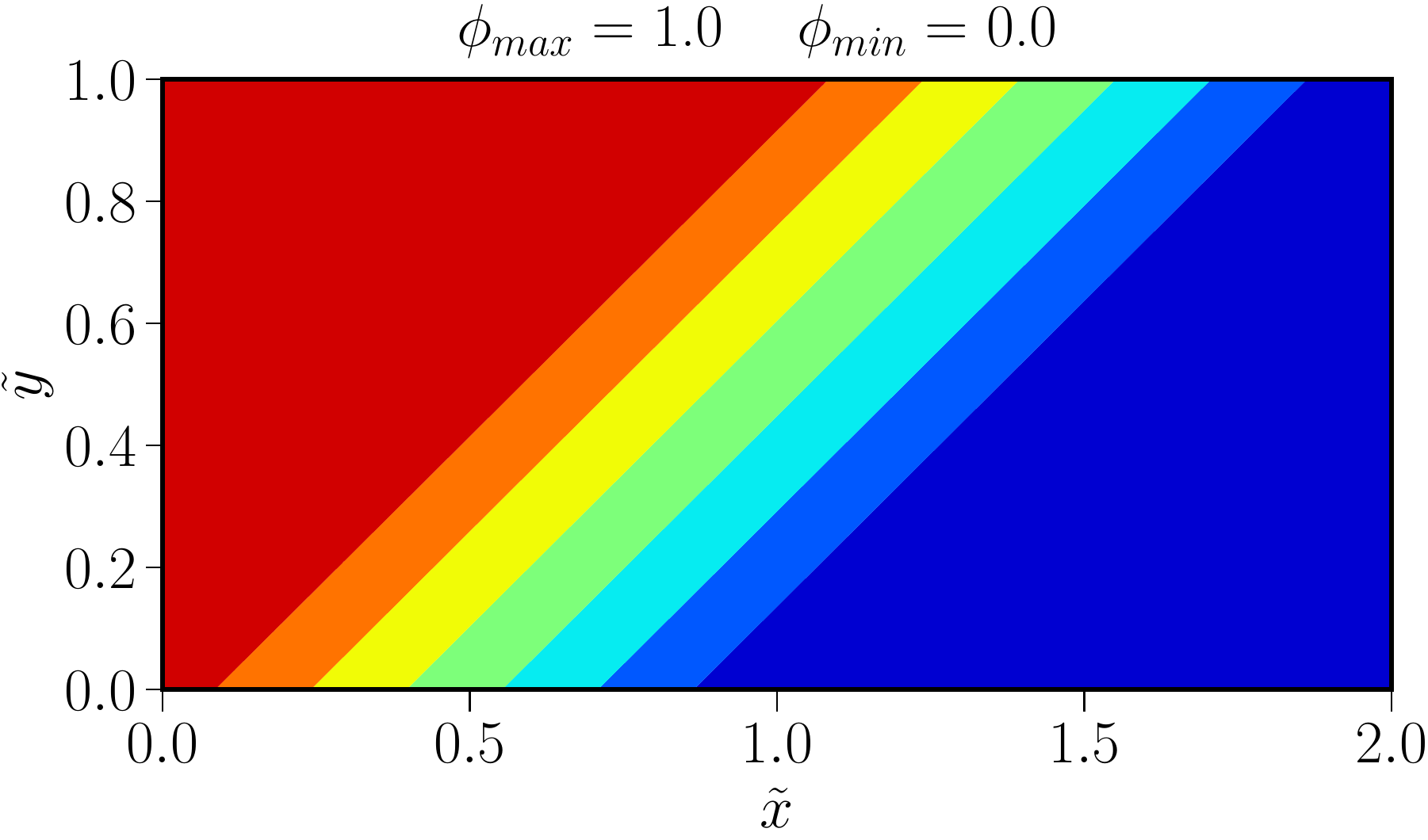}
\label{fig:space_normal}}
\subfigure[]{\includegraphics[width=0.48\textwidth]{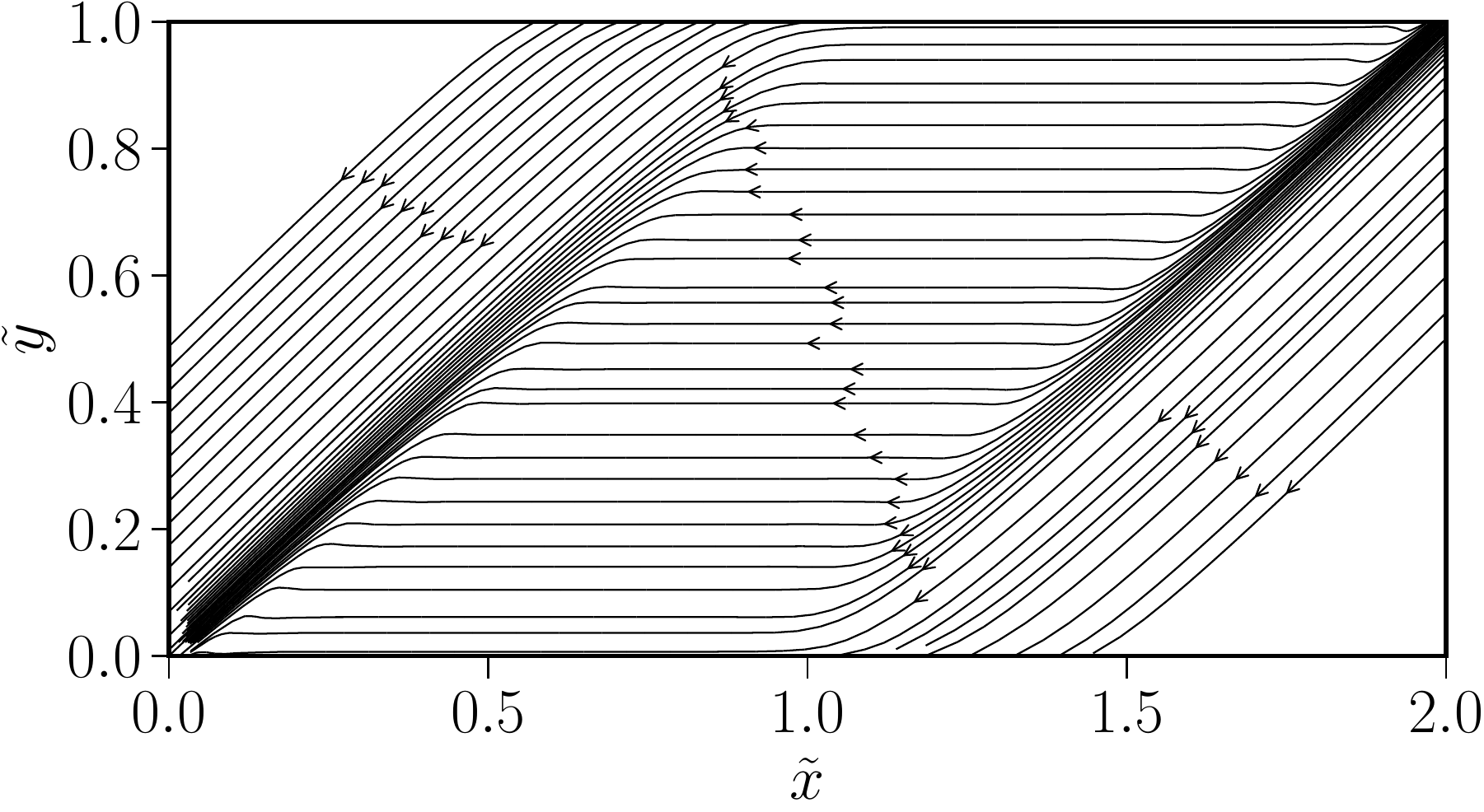}
\label{fig:ele_normal}}
\subfigure[]{\includegraphics[width=0.48\textwidth]{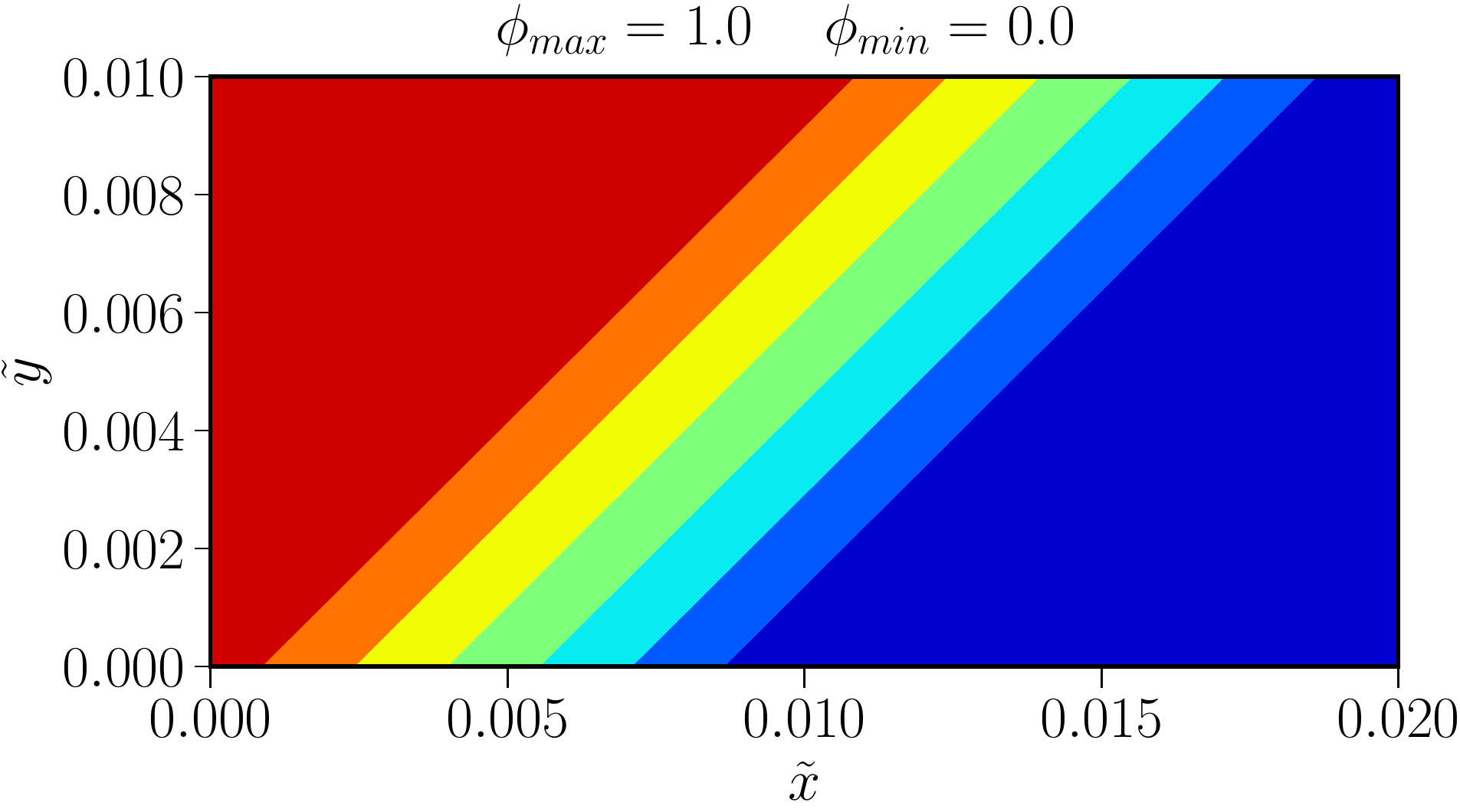}
\label{fig:space_small}}
\subfigure[]{\includegraphics[width=0.48\textwidth]{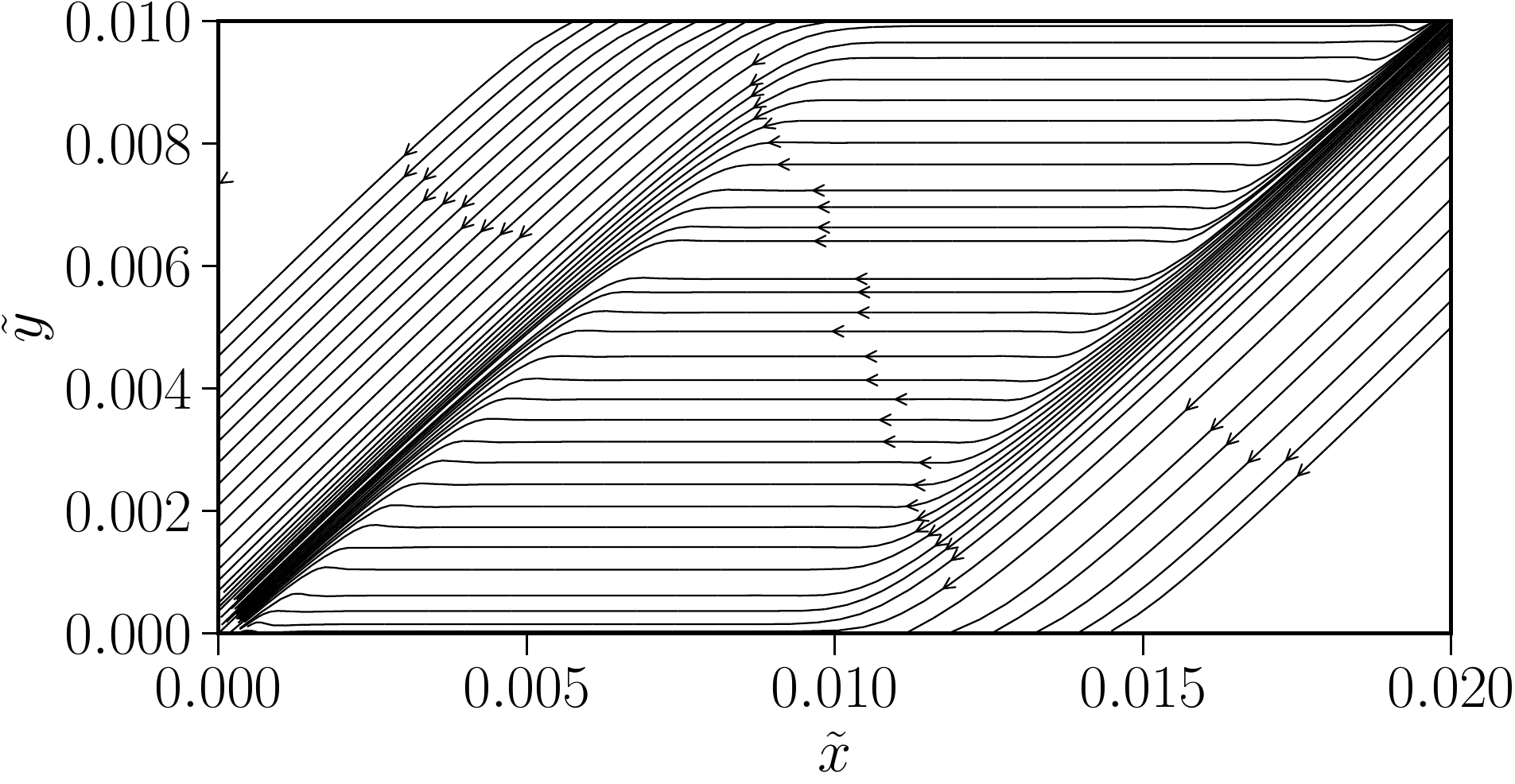}
\label{fig:ele_small}}
  \caption{Numerical results obtained by WCNS-Z for problem for $\mu_{||}/\mu_{\perp}=10^3$ on a grid size of $96 \times 96$.}
    \label{fig:new-c1}
    \end{onehalfspacing}
\end{figure}

{\color{black} We consider the test case calculated by Kawashima et al. \cite{Kawashima2015} with uniformly angled magnetic lines of force at {\color{black} $\theta = 45^{\circ}$} from the vertical, shown in Fig. \ref{fig:45}. Uniform magnetic confinement ratio, ${\mu}_{||}/{\mu}_{\perp}$ of 1000 is considered throughout the domain.  Dirichlet boundary conditions for the non-dimensional space potential are defined at the left and right side boundaries and zero-flux conditions, which are also Dirichlet, are used for the top and bottom boundaries, shown in Eq.(\ref{dirvel}).

\begin{equation}
    \tilde\phi= 
\begin{cases}
   1,&  x= 0.0\\
    0,&  x = 0.02, 2.0, 200.0
\end{cases}
\quad 
\tilde u_y = 
\begin{cases}
   0,&  y= 0.0\\
    0,&  y = 0.01, 1.0, 100.0
\end{cases}  
\label{dirvel}
\end{equation}
}
The simulations are carried out by WCNS-Z scheme, as it is shown in Ref. \cite{chamarthi2018high} that linear schemes can cause spurious oscillations. The numerical results computed on a 96 $\times$ 96 grid for a mobility ratio of $10^3$. Once again, WCNS-Z scheme along WENO extrapolation (see \cite{chamarthi2018high,Tan2010, Tan2011} for further details) are found suitable to reduce the spurious oscillations, as seen in Figs. \ref{fig:new-c1}. The numerical results are invariant of the dimensions, and also the positivity of the space potential is also preserved. {\color{black} The domain size of $[0,200] \times [0,100]$ is 
considered in Refs.\cite{Kawashima2015, chamarthi2018high, Kawashima2016}. Our scheme yields the same solution independent of the domain size.}

%\begin{figure}[H]
%\centering
%\begin{onehalfspacing}
%\subfigure[Velocity streamlines]{\includegraphics[width=0.44\textwidth]{Electron_small.pdf}
%\label{fig:new_stream_num1}}
%\subfigure[Space potential]{\includegraphics[width=0.44\textwidth]{Electron_small.pdf}
%\label{fig:New_potential_ex1}}
%  \caption{ $\beta$ =  $45^{\circ}$ Numerical results obtained by WENO-5Z-W for problem for $\mu_{||}/\mu_{\perp}=10^3$ on a grid size of $96 \times 96$.}
%    \label{fig:new-c}
%    \end{onehalfspacing}
%\end{figure}
%

%\begin{figure}[H]
%\centering
%\begin{onehalfspacing}
%
%
%
%  \caption{$\beta$ =  $0^{\circ}$ Numerical results obtained by WENO-5Z-W for problem for $\mu_{||}/\mu_{\perp}=10^3$ on a grid size of $96 \times 96$.}
%    \label{fig:new-c2}
%    \end{onehalfspacing}
%\end{figure}

\section{Conclusions}
The important findings of the paper are summarized as follows

\begin{enumerate}
\item Optimal $T_r$ and $L_r$, preconditioning matrix, are derived through Fourier analysis, and the numerical scheme found to be independent of the degree of anisotropy as well as the angle of misalignment. The preconditioned formulation is suitable for solving variable-coefficient and nonlinear anisotropic diffusion equation. 
\item Consistent third and fifth order accuracies are obtained for anisotropic diffusion by the hyperbolic approach using finite difference schemes. Very high-order compact schemes are found to improve the accuracy for most of the test cases significantly. 
\item  Results for magnetized electron test case, with the updated preconditioning matrix along with the optimal length scale formula, are oscillation free for wide-ranging dimensions. It also reiterates the conclusions given in \cite{chamarthi2018high} that WENO interpolation approach along with WENO extrapolation for boundary conditions can result in oscillation free results. Dimensionally consistent formulation given in \cite{nishikawa2018dimensional} is also required for scale-invariant results.
\end{enumerate}
{\color{black}The extension to curvilinear coordinates using WCNS\cite{Nonomura2010} and field aligned unstructured meshes by discontinuous galerkin methods\cite{Lou2019,huynh2007flux} is currently ongoing and the results will be discussed elsewhere.}

\section*{Acknowledgements}
A.S. is supported by Monbukagakusho fellowship during this work.

\section*{Appendix A}

In this appendix, the fifth order weighted nonlinear compact scheme is explained. For simplicity,  the interpolation polynomials to the left side of the cell interface at $x_{j+\frac{1}{2}}$ are only presented here. The WCNS numerical flux is constructed as follows:
\begin{equation}
{Q}^L_{j+\frac{1}{2}} = \sum\limits_{k=0}^{2} \omega_k {Q}_{j+\frac{1}{2}}^{(k)},
\end{equation}
where
\begin{equation}
\begin{aligned} \label{eq:upwind_biased_stencils}
{Q}_{j+\frac{1}{2}}^{(0)} &= \frac{1}{8}\left(3Q_{j-2} - 10Q_{j-1} + 15Q_{j} \right),\\
{Q}_{j+\frac{1}{2}}^{(1)} &= \frac{1}{8}\left(-Q_{j-1} + 6Q_{j} + 3Q_{j+1} \right), \\
{Q}_{j+\frac{1}{2}}^{(2)} &= \frac{1}{8}\left(3Q_{j} + 6Q_{j+1} - Q_{j+2} \right),
\end{aligned}
\end{equation}
and $\omega_k$ are the nonlinear weights. These nonlinear weights are computed as follows:
\begin{equation} \label{eq:JS5_nonlinear_weights}
	\omega_k = \frac{\alpha_k}{\sum\limits_{k=0}^{2}\alpha_k}, \:
   \textrm{where} \ \alpha_k = \frac{C_k}{\left(\beta_k + \epsilon \right)^m}, \: k = 0, 1, 2,
\end{equation}
and
\begin{equation} \label{eq:linear_upwind_weights}
C_0 = \frac{1}{16}, \:
C_1 = \frac{10}{16}, \:
C_2 = \frac{5}{16},  
\end{equation}
\begin{equation}
\begin{aligned} \label{eq:smoothness}
\beta_0 &= \frac{1}{4} \left( Q_{i-2} - 4 Q_{i-1} +3 Q_{i} \right)^2 + \frac{13}{12} \left( Q_{i-2} - 2 Q_{i-1} + Q_{i} \right)^2\\
\beta_1 &= \frac{1}{4} \left( Q_{i-1} - Q_{i+1} \right)^2  + \frac{13}{12} \left( Q_{i-1} - 2 Q_{i} + Q_{i+1} \right)^2 \\
\beta_2 &= \frac{1}{4} \left( 3Q_{i} - 4 Q_{i+1} +Q_{i+2} \right)^2+ \frac{13}{12} \left( Q_{i} - 2 Q_{i+1} + Q_{i+2} \right)^2.
\end{aligned}
\end{equation}
Here $\epsilon=10^{-6}$ is a small constant to prevent division by zero, and $m$ is equal to 2. In this paper, we considered an improved version of WCNS proposed by Borges et al. in \cite{Borges2008}. The improved nonlinear weights are as follows:

\begin{align}\label{om_z}
\omega^z_k=\dfrac{\alpha^z_k}{\sum_{k=0}^2\alpha^z_k},\quad
\alpha^z_{k}=\gamma_k\left(1+\left(\dfrac{\tau}{\epsilon+\beta_k}\right)^p\right),
\end{align}
 \begin{align}\label{tau_z}
\tau=|\beta_0-\beta_2|,
\end{align}
where the smoothness indicators $\beta_k$'s are the same as those given in Equations (\ref{eq:smoothness}), $\epsilon=10^{-40}$, and $\tau$ is the smoothness indicator of the large stencil. The variable $p$ is used to tune the dispersive and dissipative properties of the scheme. It is reported by Borges et al. \cite{Borges2008} that the scheme becomes more dissipative when $p$ is increased. In this paper, denoted as WCNS-Z, \(p\)= 2 is employed for the magnetized test case for optimal fifth order accuracy.

\section{References}
\bibliographystyle{elsarticle-num}
\bibliography{Anisotropic_Diffusion_final.bbl}
\end{document}